\documentclass[multphys,vecphys]{svmult}
\usepackage{makeidx}     % allows index generation
\usepackage{graphicx}    % standard LaTeX graphics tool
\usepackage{multicol}    % used for the two-column index
\begin{document}
%
%%%%%%%%%%%%%%%%%%%%%%%%%%%%%%%%%%%%make title here
%\include{tableofcontents}
%***************************************
%\newcommand{\greeksym}[1]{{\usefont{U}{psy}{m}{n}#1}}
%\newcommand{\umu}{\mbox{\greeksym{m}}}
%\newcommand{\udelta}{\mbox{\greeksym{d}}}
\newcommand{\uDelta}{\mbox{\greeksym{D}}}
\newcommand{\uPi}{\mbox{\greeksym{P}}}
\newcommand{\mug}{\mbox{\greeksym{m}}}
\newcommand{\mugp}{\mbox{\greeksym{m}}}
\newcommand{\Omg}{\mbox{\greeksym{W}}}
\setcounter{secnumdepth}{3} %%%\subsubsection-Zaehler
%%Fuer LaTex --> geraden Seitenbeginn
\catcode`@=11
\def\chapter{\thispagestyle{empty}
   \global\@topnum\z@\@afterindentfalse
   \secdef\@chapter\@schapter}
\catcode`@=12
%%%%%
\def\fract#1#2{{\textstyle\frac{#1}{#2}}}
\def\fracd#1#2{{\displaystyle\frac{#1}{#2}}}
\def\beq{\begin{eqnarray*}}
\def\eeq{\end{eqnarray*}}
\def\lz{\vspace{\baselineskip}}
\def\hlz{\vspace{0.5\baselineskip}}
\def\hhlz{\vspace{0.25\baselineskip}}
\def\lzn{\vspace{\baselineskip}\noindent}
\def\hlzn{\vspace{0.5\baselineskip}\noindent}
\def\hhlzn{\vspace{0.25\baselineskip}\noindent}
\def\rlz{\vspace{-\baselineskip}}
\def\rhlz{\vspace{-0.5\baselineskip}}
\def\rhhlz{\vspace{-0.25\baselineskip}}
\def\rlzn{\vspace{-\baselineskip}\noindent}
\def\rhlzn{\vspace{-0.5\baselineskip}\noindent}
\def\rhhlzn{\vspace{-0.25\baselineskip}\noindent}
\def\tbee{\hspace*{22.25pt}} %%Einzug fuer z.B. 1.2
\def\tbez{\hspace*{29pt}}    %%Einzug fuer z.B.
\def\tbze{\hspace*{29pt}}    %%Einzug fuer z.B. 1.2
\def\tbec{\hspace*{17.60pt}} %%Einzug fuer z.B. 1.
\def\tce{\hspace*{34.389pt}} %%Einzug fuer z.B. 11.2.1
\def\tcz{\hspace*{40.139pt}} %%Einzug fuer z.B. 11.12.1
\def\endd{\end{document}}
%uebernommen aus math.def
%\input fig.def%% fuer a,b usw.
%
%\textheight=19cm
%\nopagecropped
%\overfullrule=4pt
\overfullrule=0pt
%%%falls Text hoeher als Bild ==> Text unterhalb Bild
\def\abfa{\newdimen\figgap\figgap=18pt}
\def\abfn{\newdimen\figgap\figgap=300pt}
\def\intl{\int\limits}
\newcommand{\solm}{M$_\odot$}
\newcommand{\MOLH}{H$_2$}
\newcommand{\TKIN}{T_{\rm k}}
\newcommand{\kms}{km\,s$^{-1}$}
\newcommand{\kmsMpc}{km\,s$^{-1}$\,Mpc$^{-1}$}
\newcommand{\WAT}{H$_{2}$O}
\newcommand{\FORM}{H$_{2}$CO}
\newcommand{\AMM}{NH$_3$}
\newcommand{\percc}{cm$^{-3}$}
\newcommand{\pers}{s$^{-1}$}
\newcommand{\HCOP}{HCO$^+$}
\newcommand{\HTHPL}{H$_3^+$}
\newcommand{\etal}{et al.}
\newcommand{\TEX}{T_{\rm ex}}
\newcommand{\TMB}{T_{\rm MB}}
\newcommand{\Tsys}{T_{\rm sys}}
\newcommand{\CEIO}{C$^{18}$O}
\newcommand{\percmsq}{cm$^{-2}$}
\newcommand{\METH}{CH$_3$OH}
\newcommand{\CYAN}{HC$_3$N}
\newcommand{\dvel}{$\Delta {\rm V}_{1/2}$}
\newcommand{\vlsr}{V$_{\rm lsr}$}
\newcommand{\TROT}{T_{\rm rot}}
\newcommand{\CHTHCCH}{CH$_3$CCH}
\newcommand{\CHTHCN}{CH$_3$CN}
\newcommand{\solmass}{M$_\odot$}
\newcommand{\DGC}{D$_{\rm GC}$}
\newcommand{\THCO}{$^{13}$CO}
\newcommand{\HI}{H\,\mbox{\sc i}}
\newcommand{\HII}{H\,\mbox{\sc ii}}
\newcommand{\hii}{H\,\mbox{\sc ii}}
\newcommand{\tb}{T_{\rm b}}
\newcommand{\ta}{T_{\rm A}}
\newcommand{\tsp}{T_{\rm s}}
\newcommand{\Tb}{T_{\rm b}}
\newcommand{\Ta}{T_{\rm A}}
\newcommand{\Tsp}{T_{\rm s}}
\newcommand{\te}{T_{\rm e}}
\newcommand{\expo}{{\rm \,e}}
\newcommand{\diff}{{\rm \, d}}
\newcommand{\im}{{\rm \,i\,}}
\renewcommand{\Re}{{\rm Re}}
\renewcommand{\Im}{{\rm Im}}
\newcommand{\pl}{\parallel}
\newcommand{\sha}{\raisebox{-0.2ex}{\kern1pt\rm III\,}}
\newcommand{\ddel}{\raisebox{-0.2ex}{\rm II\,}}
\newcommand{\la}{\lesssim}
\newcommand{\ga}{\gtrapprox}
\newcommand{\gequal}{\mathrel{\stackrel{>}{=}}}
\newcommand{\lequal}{\mathrel{\stackrel{<}{=}}}
\newcommand{\longpage}{\enlargethispage{\baselineskip}}
\newcommand{\shortpage}{\enlargethispage{-\baselineskip}}
\def\endd{\end{document}}
\def\enda{\end{document}}
\def\epsilon{\varepsilon}
\def\degr{\hbox{$^\circ$}}
\def\arcmin{\hbox{$^\prime$}}
\def\arcsec{\hbox{$^{\prime\prime}$}}
\let\ts=\thinspace
\let\picplace=\vspace
%\numberlikebook
%****************************************************************************
%
\title*{Introduction to Millimeter/Sub-Millimeter Astronomy}
 \titlerunning{Millimeter/Sub-Millimeter Astronomy} %for an abbreviated version of  your contribution title if the original one is too long
\author{T.~L.~Wilson\inst{1,2}}
\institute{European Southern Observatory, K-Schwarzschild-Str. 2, D85748 Garching, Germany\texttt{twilson@eso.org \and Max-Planck-Institut-f-Radioastronomie}}
%\and Name and Address of your Institute \texttt{name@email.address}}
\maketitle
%\input title.tex
%\tableofcontents
%\chapter{}                     
\section{Introduction}
This chapter  provides an introduction to the following  lectures. It is based on  nine lectures given at the 
Saas Fee school in March 2008. Much, but not all of this material is contained in  \cite{RW4} and \cite{RW5}, but with more  more   emphasis on  
 current research  in millimeter/sub-millimeter astronomy. In the following, the basics of radiative transfer, receivers, antennas, interferometry  
radiation mechanisms and molecules are presented. 
The field of mm/sub-mm astronomy 
has become very large, so this presentation 
contains 
only a few derivations. The astute reader will note that Maxwell's Equations do not appear at all, although  the results are based on these.  
In some cases, the approach is to quote a  result  followed by an example.  This contribution  is aimed at graduate students with a good background in physics and astronomy.   
Some common terms are used but $''$jargon$''$ has been avoided as much as possible. References are given, for the most part, to more recent work where citations to earlier work can 
be found.  
The units are mostly CGS with some SI units. This follows the usage in the astronomy literature. 
One topic {\em not} covered here is polarization; see  \cite{RW4} and \cite{THUM} for an introduction. 

The field of mm/sub-mm astronomy began only in the 1960's, but the richness of the results has 
justified this series of lectures. Millimeter/sub-mm measurements  require excellent weather, very accurate antennas, and sensitive receivers. 
The interpretation of these data requires a knowledge of atomic and molecular physics, radiative transfer and interstellar chemistry. By number, about 90\% 
of molecular clouds consist of  H$_2$. However, the determination of local densities and column
densities of molecular hydrogen, H$_2$ must be indirect  since H$_2$ does not emit spectral lines in cooler clouds.   The Schr\"o\-dinger \index{Schr\"odinger}
equation governs all chemistry, but the conditions in the interstellar medium are very different (and much more varied) than those on earth. 
Earth-bound chemistry is only a subset of the more general interstellar chemistry; it  is dominated by non-equilibrium 
processes that occur at low temperatures and densities, so  determinations of collision and reaction rates are needed. A project devoted to this goal is the $''$Molecular Universe$''$ 
(see the web site for details). For an overview of interstellar chemistry, the reader is referred to  the short presentation  \cite{HERBST} or the monograph \cite{TIELENS}. 
 Although  interstellar 
chemistry plays a very important role in molecular line astronomy, it is {\em not} included due to  space limitations. Another glaring omission is a treatment of the Cosmic Microwave 
Background (CMB), since this is not treated in the following two lectures. Technically, the CMB emission is not weak, but does fill the entire sky, so special techniques must be employed.

Mm/sub-mm astronomy is one of the 
most important tools to study the birth of stars \cite{STARFORM}, \cite{PPV} and star formation in galaxies \cite{LINDA}. Stars form in molecular clouds where
extinction is large. Thus, near infrared and optical studies are of limited value. At high red shifts,  many active star forming galaxies and  Active Galactic Nuclei (AGN's) are 
enshrouded by dust \cite{PHILSOL}. 
Synchrotron emission at cm wavelengths from AGN's allows sub-arcsecond  
resolution images of relativistic electrons moving in $\vec{B}$ fields \cite{SCHEUER}.  
However, the imaging  of dust and molecules on comparable scales  is just beginning, so this is an area where  important contributions can be made.  

The closest example of a supermassive Black Hole  is Sgr~A$^*$,  8.5 kpc (1 kpc=3.08 \, $\times \, 10^{21}$ cm) from the Sun. Sgr~A$^*$ is  quiescent at  present, but is 
thought to be  similar to the $''$engines$''$ that power AGN's. The 
radio emission from Sgr A$^*$ is interpreted as  optically thick synchrotron emission, so with  mm/sub-mm Very long Baseline interferometry, 
gas on Astronomical Unit (1 AU=1.5 \,  $\times \, 10^{13}$ cm) scales, close to the Schwarzschild radius, can be sampled. 

 In the field of star formation, we believe that all of the physical principles are known. However,  star formation  is complex and measurements are needed to determine   
which processes are dominant and which can be neglected. In contrast,  for studies of the early universe, Black Holes and AGN's  
not  all of the relevant  physical laws  may be known. As in all of astronomy, measurements often lead to unexpected results that  
may have  far reaching effects on our understanding of fundamental physical laws.  

The dominant continuum radiation mechanism in the mm/sub-mm wavelength range is  dust emission. This  differs from free-free and synchrotron emission, which is commonly encountered at 
centimeter wavelengths \cite{ENDRIK}. Spectral line radiation is dominated by thermal and quasi-thermal 
molecular radiation, although there are a few important atomic lines of carbon, oxygen and nitrogen. 

The sensitivity at mm/sub-mm wavelengths has become  about 100 times better than was the case  in the 1960's \cite{GURV}. 
However sensitivity alone  is not enough to  transform the field. Rather, high resolution images are needed. 
 This will change when the {\em Atacama Large Millimeter Array} (ALMA) 
begins operation. ALMA will initially operate  
between 3.5 mm and 0.4 mm. ALMA has  a unique combination of high angular resolution and high sensitivity. This will allow ALMA to transform the field of mm/sub-mm astronomy. 
A summary of the science planned for ALMA is to be found  in \cite{ALMA-SCI}.

\section{Some Background}
    \label{sbd}
 
In this section, we review the basics needed in following sections (see \cite{RW4}). Electromagnetic radiation in the radio window can be interpreted as a wave 
phenomenon, i.~e. in term of classical physics. When the scale of the system involved is much larger than a 
wavelength, we can consider the radiation to travel in straight lines
or {\em rays}. The  power, ${\rm d}P$, intercepted by a
infinitesimal surface ${\rm d}\sigma$ 
 is
\begin{equation}
  \label{inf-power}
 {\rm d}P = I_\nu \cos \theta\, {\rm d}\Omega\, {\rm d}\sigma\,
{\rm d}\nu
\end{equation}
where\\[1.5ex]
 \begin{tabular}{@{}lll}

 \tabcolsep0.5mm
 ${\rm d} P$&=& power, in watts, \\
 ${\rm d }\sigma$&=& area of surface, m$^2 $, \\
 ${\rm d }\nu$&=& bandwidth, in Hz, \\
 $\theta$&=&angle between the normal to ${\rm d}\sigma$ and the
direction
 to ${\rm d}\Omega$, \\
 ${\rm I_{\nu}}$&=&brightness or specific intensity, in
  W\,m$^{-2}$\,Hz$^{-1}$\,sr$^{-1}$. \\[1.5ex]
 \end{tabular}

Equation (\ref{inf-power}) is the definition of
the brightness $I_{\rm \nu}$. Quite often the term {\it
intensity}\index{Intensity} or {\it specific intensity} $I_{\rm \nu}$ is
used instead of the term {\em brightness\index{Brightness}}. We will use
all three designations interchangeably.
 
 The total\index{Flux!total} flux of a source is obtained by
integrating (\ref{inf-power})
over the total solid angle $\Omega_{\rm s}$ subtended by the source
\begin{equation}
 \label{total-flux}
  S_\nu = \int\limits_{\Omega_{\rm s}} I_\nu (\theta,\varphi)
                   \cos \theta \, {\rm d}\Omega ,
\end{equation}
and this flux density is measured in units of W\,m$^{-2}$\,Hz$^{-1}$.
Since the flux density of astronomical sources is usually very small, a special
 unit, the Jansky (hereafter Jy) has
been introduced
\begin{equation}
 \label{jansky}
  1\,{\rm Jy} = 10^{-26} \,{\rm W \, m}^{-2} {\rm Hz}^{-1}
       = 10^{-23} \, {\rm erg\,s^{-1}} {\rm cm}^{-2} {\rm Hz}^{-1}\ts.
\end{equation}

  As long as the surface element
${\rm d}\sigma$ covers the ray bundle completely, the power remains constant:
\begin{equation}
 \label{dw}
  {\rm d}P_1 = {\rm d}P_2\ts .
\end{equation}
From this, we can easily obtain
 \begin{equation}
 \label{bnu}
 I_{\rm \nu_1} = I_{\rm \nu_2}
\end{equation}
so that the brightness is independent of the distance. The total\index{Flux!total} flux $ S_{\rm \nu}$ density
shows the expected dependence of  $ 1 / r^2$.

   Another useful quantity related to the brightness is the radiation
energy\index{Radiation!energy} density $u_{\rm \nu}$ in units of erg cm$^{-3}$. 
From dimensional analysis, $u_{\rm \nu}$ is intensity divided by speed. Since radiation propagates with
the velocity of light $c$, we have for the {\it spectral energy density
per solid angle}
\begin{equation}
 \label{energy-den}
  u_{\rm \nu}(\Omega) = \frac{1}{c} \, I_{\rm \nu}\ts .
\end{equation}
If integrated over the whole sphere, $4 \pi$ steradian,
(Eq.~\ref{energy-den}) will yeild   the {\it total spectral energy density}
\begin{equation}
 \label{tot-en-den}
   u_{\rm \nu} = \int\limits_{(4\pi)} u_{\rm \nu}(\Omega) \,{\rm d}\Omega =
                 \frac{1}{c} \int\limits_{(4\pi)} I_{\rm \nu} \,{\rm
d}\Omega\ts .
\end{equation}
 
\rhlz\rhhlz
 
\subsection{Radiative Transfer}
  \label{radtran}

The {\em equation of transfer\index{Transfer!equation}} is 
\begin{equation}
 \label{eq-of-transfer}
 \fboxsep2mm
 \fbox{$ \displaystyle
 \frac{{\rm d}I_\nu}{{\rm d}s} = - \kappa_\nu I_\nu + \epsilon_\nu $}
\quad .
\end{equation}
 
The linear absorption coefficient  $\kappa_\nu$ and the emissivity $\epsilon_\nu$ are independent of the
intensity $I_\nu$ leading to  the above form for ${\rm
d}I_{\nu }$.

In  Thermodynamic\index{Equilibrium!thermodynamic} Equilibrium (TE) there is complete equilibrium
of the radiation with its surroundings, the
brightness distribution is described by the Planck\index{Planck
function}
        function, which depends only on the temperature
$T$, of the surroundings. The properties of the Planck function will be described in the next section. 
     
%\begin{equation}
%        \label{inu-te}
%         \frac{{\rm d}I_\nu}{{\rm d}s} = 0\ts , \qquad I_\nu = B_\nu(T)
%=  \epsilon_\nu / \kappa_\nu
%       \end{equation}
%       \begin{equation}
%         \label{bnu-te}
%         \fboxsep2mm
%         \fbox{$ \displaystyle
%        B_\nu(T) = \frac{2 h\nu^3}{c^2} \frac{1}{{\rm e}^{h\nu / kT}
%- 1}
%$} \quad .
%       \end{equation}

 In  Local Thermodynamic\index{Equilibrium!thermodynamic} Equilibrium (LTE) %: Full
%thermodynamic
%        equilibrium will be realized only in very special circumstances.
        {\em Kirchhoff's law} holds  
       \begin{equation}
        \label{kirchhoff}
        \fboxsep2mm
        \fbox{$ \displaystyle
          \frac{\epsilon_\nu}{\kappa_\nu} = B_\nu(T) $}
       \end{equation}
 This is  independent of the material, as is the case with complete  thermodynamic  equilibrium. In general however, $I_\nu$  will
       differ from $B_\nu(T)$.
%\end{description}
%\enlargethispage{\baselineskip}
%\enlargethispage{\baselineskip}

If we define the {\it optical\index{Depth!optical} depth}
${\rm d}\tau_\nu$ 
by
 \begin{equation}
   \label{dtau-nu}
    {\rm d}\tau_\nu = - \kappa_\nu \,{\rm d}s
  \end{equation}
or
 \begin{equation}
  \label{tau-nu}
 \tau_\nu(s) = \int\limits_{s_0}^s \kappa_\nu (s) \,{\rm d}s \ts ,
 \end{equation}
then the equation of transfer\index{Transfer!equation}
(\ref{eq-of-transfer}) can be written as
 \begin{equation}
  \label{eq-inu-bnu}
   \fboxsep2mm
   \fbox{$ \displaystyle
      - \frac{1}{\kappa_\nu} \frac{{\rm d}I_\nu}{{\rm d}s} = \frac{{\rm d}I_\nu}{{\rm d}\tau_\nu} =
       I_\nu - B_\nu(T)  $} \quad .
 \end{equation}
The solution of (\ref{eq-inu-bnu}) is %obtained by first multiplying
%(\ref{eq-inu-bnu}) by $\exp (- \tau_\nu)$ and then integrating
%$\tau_\nu$ by parts:
%\beq
% \int\limits_0^{\tau_\nu(s)} \,{\rm e}^{-\tau} \frac{{\rm d}I_\nu}{{\rm d}\tau} {\rm d}\tau =
% I_\nu \,{\rm e}^{- \tau} \Bigg|_0^{\tau_\nu(s)} + \int\limits_0^{\tau_\nu(s)}
% I_\nu \,{\rm e}^{- \tau} {\rm d}\tau =
% \int\limits_0^{\tau_\nu(s)} (I_\nu - B_\nu) \,{\rm e}^{- \tau} {\rm d}\tau
% \eeq
% \beq
% I_\nu(\tau_\nu(s)) \,{\rm e}^{- \tau_\nu(s)} - I_\nu(\tau_\nu(s_0))
%     \,{\rm e}^{0} =
% - \int\limits_0^{\tau_\nu(s)} B_\nu(T(\tau)) \,{\rm e}^{- \tau} {\rm d}\tau
% \eeq
%or finally
 \begin{equation}
  \label{inu-s}
  \fboxsep2mm
  \fbox{$ \displaystyle
   I_\nu(s) = I_\nu(0) \,{\rm e}^{-\tau_\nu(s)} + \int\limits_0^{\tau_\nu(s)}
               B_\nu(T(\tau)) \,{\rm e}^{- \tau}
              {\rm d}\tau
  $} \quad .
 \end{equation}
%Due to the definition (\ref{dtau-nu}), $s$ and $\tau$ increase in opposite
%directions as indicated in Fig.\ts \ref{eq-of-tr}.
 
   If the medium is isothermal, %that is, if
\beq
  T(\tau) = T(s) = T = \mbox{const.}
\eeq
the integral is % in (\ref{inu-s}) can be computed explicitly resulting in
 \begin{equation}
  \label{sol-isoth}
   \fboxsep2mm
   \fbox{$ \displaystyle
     I_\nu(s) = I_\nu(0) \,{\rm e}^{- \tau_\nu(s)} + B_\nu(T)\,
                (1 - \,{\rm e}^{- \tau_\nu(s)}) $} \quad .
  \end{equation}
For a large optical\index{Optical depth} depth, that is for $\tau_\nu(0)
\rightarrow \infty$,
(\ref{sol-isoth}) in LTE approaches the limit
 \begin{equation}
  \label{large-opt-d}
   I_\nu = B_\nu(T)\ts .
 \end{equation}
This is case for planets and the 2.7 K microwave background. 
%The observed brightness $I_\nu$ for the optically thick case is equal to
%the Planck black-body brightness distribution independent of the material.
The difference between $I_\nu(s)$ and  $I_\nu(0)$ gives
\begin{equation}
 \label{I-comp}
 \Delta I_\nu(s)=I_\nu(s) - I_\nu(0) =
   (B_\nu(T)-I_\nu(0))(1-\expo^{-\tau}) \; .
\end{equation}
Eq.~\ref{I-comp} represents  an on-source minus an off-source measurement. 

\subsection[Black Body Radiation and  Brightness Temperature]
{Black Body Radiation\\
\tbee{}and  Brightness Temperature}
\sectionmark{Black Body Radiation and the Brightness Temperature}
 
\enlargethispage{\baselineskip}
The spectral distribution of the radiation of a black\index{Black body!radiation} body in thermodynamic equilibrium is given by the
Planck\index{Planck function} law 
\beq
   \fboxsep2mm
   \fbox{$ \displaystyle
     B_\nu(T) = \frac{2 h \nu^3}{c^2} \frac{1}{{\rm e}^{h\nu / kT} -1}
        $} \quad .
\eeq
If $ h \nu \ll k T$, one obtains {\em Rayleigh\index{Rayleigh-Jeans law}-Jeans Law}.  \vspace{-0.5mm}
\enlargethispage{2\baselineskip}%
\noindent
 \begin{equation}
  \label{b-jeans}
   \fboxsep2mm
   \fbox{$ \displaystyle
    B_{\rm RJ}(\nu, T) = \frac{2 \nu^2}{c^2} k T $} \quad .
  \end{equation}
This is an expression of  classical physics in which there   is an energy  $kT$ per mode. The term $\frac{2 \nu^2}{c^2}$ is  the {\em density of states} for 3 
dimensions. 
 In the millimeter and submillimeter range, one
frequently defines a {\em radiation temperature},  $J(T)$ as
 \begin{equation}
 J(T) = \frac{c^2}{2 k \nu^2}\, I =
 \frac{h \nu}{ k} \frac{1}{
     \expo^{{h \nu}/{k T }} -1
     } \; .
\end{equation}
 
   Inserting numerical values for $k$ and $h$, we find  that the Rayleigh-Jeans
relation holds for frequencies
 \begin{equation}
  \label{nu-jeans}
    \frac{\nu}{\rm GHz} \ll 20.84 \left( \frac{T}{\rm K} \right) \, .
 \end{equation}
It can thus be used for all thermal radio sources except perhaps for low
temperatures\index{Brightness temperature}
 in the mm/sub-mm range.

%\item[2)]  $h \nu \gg k T$: {\em Wien\index{Wien's law}'s Law}. In this
%case ${\rm e}^x \gg 1$,
% so that
%  \begin{equation}
%   \label{b-wien}
%    \fboxsep2mm
%    \fbox{$ \displaystyle
%     B_{\rm W}(\nu, T) = \frac{2 h \nu^3}{c^2} \,{\rm e}^{- h\nu / k T}
%         $}  \quad .
%  \end{equation}
%While this limit is quite useful for stellar measurements in the visual and
%ultraviolet range, it plays no role in radio astronomy.
%\end{description}
 
In the Rayleigh-Jeans relation,  the brightness and the thermodynamic temperature of the
black body that emits this radiation are strictly proportional
(\ref{b-jeans}). This feature is so useful that it has become the custom
in radio astronomy to measure the brightness of an extended source by
its {\em brightness temperature\index{Temperature!brightness}} $\tb$.
This is the temperature which would result in the given brightness if
inserted into the Rayleigh-Jeans law
 \begin{equation}
  \label{t-b}
    \tb = \frac{c^2}{2 k} \frac{1}{\nu^2}\, I_\nu =
   \frac{\lambda^2}{2 k} \, I_\nu \ts .
 \end{equation}

%******************

Combining (\ref{total-flux}) with (\ref{t-b}), we have 
\begin{equation}
 \label{flux-den-tb1}
    \fboxsep2mm
    \fbox{$ \displaystyle
S_{\nu}=\frac{2\, k \, \nu^2}{c^2} \tb \, \uDelta  \Omega  
        $}  \quad .
\end{equation}
For a Gaussian source, this relation is 
 \begin{equation}
   \label{flux-den-tb2}
    \left[ \frac{S_{\nu}}{\rm Jy} \right] = 0.0736 \,  \tb \,  \left[ \frac{\theta}{\rm arc \, seconds} \right]^{2}  \left[  \frac{\lambda}{\rm mm} \right]^{-2}
          \end{equation}
That is, if the flux density $S_\nu$ and  the source size are known, then the true brightness temperature, $\tb$, of the source can be determined.
The concept of temperature in radio astronomy has given rise to confusion. If one measures $S_\nu$ and the {\em apparent} source size, Eq.~\ref{flux-den-tb2} 
allows one to calculate the {\em main beam brightness temperature}, T$_{\rm MB}$. Details of this procedure will be given in Section \ref{desap}. 
The performance of coherent  receivers is   characterized {\em receiver noise temperature}; see Section \ref{cal-proc} for details. The combination of receiver 
and atmosphere is characterized by {\em system noise temperature}; this is discussed in  Section \ref{observations} and following. 

%***************new equation
If $I_\nu$ is emitted by a black body and $h \nu \ll k T$ then
(\ref{t-b}) gives the thermodynamic temperature of the source, a
value that is independent of $\nu$. If other processes are responsible for the emission of the
 radiation, $T_{\rm b}$ will depend on the
 frequency; it is, however,
still a useful quantity and is commonly used in  practical work.

 \enlargethispage{2\baselineskip}
This is the case even if the frequency is so high that condition
(\ref{nu-jeans}) is not valid. Then (\ref{t-b}) can still be applied,
but  $T_{\rm b}$ is different
from the thermodynamic temperature of a black body. However, it is 
rather simple to obtain the appropriate correction factors.

It is also convenient to introduce the concept of brightness
temperature into the radiative
transfer equation (\ref{I-comp}).  Formally one can obtain
\beq
 J(T)=\frac{c^2}{2 k \nu^2}(B_\nu(T)-I_\nu(0))(1-\expo^{-\tau_\nu(s)})
\; .
\eeq
Usually calibration procedures  allow one to express
$J(T)$ as $T$. This measured quantity is referred to as $T^*_{\rm R}$,
the
{\em radiation temperature}, or the {\em brightness temperature},
$T_{\rm b}$. In the cm wavelength range, one can apply
(\ref{t-b}) to
(\ref{eq-inu-bnu}) and one obtains
 \begin{equation}
  \label{eq-tb}
   \fboxsep2mm
   \fbox{$ \displaystyle
       \frac{{\rm d}T_{\rm b}(s)}{{\rm d}\tau_\nu} =
           T_{\rm b}(s) - T(s) $} \quad ,
 \end{equation}
where $T(s)$ is the thermodynamic temperature of the medium at the
po\-si\-tion~$s$.
The general solution is
 \begin{equation}
  \label{sol-tb}
  \fboxsep2mm
  \fbox{$ \displaystyle
     T_{\rm b}(s) = T_{\rm b}(0) \,{\rm e}^{- \tau_\nu(s)} +
                    \int\limits_0^{\tau_\nu(s)} T(s) \,{\rm e}^{- \tau}
                    {\rm d}\tau   $} \quad .
 \end{equation}
If the medium is isothermal\index{Isothermal!medium}, this becomes
 \begin{equation}
  \label{tb-isoth}
   \fboxsep2mm
   \fbox{$ \displaystyle
     T_{\rm b}(s) = T_{\rm b}(0) \,{\rm e}^{- \tau_\nu(s)} +
          T \, (1- \,{\rm e}^{- \tau_\nu(s)})   $} \quad .
  \end{equation}

\subsection{The Nyquist\index{Nyquist theorem} Theorem and the Noise Temperature}

We now relate voltage and temperature; this is essential for the analysis of receiver systems limited by noise. 
The average power per unit bandwidth produced by a  resistor $R$  is
\begin{equation}
 \label{pow}
 P_\nu = \langle iv \rangle = \frac{\langle v^2 \rangle}{2R} =
  \frac{1}{4R}\langle v_{\rm N}^2 \rangle\ts ,
\end{equation}
where $v(t)$ is the voltage that is produced by $i$ across $R$, and
$\langle\cdots\rangle$ indicates a time average. The first factor $\frac{1}{2}$ arises from the condition for the  transfer of maximum power from $R$ over a  broad 
range of frequencies. The second factor
$\frac{1}{2}$ arises from the time average of $v^2$.
 %*************************************************************************************
%\abfa
%%1.8%
%\begi%n{figure}
%\sidecaption
% %\mpicplace{3.9cm}{2.6cm}
% \includegraphics[width=3.9cm,height=2.6cm]{kap1f8.eps}
%    \caption[The origin of Johnson noise]
%            { \label{Johnson noise}
%             A sketch of a circuit containing a resistor $R$, to illustrate
%  the origin of Johnson noise. The resistor $R$, on the left,
%  at a temperature $T$, provides a power $k\,T$ to a matched
%  load $R_X$, on the right}
%\end{figure}
 %************************************************************************************* 
An analysis of the random walk process shows that
\begin{equation}
 \langle v_{\rm N}^2 \rangle = 4 R \,k \,T \; .
\end{equation}
Inserting this into (\ref{pow}) we obtain
\begin{equation}
\label{Nyquist}
 P_\nu = k\, T  \; .
\end{equation}
Eq.\,(\ref{Nyquist}) can also be obtained by a formulation of the Planck law for one dimension and the Rayleigh-Jeans limit. Then, the available noise power of a
resistor is proportional to its temperature, the {\em
noise\index{Temperature!noise} temperature} $T_{\rm N}$, and independent
of the value of $R$. Throughout the whole radio range, from the longest
waves to the far infrared region the noise spectrum is white, that is,
its power is independent of  frequency. Since the impedance of
a noise source should  be matched to that of the receiver, such a
noise source can only be matched over some finite bandwidth.
 
%Not all circuit elements can be characterized by thermal noise. For
%example a microwave oscillator can deliver the equivalent of more than 
%$10^{16}$\,K, although the physical
%temperature is only 300\,K. Clearly this is a very nonthermal process, and in this case temperature 
%is not a useful concept. 

\subsubsection{Hertz Dipole and Larmor Formula}\label{hertz-larmor}
 In contrast to thermal noise, radiation from an antenna  has a definite polarization and direction.  One  example (in electromagnetism there 
are {\em no} simple examples!) is  a Hertz dipole.
The total power radiated from a Hertz dipole\index{Antenna!Hertz dipole} carrying an oscillating current $I$ at a wavelength $\lambda$ is 
 \begin{equation}
   \label{tot-rad-pow}
   \fboxsep2mm\fbox{$ \displaystyle
    P = \frac{2 c}{3}  \left(
                \frac{I \Delta l}{2 \lambda} \right)^2  $} \quad .
 \end{equation}
For the Hertz dipole, the radiation is linearly polarized with the $\vec{E}$ field along the direction of the dipole. The radiation pattern has a donut shape, with a 
cylindrically symmetric maximum perpendicular to the 
axis of the dipole. Along the direction of the dipole, the radiation field is zero.  One can use collections of  dipoles, driven in phase, 
to restrict the direction of 
radiation. As a rule of thumb,  Hertz dipole radiators have  the best radiative efficiency when the wavelength of the radiation is  roughly 
the size of the dipole. Following Planck, one can produce the the Black Body law from  a collection of (quantized)  Hertz  dipole oscillators  with random phases.
%From (\ref{tot-rad-pow}), the

 There are similarities between  Hertz dipole radiation and radiation from atoms.  Following Larmor, the power radiated by a single  electron  oscillating 
with a velocity $v(t)$ is 
 \begin{equation}
    P(t) = \frac{2}{3}\frac{e^2 \dot{v}(t)^2}{c^3} \,.
 \end{equation}
Expressing  $\dot{v} = \ddot{x}$, we obtain
an average power, emitted over one period of sinusoidal oscillation, $x=\sin{2\pi \nu t}$. 
 \begin{equation}
  \label{aveped-1}
  \langle P \rangle = \frac{64 \pi^4}{3 \,c^3} \,\nu_{mn}^4
            \left(\frac{e \,x_0}{2}\right)^2 \,.
 \end{equation}
Using $\frac{e \,x_0}{2} = |\mu|$ one arrives at the expression for spontaneous emission from a quantum mechanical system. This is presented again in Eq.~\ref{amn}. It 
is often the case that quantum mechanical expressions and  classical  correspond to within factors of a few. 
%\addtocontents{toc}   
%****************************************************************************************************************************************************************rec-theory
\section[Theory of Receivers]
{Signal Processing\newline
\tbee{}and Stationary Stochastic Processes}
\sectionmark{Signal Processing and Stationary Stochastic Processes}
 \label{stoch-proc}

Next, we present the basics of  signal processing and noise
analysis  needed to understand the 
properties of radiometers \cite{RW4}. Fourier methods are of great value in analyzing receiver properties \cite{BRACE1}. 
A review of submillimeter receiver systems is to be found in \cite{RIEKE}.
 
The concept of spectral\index{Power density!spectral} power density was
introduced in Eq.\,\ref{pow} in a practical example. Radio receivers are devices that measure spectral power density\index{Receiver!spectral power density}. 
Since the signals are dominated by noise, statistical analyses are needed.  
The most important of these statistical quantities is the
probability density function, $p(x)$, which gives the probability that
at any arbitrary moment of time the value of the process $x(t)$ falls
within an interval $(x - \frac{1}{2} \diff x, x + \frac{1}{2} \diff x)$.
For a stationary random process, $p(x)$ will be independent of the time
$t$.

The {\em expected\index{Value!expected} value} $E\{ x\}$ or {\em
mean\index{Value!mean} value} of the random
variable $x$ is given by the integral
 \begin{equation}
   \label{ex}
   \fboxsep2mm\fbox{$ \displaystyle
     E \{ x \} = \int\limits_{-\infty}^\infty x \,p(x) \diff x $}
 \end{equation}
and, by analogy, the expectation value $E \{ f(x)\}$ of a function
$f(x)$ is given by
 \begin{equation}
   \label{efx}
   \fboxsep2mm\fbox{$ \displaystyle
     E \{ f(x) \} = \int\limits_{-\infty}^\infty f(x) \,p(x) \diff x $}
\quad .
 \end{equation}

A digression: the trend in {\em all} forms of communications (including radio astronomy!) is toward digital processing. In general, 
a digitized function must be {\em sampled} at regular intervals. Assume that the input signal extends from zero Hz to $\nu_0$ Hz (this is referred to the video
\index{Receiver!video band} band). Then the bandwidth, $\Delta \nu_0$, is $\nu_0$ and maximum frequency is $\nu_0$. If we picture the input as a collection of sine waves, 
it is clear that the sampling rate {\em must} be at least $\nu_0= 2 \Delta \nu$ to  characterize the sinusoid with the highest frequency, $\sin{2\pi \nu_0 t}$. 
Thus the sample rate must be twice the bandwidth  of the video 
input. This is referred to as the  {\em Nyquist Sampling Rate\index{Receiver!Nyquist Sampling Rate}}. This is a minimum; a higher sampling 
rate can only improve the characterization of the input.     A 
higher sampling rate ($''$oversampling$''$\index{Digital!oversampling}) 
will allow the input to be better characterized, thus giving a better   Signal-to-Noise ratio (S/N) ratio. The sampling functions must
occupy an extremely small time interval compared to the time between samples. 
If only a portion of the input function is retained in the quantization and sampling process,
information is lost. This results in a lowering of the S/N ratio. 

At present, commercially available digitizers can sample a 1.5 GHz bandwidth with  8 bit quantization. This allows digital systems to reach within a few percent of 
the sensitivity of analog systems, but with much 
greater stability. After digitization, one can maintain that all of the following processes are  $''$just arithmetic$''$. 
The quality of the data (i.~e.~the S/N)  will depend on the analog receiver
elements. The digital parts of a receiver can lower  S/N ratios, but not raise them!

\subsection{Square Law Detectors}

In radio receivers,  noise  is passed through a device 
that produces an output signal $y(t)$
which is proportional to the power in  a given input  $v(t)$:
 \begin{equation}
   \label{yax}
     y(t) = a \, v^2(t) \, .
 \end{equation}

 This involves an evaluation of the integral 
\beq 
E\{ y(t) \}  &=& E \{a\, v^2(t) \}=
\frac{a}{\sigma \sqrt{\displaystyle 2\pi}}\int_{-\infty}^{+\infty} v^2 {\rm e}^{-v^2/2\sigma^2}{\rm d}v
\eeq
There are  standard approachs  used to evaluate this expression.  The result is\\
\begin{equation}
E\{ y(t) \}  =  E \{ v^2(t) \} = a \, \sigma_v^2 
\end{equation}

For the evaluation of $E\{y^2(t)\}$, 
one must calculate 
\beq 
E\{ y(t)^2 \}  &=& E \{ v^4(t) \}=
\frac{1}{\sigma \sqrt{\displaystyle 2\pi}}\int_{-\infty}^{+\infty} v^4 {\rm e}^{-v^4/2\sigma^2}{\rm d}v
\eeq 
%This is best done using an 
%integration by parts,
%with $u=x^3$, and $dv=x{\rm e}^{-x^2/2\sigma^2}$. 
The result of this integration is 
\begin{equation}
  \label{erw}
E \{y^2(t) \} = 3 \, a^2 \, \sigma_v^4 
\end{equation}
and hence
 \begin{equation}
  \label{sigy}
    \sigma_y^2 = E \{ y^2(t) \} - E^2 \{ y(t) \} = 
                2 \,E^2 \{ y(t) \} \, .
 \end{equation}
%In contrast to linear systems, the mean value of the output signal of a
%square-law detector does not equal zero, even if the input signal has
%a zero expected mean value.

\subsection{Limiting Receiver Sensitivity}
  \label{lrs}

A receiver must be {\em sensitive}, that is,  be able to detect faint
signals in the presence of noise. There are
limits for this sensitivity, since  the receiver input and the receiver itself are 
affected by  noise.  Even when no
input source is connected to a receiver, there is an output
signal, since  any receiver generates thermal noise.
This noise is amplified together with the signal. Since signal and noise have the same statistical properties, these cannot be distinguished. To analyze the performance of a
receiver we will use the model of an ideal receiver producing no internal noise, but connected simultaneously to two noise sources, one
for the external source noise and a second for the  receiver noise. To be useful,
receivers must increase the input power level. The power per unit bandwidth, $P_\nu$,  entering a receiver can be characterized by a temperature, as given by Eq.\,\ref{Nyquist}, 
$P_\nu=kT$. 
Furthermore, it is {\em always} the case that the  noise contributions from source, atmosphere, ground and receiver,  $T_i$, are additive, 
\beq
   T_{\rm sys} = \sum T_i
\eeq
%We  apply these concepts to a 2 port system, as shown in Fig.\,\ref{two-port}. 
% we relate noise power to temperatures, and ultimately to measurement uncertainties. 
%The signal input, $S_1$, and output $S_2$ are related by the system gain, $G$. The noise output, $N_2$ is the noise input, $N_1$,  multiplied by the gain, plus the noise added by the system,
%$N_{\rm int}$. 
An often-used figure of merit is the {\em Noise Factor\index{Receiver!noise factor}}, $F$. This is defined as 
\begin{equation}
\label{nois-fact}
   F = \frac{S_1/N_1}{S_2/N_2} = \frac{N_2}{G\,N_1}=1 + \frac{T_{\rm R}}{T_1}
\end{equation}
that is, any additional noise generated in the receiver contributes to $N_2$. For direct detection systems, such as a {\em Bolometers}, $G=1$. 
If $T_1$ is set equal to $T_0=290$K, we have 
\beq
   T_{\rm R} = F-1 \times 290 
\eeq
Given a value of $F$, one can determine the receiver noise temperature. If for $\nu_0=115$ GHz,  $F=3$db (=10$^{0.1\, F {\rm db}}$), $T_{\rm R}$=290\,K, a lousy receiver noise temperature. 
 %The section relates receiver properties  to the minimum uncertainty in a measurement. 

%*********************************************************************************************************************************************move 31 July 

\subsubsection{Receiver Calibration}
 \label{cal-proc}

Our goal is  to  characterize receiver noise performance  in   degrees Kelvin.  In the calibration\index{Receiver!calibration procedure} process,
a noise power scale (spectral power density) is established at the receiver input.  In radio
astronomy the noise power of coherent receivers (i.~e.~those which preserve the phase of the input)  is usually measured 
in terms of the
noise\index{Temperature!noise} temperature. To calibrate a
receiver, one relates the noise temperature increment $\Delta T$
at the receiver input to a given measured receiver output
increment $\Delta z$ (this applies to coherent  receivers which have a wide dynamic range and a total power or $''$DC$''$ response). 
 In principle, the receiver 
noise temperature, $ T_{\rm R}$, could be computed from the output signal $z$ provided
the
detector characteristics are known. In practice 
the receiver is calibrated by connecting two or more  known 
sources to the input. Usually matched resistive loads at  known 
(thermodynamic)
temperatures $T_{\rm L}$ and $T_{\rm H}$ are used. To within a constant, the receiver outputs
are 
 \begin{eqnarray*}
    z_{\rm L} &=& (T_{\rm L} + T_{\rm R}) \, G \ts,   \\
    z_{\rm H} &=& (T_{\rm H} + T_{\rm R}) \, G \ts,
 \end{eqnarray*}
from which
 \begin{equation}
    \label{tsrc}
    \fboxsep2mm\fbox{$ \displaystyle
      T_{\rm rx}= \frac{T_{\rm H} - T_{\rm L} \, y}{y - 1} $}
      \quad ,
 \end{equation}
where
 \begin{equation}
   \label{yhl}
     y = z_{\rm H} / z_{\rm L} \ts .
 \end{equation}
This is known as the $''$y-factor$''$\index{Receiver!y-factor}; the procedure is a $''$hot-cold$''$ measurement. Note that the {\bf y} factor as presented here is 
determined in the Rayleigh-Jeans limit. 
The  temperatures $T_{\rm H}$ and $T_{\rm L}$ are usually produced
by absorbers in the mm/sub-mm  range. Usually these are chosen to be 
at the ambient temperature ($T_{\rm H} \cong
293$\,K or $20\degr$\,C) and at the temperature of liquid nitrogen
($T_{\rm L} \cong 78$\,K or $-195\degr$\,C). In rare cases, one might use  liquid
helium, which has a boiling point $T_{\rm L} \cong 4.2$\,K. In this
process, the receiver is assumed to be a linear power measuring device (i.~e.~we assume that any  non-linearity of the receiver is a small quantity).
Usually such a   $''$hot-cold$''$ calibration  is  done  infrequently. 
As will be discussed in Section  \ref{mm-cal-proc}    in the mm/sub-mm wavelength range, measurements of the emission from the atmosphere and 
then from an ambient resistive load are combined 
with models to provide an estimate of the 
atmospheric transmission. For a determination of the receiver noise, an additional measurement, usually with a cooled resistive load is needed.

Bolometers (Section \ref{bolo-rx}) do not preserve phase, so are incoherent receivers. Their 
performance is strongly dependent on the bias of the detector element. The Bolometer performance  
is characterized by the {\em Noise Equivalent Power}, or NEP. The NEP is given in units of Watts Hz$^{-1/2}$.  NEP is the input power level  
which doubles the output power. Usually bolometers are $''$AC$''$ coupled, that is, the output responds to {\em differences} in the input power, so 
hot-cold measurements are not useful for characterizing bolometers. 

%*****************************************************************************************************************************************************moved to here July 31
%use simpler derivation of reciever noise(from Moran/Staelin)

%\noindent% 

\subsubsection{Noise Uncertainties due to Random Processes}
\label{staelin}

%*************************************************************************
%%4.7%FIG 4.7 AUG 9, 07
\begin{figure}[b]
%\sidecaption
%   \mpicplace{3.6cm}{1.2cm}
  \includegraphics[width=6.0cm,height=1.70cm]{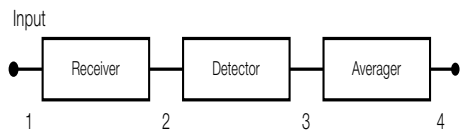}
\caption[ The principal parts of a receiver]{\label{rec-parts}
  The principal parts of a receiver. In the text, $v_1$ is at point 1, $v_2$ at point 2, etc. }
%\vglue6pt
\end{figure}
%*************************************************************************

It has been found that both source and receiver noise has a gaussian distribution\cite{RW4}. 
We assume that the signal is a 
Gaussian random variable with mean zero
 \begin{equation}
   \label{gdd}
   p(x) = \frac{1}{\sigma \sqrt{\displaystyle 2\pi}}
           \expo^{-x^2 / 2\sigma^2} \, .
 \end{equation}      
which is sampled at a rate equal to twice the bandwidth.\\ 

\noindent  
Refer to Fig. \ref{rec-parts}.  By assumption $E(v_1)=0$. The input, $v_1$ has a much larger bandwidth, $B$, 
than the bandwidth of the receiver, that is,  $\uDelta \nu \ll B$.The output of the receiver is $v_2$, with  a bandwidth $\uDelta \nu$. The power\index{Receiver Noise!square law detector} 
corresponding to the voltage $v_2$ is $\left\langle{}v_2^2\right\rangle$. 
\noindent
\begin{equation}     
P_2=v_2^2=  \sigma^2 = k \,  T_{\rm sys}\,  G \uDelta \nu\;,   
\end{equation}
where $\uDelta \nu$ is the receiver bandwidth, $G$ is the gain, and $ T_{\rm sys}$ is the total noise from the input $T_{\rm A}$ and the receiver $T_{\rm R}$. 
The contributions to  $T_{\rm A}$ are the external inputs from the source,
ground and atmosphere. 
\noindent  
Given that the output of the square law detector is $v_3$
\begin{equation}
\left\langle{}v_3\right\rangle=\left\langle{}v_2^2\right\rangle
\end{equation}
\noindent
then after square-law detection we have 
\begin{equation}
\label{sq-out}
 \left\langle{}v_3\right\rangle =  \left\langle{}v_2^2\right\rangle =  
\sigma^2 = k T_{\rm sys} G \uDelta \nu\;.       
\end{equation}
Crucial to a determination of the noise is the mean value and 
variance of $\left\langle{}v_3\right\rangle$
From (\ref{erw}) the result is \\ 
\begin{equation}
\left\langle{}v_3^2\right\rangle=\left\langle{}v_2^4\right\rangle=3\left\langle{}v_2^2\right\rangle
\end{equation}
this is needed to determine $\left\langle{} \sigma_3^2\right\rangle$. 
Then, \\ 
\begin{equation}
 \sigma_3^2=\left\langle{}v_3^2\right\rangle-
\left\langle{}v_3\right\rangle^2
\end{equation} 
$ \left\langle{}v_3^2\right\rangle $ is the total   noise power (= receiver plus  input signal). Using the Nyquist sampling rate, 
 the averaged output, $v_4$, 
is $({1}/{N}) \Sigma v_3$ where $N=2  \uDelta \nu \, \tau$. \\

\noindent
From  $v_4$ and $\sigma_4^2=\sigma_3^2/ N$, we obtain 
the result \\ 
\begin{equation}
\sigma_4=k \uDelta\nu G (T_{\rm A} + 
T_{\rm R})/\sqrt{ \uDelta \nu \, \tau}
\end{equation} 
We have explicitly separated  $T_{\rm sys}$ into the sum $T_{\rm A} + T_{\rm R}$. 
Finally, we  use the calibration procedure in Sect.\, \ref{cal-proc}  
to eliminate the term $kG \uDelta \nu $:    
 \begin{equation}
   \label{dicke}
   \fboxsep2mm\fbox{$ \displaystyle
    \frac{\uDelta T}{T_{\rm sys}} = \frac{1}{\displaystyle
  \sqrt{\uDelta \nu \, \tau}} $} \quad .
 \end{equation}
 The calibration process allows us to 
specify  the receiver output in degrees Kelvin instead of in Watts per Hz. We therefore characterize the receiver 
quality by the system noise\index{System Noise!square law detector}
temperature $T_{\rm sys}=T_{\rm A} + T_{\rm R}$.

For a given 
system, the improvement in the RMS noise {\em cannot} be better than as given in Eq.\,(\ref{dicke}). Systematic errors will only increase $\Delta T$, 
although the time behavior may follow the behavior described by (Eq.~\ref{dicke})\cite{DICKE}. 
We repeat for emphasis: $T_{\rm sys}$ is the noise from the {\em entire} system. That is, it includes the noise from the receiver, atmosphere,
ground, and {\em and the source}. Therefore $\Delta T$ is larger for an
intense source. Except for some planets, however, this situation is rare in the mm/sub-mm range.

\subsubsection{Receiver Stability}
  \label{rec-stab}
 
Sensitive receivers are  designed to achieve
a low value for $T_{\rm sys}$. Since the signals received are of exceedingly low
power,  receivers must also provide sufficient output power. This requires a large receiver gain, so  even very small gain instabilities
can dominate the thermal receiver noise. Therefore receiver
stability\index{Receiver!stability} considerations are also  of prime importance. Great advances have been made in improving receiver stability. However in the 
mm/sub-mm range, the atmosphere plays an important role. To insure that the noise decreases following  Eq.\,\ref{dicke}, systematic effects from atmospheric and receiver 
instabilities are minimized. Atmospheric {\em changes} are of crucial importance. These can be compensated for by rapidly taking the difference between  
the measurement of the source of interest 
and a  reference region or a nearby calibration source. Such {\em comparison switched} measurements are necessary for all  ground based observations.   R.~H.~Dicke was the first to 
apply comparison  switching  was first applied to radio astronomical  receivers \cite{DICKE}.

 The time spent measuring references or performing calibrations will {\em not} contribute to an improvement in the S/N ratio.  In fact, the subtraction of two noisy inputs {\em worsens} the 
difference, but are needed to reduce instabilities that give rise to systematic errors. 
The time $\tau$ is the {\em total} time taken for the measurement (i.~e.~on-source and off-source).  

Even for the output of a total power receiver there will be additional noise in excess of that given by (\ref{dicke}) since  the signals to be
differenced  are $\uDelta T + T_{\rm sys}$ and $T_{\rm sys}$. This is needed since  $\uDelta T << T_{\rm sys}$. For example, if one-half the total time is spent on the reference, the 
$\uDelta T$ for difference of on-source minus off-source in  (\ref{dicke}) is a factor of 2 larger. 
%The error of this difference signal  is given by (\ref{redd}).  

%\addtocontents{toc}
\section{Practical Receivers}
This section is concerned with the practical aspects of receivers that are currently in use \cite{GOLDSMITHED}, \cite{RIEKE}, with some  background material  from \cite{RW4}.
%*******************************************************************************************************************************************************************rec-practical-saas
\subsection{Bolometer Radiometers}\label{bolo-rx}
%************************************************************************** 

%\abfn
%%4.10
%OK TLW Aug 5, 07---this is Fig 5.1
\begin{figure}[b]
%\sidecaption
  %\mpicplace{6.2cm}{2.7cm}
  \includegraphics[width=6.2cm,height=2.7cm]{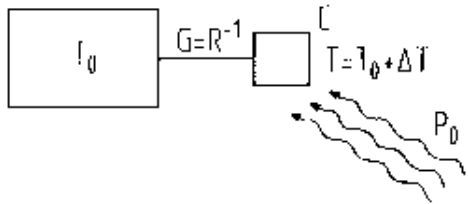}
  \caption[A sketch of a bolometer]
 { \label{bolometer}
A bolometer is represented by the smaller square to the right. The power from an astronomical
 source, $P_0$, raises the temperature of the bolometer element
 by $\Delta \, T$, which is much smaller than the temperature $T_0$ of the
 heat sink. Heat capacity, ${\cal C}$, is analogous to
 capacitance. The conductance, ${\cal G}$ is
 analogous to electrical conductance, $G$, which is $1/R$. The
 noise performance of bolometers depends critically on the
 thermodynamic temperature, $T_0$, and on the conductance
 ${\cal G}$. The temperature change causes a change in the
 voltage drop across the bolometer (the details of the electric circuit are not
 shown)}
\end{figure}
\abfa
%************************************************************
 
The operation of  bolometers\index{Bolometer} makes use of the effect
that the resistance, $R$, of a material varies with the temperature.
When radiation is absorbed by the bolometer material, the temperature
varies; this temperature change is a measure of the intensity of the
incident radiation. Because this thermal effect is rather independent of
the frequency of the radiation absorbed, bolometers are intrinsically
broadband devices. Frequency discrimination  must be provided by external filters. 

The Noise Equivalent Power {\em (NEP)} quoted for a bolometer is the input power that doubles the output of this device. The expression for NEP is  
\begin{equation}
  \label{nepph}
 \fboxsep2mm
    \fbox{$ \displaystyle
 {\rm NEP}_{\rm ph}= 2 \epsilon \,k \,T_{\rm BG} \,\sqrt{\Delta \nu}
 $} \quad .
\end{equation}
Where  $\epsilon$ is the emissivity of the background, and T$_{\rm BG}$ is temperature of the background. 
Typical values for ground based bolometers are $\epsilon = 0.5, T_{\rm BG} = 300$ K  and
$\Delta \nu= 100$ GHz. For these values
$ NEP_{ \rm ph}= 1.3 \times 10^{-15}$ \, Watts\,Hz$^{-1/2}$.  
With the collecting area of the 30\,m IRAM telescope and a
100\,GHz bandwidth one can easily detect mJy sources.

\subsection{Currently Used  Bolometer Systems}

Bolometers mounted on ground based radio telescopes are
background noise limited, so the only way to substantially
increase mapping speed for extended sources is to construct
large arrays consisting of many pixels. In present systems,
the pixels are separated by 2 beamwidths, because of the
size of individual bolometer feeds.  The systems which best 
cancel atmospheric fluctuations 
are composed of rings of  close-packed 
detectors surrounding  a single detector placed in the center of the array. 
Two large
bolometer arrays have produced many significant published results. The first is MAMBO2 (MAx-Planck-Millimeter Bolometer)\index{Bolometer!MAMBO2}. This is 
 a 117 element
array used at the IRAM 30-m telescope. This system operates at 1.3 mm, and
provides an angular resolution of 11$''$. The portion of the sky that is measured at one instant is the  {\em  field of view\index{Field of View}}, (FOV). The FOV of
MAMBO2 is 240$''$.  The second system is SCUBA (Submillimeter Common User
Bolometer Array)\index{Bolometer!SCUBA}\cite{HOLLAN}. This is used  on the James-Clerk-Maxwell (JCMT) 15-m sub-mm
telescope at Mauna Kea, Hawaii. SCUBA consists of a 37 element array
operating at 0.87 mm, with an angular resolution of 14$''$ and a 91 element array operating at
0.45 mm with an angular resolution of 7.5$''$; both have a FOV of about 2.3$'$.  The LABOCA (LArge Bolometer CAmera)\index{Bolometer!LABOCA} array operates on the APEX 12 meter telescope. 
APEX is on the 5.1 km high Chaijnantor plateau, the ALMA site in northern Chile. The LABOCA camera operates at 
0.87 mm waqvelength, with 295 bolometer elements. These are arranged in 9 concentric 
hexagons around a center element. The angular resolution of each element is 18.6$''$, the FOV is 11.4$'$. Such an arrangement is ideal for the measurement of small sources since the outer 
rings of detectors can be used to subtract the emission from the sky, while the central elements are measuring the source.

\subsubsection{Superconducting Bolometers}
A promising new development in bolometer receivers is
{\em Transition Edge Sensors} referred to as  TES bolometers.
These superconducting devices may allow more than an order
of magnitude increase in sensitivity, if the bolometer is
not background limited. For broadband  
bolometers used on  earth-bound telescopes,  
 the warm background 
limits the performance. 
With no background, the noise improvement with TES systems is 
limited by photon noise; in a background noise limited 
situation, TES's should be  
$\sim$2--3 times more sensitive than semiconductor bolometers. 
For ground based telescopes, TES's greatest advantage is multiplexing many detectors with a
superconducting readout device, so one can 
construct even larger arrays of bolometers. SCUBA  will be replaced  with
SCUBA-2 now being constructed at the U.~K.~Astronomy Technology Center. SCUBA-2 is an array of 2 TES bolometers, each
consisting of 6400 elements operating at 0.87 mm and 0.45
mm. The FOV of SCUBA-2 will be 8$'$. The SCUBA-2 design is
based on photo-deposition technology similar to that used
for integrated circuits. This type of construction allows
for a closer packing of the individual bolometer pixels. In
SCUBA-2 these will be separated by 1/2 of a beam, instead of
the usual 2 beam spacing\index{Bolometer!SCUBA-2}.

\subsubsection{Polarization \& Spectral Line Measurements}
In addition to measuring the continuum total power, one can
mount a polarization-sensitive device in front of the
bolometer and thereby measure the direction and degree of
linear polarization. These devices have been  used with SCUBA on the James-Clerk-Maxwell Telescope in Hawaii and (in the past) 
with a 19 beam bolometer at the Heinrich-Hertz-Telescope in Arizona. 

 It is possible to also
carry out  spectroscopy, if frequency sensitive
elements, either Michelson or Fabry-Perot interferometers,
are placed before the bolometer
element\index{Bolometer!Spectral Line}. Since these spectrometers operate at the sky frequency, the fractional resolution ($\uDelta \nu/\nu$) is limited. %One such instrument  is 
% the South Pole Imaging
%Fabry-Perot Interferometer, SPIFI (Stacey et al.~2002). 

\subsection{Coherent Receivers}
 Coherent receivers are those which preserve the phase of the signal.  Usually, coherent receivers make use of the 
superheterodyne\index{receiver!heterodyne}  (or more commonly $''$heterodyne$''$) principle to shift 
the signal input frequency  without changing  other properties; in practice, this  is carried out by the use of 
mixers (Section \ref{mixer-rx}). Heterodyne is  commonly used in all branches of 
communications technology; its use allows measurements with unlimited spectral resolution, $\Delta \nu/\nu$.

 \subsubsection{The Minimum Noise in a Coherent System}

%**************************************************************************************************
%new fig for mixersidentical 
%OK TLW Aug 5, 07---this is now Fig 5.4
\begin{figure}[b]%
%\sidecaptionab
%  \mpicplace{5cm}{10cm}
  \includegraphics[width=7cm,height=7cm]{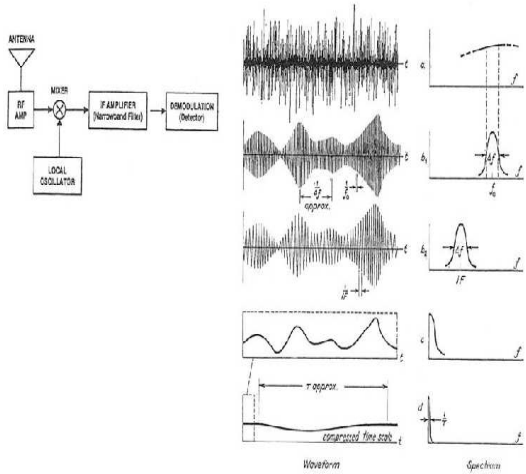}
  \caption[ The parts of a coherent receiver]
 {\label{coh-rx-w-noise}
On the left are the  parts of a coherent receiver. For $\nu_0 < $120 GHz one can amplify before  mixing while for higher frequencies the first 
circuit element is the mixer.  In the middle are sketches of the time behavior of broadband  noise after 
passing through these parts of the receiver; on the right are sketches of the corresponding frequency behavior. The mixer shifts the signal to a 
lower freuqency while the amplifier increases the output over a narrow band. The square law detector converts  rapidly oscillating signals into a 
smooth response that has positive values. The middle and right sketches are taken from  \cite{PAWSEY}. }
\end{figure}
%********************************************************************************************************************

     The ultimate limit for  coherent receivers or amplifiers is obtained
by application of the {\em Heisenberg\index{Heisenberg uncertainty principle}
uncertainty principle}.  The $minimum\index{Noise!minimum}$ noise of a coherent  amplifier 
     results in a receiver noise temperature of
\begin{equation}
     \label{heisenberg6}
       \fboxsep2mm\fbox{$ \displaystyle
            T_{\rm rx}({\rm minimum}) = \frac{h \nu}{ k}
        $}\quad .
     \end{equation}
     For {\em incoherent} detectors, such as bolometers, phase is not
     preserved, so this limit does {\em not} exist. In the 
     millimeter wavelength regions, this noise temperature limit is
     quite small. At $\lambda$=2.6 mm ($\nu$=115 GHz), this limit  is 5.5 K. 
% This derivation is valid for receiver noise temperatures that are rather far above the quantum limits. 
%added Aug 9, 07

\subsubsection{ Elements of Coherent Receivers}
The noise in the first element dominates the system noise. The exact expression is given by the {\em Friis } relation 
\index{Friis formula} which takes into account the effect of 
having cascaded amplifiers\index{Noise!cascaded amplifiers} :
 \begin{equation}
   \label{ampst}
   \fboxsep2mm\fbox{$ \displaystyle
      T_{\rm S} = T_{\rm S1} + \frac{1}{G_1} T_{\rm S2} + \frac{1}{G_1 G_2}
           T_{\rm S3} + \dots + \frac{1}{G_1 G_2 \dots G_{n-1}} T_{{\rm S}n}
       $} \quad .
 \end{equation} 
Where $G_1$ is the gain of the first element, and $ T_{\rm S1}$ is the noise temperature of this element. The corresponding values apply to the following elements in a receiver. 
%******************************************************************************************************************************
For $\lambda <$3 mm, cooled first elements typically have $G_1=10^3$ and $T_{\rm S1}=50$K; for 
$\lambda <$0.3 mm, cooled first elements typically have $G_1=1$ and $T_{\rm S1}=500$K. 
%******************************************************************************************************************************

 \subsubsection{Mixers}\label{mixer-rx}
%**************************************************************************************************
%new fig for mixersidentical 
%OK TLW Aug 5, 07---this is now Fig 5.4
\begin{figure}[b]%
%\sidecaptionab
%  \mpicplace{5cm}{10cm}
  \includegraphics[width=5cm,height=4.5cm]{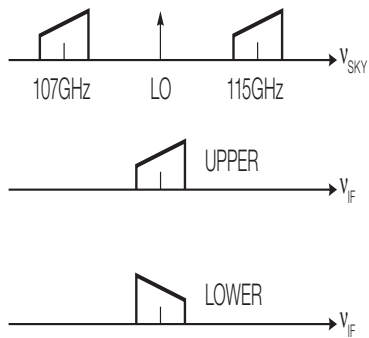}
  \caption[frequencies shifted from the sky frequency to the output  of a double sideband mixer]
 {\label{sidebands-mix}
A sketch of the frequencies shifted from the sky frequency  (top) to the output (lower) of a double sideband mixer. In this example,
the input is at the sky frequencies for the Upper Side Band (USB) of 115 GHz, and Lower Side Band (LSB) of 107 GHz
 while the output frequency is 4 GHz. The slanted boxes represent the passbands; the 
direction of the slant in the boxes  indicate the upper (higher) and lower (lower) edge of the bandpass in 
frequency \cite{RW5}. ALMA mm mixers are SSB mixers so these effects are avoided. }
\end{figure}
%********************************************************************************************************************

 Mixers allow the signal frequency to be changed without altering the characteristics of the signal.  
In the mixing process, one  multiplies  the input signal with an intense monochromatic signal from a {\em local oscillator}, LO\index{Oscillator!local},  in a non-linear circuit element. 
 In principle a  mixer produces a shift in frequency of an input signal with no other effect on the signal properties. For a single mixer,  two frequency bands, at equal 
separations from  the LO frequency 
are shifted into   intermediate (IF) frequency band. This is Double Sideband (DSB) mixer operation. %For spectral lines, usually one sideband is wanted, but the other not. 
These are referred to as the {\em signal} and {\em image} bands\index{Image!frequency}. 
In the mm/sub-mm wavelength ranges, such mixers are still
commonly used as the first stage of a receiver. For single dish {\em continuum} measurements,
both sidebands contain the signal, so  DSB operation does not decrease the signal-to-noise (S/N) ratio. However, for single dish spectral line measurements, the spectral
line of interest is in one sideband only. The other sideband is then a
source of extra noise (i.~e.~lower S/N ratio) and perhaps confusing lines.  Therefore single\index{Receiver!single sideband} sideband (SSB)
operation is desired. If the image sideband is eliminated, the mixer is said to operate in  SSB mode. This can be  accomplished by inserting a 
filter before the mixer. However, filters cause a loss of signal, so lower the S/N ratio. For low noise  applications  a more complex arrangement is needed. 
In many cases,  a {\em single sideband } mixer is used. This  consists of two identical mixers driven by a single local oscillator through a phase shifting device.

%**************************************************************************************************************
For example, the ALMA Band 7 mm/sub-mm receiver shifts a signal centered at 345 GHz from the sky frequency, 341 to 349 GHz  
to  4-12 GHz. At 4 to 12 GHz  the signal is  easily amplified. This is an SSB mixer; the unwanted sideband is not accepted.

Noise in mixers has 3 causes. The first is the mixer itself. Since one half of the input signal at $\nu_{\rm sky}$ is shifted to an unwanted frequency $\nu_{\rm LO} + \nu_{\rm sky}$, 
the signal suffers 
 a factor-of-two (3 db) loss. This is referred to as {\em conversion\index{Loss!conversion} loss}.Classical mixers typically have  3 db (=factor of 2) loss.  In Eq.~\ref{ampst}, $G_1=1/2$. 
In addition there will be an additional noise contribution from the mixer itself ( $T_{\rm S1}$ in (\ref{ampst})). Second, the LO may have $''$phase noise$''$, that is a 
rapid change of phase, which will affect signal properties, so adds to the  uncertainties. 
Third, the amplitude of the LO may vary. This  effect can be minimized since    the mixer LO power is adjusted so that the
mixer output is saturated. Then there is  no variation of the output signal power if LO power  varies.

\subsection{HEMTs/MMICs}
 Within a highly ordered crystal made of identical atoms, free electrons can  move
only within certain {\em energy bands}\index{Band!energy}.
By varying the material, both the width of the band, the
{\em band gap}\index{Band!gap}, and the energy to reach a conduction band can be varied.
                 The crucial part of any semiconductor device is the 
junction\index{Semiconductor!junction}. On the one side there is an
excess of material with negative carriers\index{Carrier!negative}, 
forming n-type
material and the other side material with a deficit of electrons, that
is p-type material. The p-type material has an excess of 
positive carriers, or {\em holes}. At the
junction of a p- and n-type material, the electrons in the n-type
material can diffuse into the p-type material (and vice-versa), so
there will be a net potential difference. The diffusion of charges, p
to n and n to p, cannot continue indefinitely, but a difference in
the charges near the boundary of the n and p material will remain,
because of the low conductivity of the semiconductor material.
                      From the potential difference at the junction, a flow
of electrons in the positive direction is easy, but a flow in the negative
direction will be hindered. 
%****************************************************************************************************************************************************************************
Typical p-n junctions have a slow  response so  are suitable only as square-law
detectors. Schottky (metal-semiconductor) junctions have a lower 
capacitance, so
are better suited to applications such as microwave mixers. 
                       The combination of three layers, p-n-p, in a so-called $''$sandwich$''$, is a simple extension of the p-n junction. 
In Field Effect Transistors, FET's, 
the electric field  controls the carrier flow
 so this is an 
amplifier\index{Amplifier!transistor}.  
                            At  frequencies of 100 GHz, unipolar devices,
which have only one type of carrier, are used as microwave amplifier front ends. 
 {\em High Electron Mobility Transistors},   HEMTs,  are an evolution of FETs. The design goals of HEMT's are: 
(1) to obtain lower intrinsic amplifier noise 
and (2) operation at higher frequency. In HEMTs, the charge carriers are 
present in a  channel of small size.  This confinement 
 gives rise to  a two dimensional electron gas, or   
  $''$2 DEG$''$, 
 where  there is less scattering and hence lower 
noise.   When cooled, there is a significant improvement in the 
noise performance, since the main contribution is from the oscillations of nuclei 
in the lattice, which are strongly temperature dependent. 
%*****************************************************************************************************************************************************************
To extend  the operation of HEMT to  
 higher frequencies, one must increase  electron mobility, $\mu$, 
and saturation
velocity $V_{\rm s}$\index{Velocity!saturation}. A  reduction in  the 
scattering by introducing impurities ($''$doping$''$)  
leads to a larger electron mobility, $\mu$,  and hence faster transit
times, in addition 
to lower amplifier noise.

For use  up to $\nu$=115\,GHz  with good noise
performance, one has turned to  modifications of  HEMTs based on advances in material-growth technology. 
The SEQUOIA receiver array of the Five College Radio Astronomy Observatory 
 uses Microwave Monolithic Integrated Circuits (MMIC's)\index{MMIC} in 32  front ends for a 16 beam, 
two polarization system (pioneered by S. Weinreb). 
The MMIC is a complete amplifier on a single semiconductor, instead of using lumped components. 
The MMIC's have excellent performance in the 80--115\,GHz region without requiring tuning adjustments. 
The simplicity makes MMIC's better suited for multi-beam systems.

For low noise IF amplifiers,  4 to 8 GHz IF systems using Gallium-Arsinide HEMTs with 5 K noise temperature 
and more than 20 db of gain have been built.  With Indium-Phosphide HEMTs on GaAs-substrates, 
even lower noise temperatures are possible. As a rule of thumb, one expects an increase of 0.7 K per GHz for GaAs, 
while the corresponding value for InP HEMTs is 0.25 K per GHz.  For front ends,  noise temperatures  
of the  amplifiers in the 18--26  GHz range are typically 12 K.

\subsubsection{Superconducting Mixers}
  \label{supercond.mix}
 
 Very general, semi-classical
considerations indicate that the slope of the current-voltage, $I$-$V$, curve for classical mixers %shown in Fig.\ts \ref{energyband} 
changes gently.  This  leads to a relatively poor noise
\index{Noise!figure} figure for classical mixers, since much of the input signal is not
converted to a lower frequency.
 
A significant improvement can be obtained if the junction is operated
in the superconducting mode. Then the  gap  between filled and empty states   is
comparable to  photon energies in the mm/sub-mm range. In addition, the LO  power requirements are $\approx 1000$ times lower than are
needed for conventional mixers. Finally, the physical layout of such
devices is simpler since the mixer is a planar device, deposited on
a substrate by lithographic techniques.
%*****************************************************************************************
 %OK TLW Aug 5, 07---this is now Fig 5.11ab
%%4.18
%\begin{figure}[b]
%\sidecaptionab
  %\mpicplace{11.7cm}{4.6cm}
 % \includegraphics[width=11.7cm,height=4.6cm]{kap4f18.eps}
  % \caption[The energy bands of a (SIS) junction]
% {\label{SIS layer}
% ({\bf a}) A sketch of the energy bands of a
%superconducting-insulating-superconducting (SIS) layer. The gap
% between the filled states ({\it below, shaded\/}) and the empty states
%({\it above\/})
% is $2\Delta$. On the left we sketch the process in which an
% electron absorbs a photon, gaining energy which allows it to
%tunnel through the insulating barrier.
% ({\bf b}) The $I$-$V$ curve of the SIS junction. The dashed line
% is the behavior when there is no local oscillator (LO) power; the
% solid line shows the behavior with LO power supplied. When DC biased at
% $2\Delta / e$, the SIS mixer efficiently converts photons to a
% lower frequency}
%\end{figure}
 %**************************************************************************************************************
SIS\index{Mixer!superconducting} mixers consist of a
superconducting layer, a thin insulating layer and another
superconducting layer.
SIS mixers  
depend on single carriers; a longer but more accurate description 
of  SIS mixers is  $''$single quasiparticle
photon assisted tunneling detectors$''$. When the SIS junction is properly 
biased, the filled states  
reach the level of the unfilled band, and the
electrons can quantum mechanically tunnel through the insulating strip.
The $I$-$V$ curve for a SIS device  
shows 
sudden jumps in the $I$-$V$ curve; these are typical of quantum-mechanical 
phenomena. For low noise operation, the SIS mixer must be DC biased 
at an appropriate voltage and current.  If, in addition to the mixer 
bias, there is a source of photons of
energy $h \nu$, then the tunneling can occur. If one then biases an SIS\index{SIS device} device and  applies an LO
signal at a frequency $\nu$,  the $I$-$V$ curve becomes very sharp, so the conversion of sky signals to the IF frequency  is much more effective than with a classical
mixer.

%*****************************************************************************************
%%OK TLW Aug 5, 07---this is now Fig 5.12
%
%%4.19
%\begin{figure}
%\sidecaption
  %\mpicplace{6.4cm}{4.6cm}
%  \includegraphics[width=6.4cm,height=4.6cm]{kap4f19.eps}
% \caption[A SIS junction in a waveguide]
% { \label{SIS mixer}
% A sketch of an SIS junction placed in a waveguide. Both the
% astronomical and LO signals enter through the waveguide; the
% difference frequency is present at the IF output. This response
% is optimized by tuning the back short}
%\end{figure}
%*************************************************************************************************************
Under certain circumstances, SIS devices can produce gain. If the SIS 
mixer is tuned to produce substantial gain, it is 
unstable. Thus, this  not useful in
radio astronomical applications. In the mixer mode, that is, 
as a frequency converter,
SIS devices can have a small amount of gain. This tends to  balance inevitable  mixer  
losses, so SIS devices have  losses 
that are lower than classical 
mixers. SIS mixers have performance that is   
unmatched in the mm/sub-mm region. In addition to  {\em single sideband} properties, improvements to existing designs 
include {\em tunerless} and SIS mixers. Tunerless mixers
have the advantage of repeatability in tuning.  For ALMA, SIS mm mixer designs  are wideband, tunerless, single sideband 
devices with extremely low mixer noise temperatures. %we show a photo of such a device in Fig.\,\ref{SRON-mixer}. 

An increase in the gap energy is needed to allow the efficient mixing at higher frequenccies. This is done with Niobium superconducting
materials; the geometric junction sizes are  $1\ts \mug$m
by $1\ts \mug$m. For frequencies above 900\,GHz, one uses
niobium nitride junctions. Variants of such devices, such as the use of 
junctions in
series, can be used to reduce the capacitance. An alternative is to
reduce the size of the individual junctions to $0.25\ts \mug$m.

\noindent
SIS mixers are the front ends of choice for operation between 150\,GHz and 900\,GHz because  these  are low-noise devices, the IF bandwidths can be $>$1 GHz,  
these are tunable over $\sim$30\% of the frequency range and  the local oscillator power needed is $<$1 $\mug{\rm W}$.

\subsubsection{Hot Electron Bolometers}

Superconducting Hot Electron Bolometer-mixers (HEB) are heterodyne 
devices, in spite of the name. These mixers make use of superconducting thin
films which have sub-micron sizes. In an HEB mixer excess
noise is removed either by diffusion of hot electrons out
the junction, or by an electron-phonon exchange.\index{Bolometer!Hot Electron}\index{HEB} 
The first HEBs operating on radio telescopes and used to take astronomical
 made use of electron-phonon exchange.
The HEB junctions were of $\umu$m size, consisting  of Niobium
Nitride (NbN), cooled to 4.2 K. Junctions of Aluminum-Titanium-Nitride, AlTiN,
have provided lower receiver noise temperatures. The IF center frequency was
1.8 GHz, and a had a full width of  1 GHz. Such a system was used to measure the $J=9-8$ carbon monoxide line
at 1.037 THz. Later, the group of  First Physical Institute at Cologne University used the Atacama Pathfinder 
EXperiment (APEX) telescope to measure the [N\,II] line at 1.5\,THz (see Table 1).

%\section{Summary of Front Ends Presently in Use}

%***************

%\section{Low-Noise Front Ends and IF Amplifier Systems}
\subsubsection{Single Pixel Receiver Systems}

In summary, devices that provide the lowest noise front ends are:

\noindent
for $\nu< 115$\,GHz, High Electron Mobility Transistors (HEMT) and Microwave Monolithic Integrated Circuits (MMIC)
 
\noindent 
for $72 < \nu< 800$\,GHz, Superconducting Mixers (SIS)
 
\noindent
 for $\nu> 900$\,GHz, Hot Electron Bolometers (HEB)
%\end{itemize}

%See  Fig.\ts \ref{rntemp} for a comparison of front end receiver noise temperatures. 

\vspace{0.2cm}
\noindent
SIS mixers provide the lowest  receiver noise 
in the mm and sub-mm range. SIS mixers are much more sensitive than classical
Schottky mixers, and require less local oscillator power,  
but  must be cooled to 4 Kelvin. All 
millimeter mixer receivers are tunable over 10 to 20 \% of the sky frequency. 
From the band gaps of junction materials,
there is a short wavelength limit to the operation of SIS
devices. For spectral line measurements at wavelengths, at
$\lambda<0.3$\,mm, superconducting Hot Electron Bolometers 
(HEB), which have no such limit, have been developed.  
At frequencies above 2 THz there
is a transition to far-infrared and optical techniques. The highest frequency heterodyne systems in radio astronomy 
are used in the Herschel-HIFI satellite. These are SIS and HEB mixers.

%In addition to the front end mixers and amplifiers,  the connections  between feed and
%receiver are also often cooled. For some receivers 
%sections of the feed horn with the coupling probes are cooled.

The SIS or HEB mixers convert the sky frequency to the fixed IF frequency,
where the signal is amplified by the IF amplifiers. Most of the amplification is done in the IF. The 
IF should only contribute a negligible part to the system noise temperature.
Because some losses are associated with frequency conversion,
the first mixer is a major source for the system noise.
 Two ways exist to decrease this contribution: (1)  by use of either an SIS or HEB mixer to convert the input to 
a lower frequency, or (2) at lower frequencies  by use of a  low-noise amplifier before the mixer.

%%*****************************************************************************************
 %OK TLW Aug 5, 07---this is now Fig 5.14
%%4.13
%\begin{figure}
%\sidecaption
  %\mpicplace{8.4cm}{6.4cm}
 % \includegraphics[width=0.8\linewidth]{kap4f13e4final.eps}
 % \caption[Receiver noise temperatures for  coherent
%amplifiers]
% { \label{rntemp}
%Receiver noise temperatures for coherent amplifier systems compared to
%the temperatures from different astronomical sources and the 
%atmosphere. The atmospheric emission is based on a model of zenith
%emission for 0.4 mm of water vapor (plot from B. Nicolic (Cambridge Univ.) from the 'AM' program of S. Paine (CfA)). This does not take
%into account the absorption corresponding to this emission. In the 1 to
%26 GHz range, the horizontal bars represent the noise temperatures of
%HEMT amplifiers (priv. comm. H. Mattes, MPIfR). The shaded region between 85 and
%115.6 GHz is the receiver noise for the SEQUOIA array which is made up
%of monolithic millimeter integrated circuits (MMIC), at Five College
%Radio Astronomy Observatory. The meaning of the other symbols is given
%in the upper left of the diagram (partially taken from Rieke 2002). The mixer noise
%temperatures given as double sideband (DSB) values were converted to single sideband (SSB) temperatures by increasing the receiver noise by a factor of 2. The ALMA mixer 
%noise temperatures  are SSB. The  HEMT values are SSB.}
%\end{figure}
%************************************************************************************************************************
%%TLW insert from April 13, 03
\subsubsection{Multibeam Systems}

At 3 mm, the SEQUOIA array receiver produced at the Five College Radio Astronomy Observatory (FCRAO) with 32  MMIC front ends
connected to 16 beams 
had been used on 14-m telescope of the FCRAO
 for the last
few years. 
Multibeam system that use SIS front ends are rare. A 9 beam Heterodyne Receiver Array of
SIS mixers at 1.3 mm, HERA,  has been installed on the IRAM 30-m
millimeter telescope  to
measure spectral line emission.   To simplify data taking and
reduction, the HERA beams are kept on a Right Ascension-Declination coordinate frame. 
HARP-B is a 16 beam SIS system in operation at the James-Clerk-Maxwell telescope. The sky frequency is 
325 to 375 GHz. The beam size of each element is
14$''$, with a beam separation of 30$''$, and a FOV of about 2$'$. 
The total 
number of spectral channels in a 
heterodyne multi-beam system will be  large. In addition,  
complex optics is  needed  to properly illuminate all of the beams. 
In the mm range this  
usually means that the receiver noise temperature of each element is somewhat 
larger than
that for a single pixel receiver system. For further details of  SEQUOIA,  HERA or  HARP, see the appropriate web sites.

For single dish continuum measurements at 
$\lambda <$ 2 mm,
multi-beam 
systems make use of bolometers. GeGa bolometers are the most common systems and the best such 
systems have a large number of beams. In the future,  TES bolometers seem 
to have great promise. 
Compared to incoherent recievers,  
heterodyne systems are still the most efficient 
for spectral lines in the range $\lambda>$0.3mm, although Fabry-Perot 
systems (such as SPIFI; see the web site)  may be competitive for
some projects.  For bolometers  on the Herschel satellite, one  uses gratings or   Fabry-Perot systems.  For SCUBA-2, an analog
Michelson\index{Spectrometer!Michelson} ( Fourier
\index{Spectrometer!Fourier}
transform interferometer) is proposed. 

\subsection[Back Ends]{Back Ends: Spectrometers}
\sectionmark{Back Ends}
% \label{crp}

%\subsubsection{Spectrometers}
 
The term $''$Back End$''$ is used to specify the devices following the IF amplifiers. Of the many different  back ends that have been designed for
specialized purposes, spectrometers\index{Spectrometer} are probably
the most widely used. Previously  this was carried out in especially designed hardware, but recently 
there have been devices based on general purpose digital computers.
 
Spectrometers analyze the spectral
information contained in the radiation field. To accomplish this these 
 must be  SSB\index{Receiver!SSB}  and the frequency resolution $\Delta
\nu$ is usually fine; perhaps in the kHz range. In addition, the time stability
must be high. 
If a resolution of $\Delta \nu$ is to be achieved for the spectrometer,
all those parts of the system that enter critically into the
frequency\index{Frequency!response} response have to be maintained to
better than $0.1 \, \Delta \nu$. An overview of the current state of spectrometers is to be found in \cite{BAKER}.

\subsubsection{Multichannel Filter Spectrometers}
 \label{mfsp}
 
The time needed to measure the power spectrum for a given celestial
position can be reduced by a factor $n$ if the IF section with the
filters defining the bandwidth $\Delta \nu$, the square-law detectors
and the integrators are built not merely once, but $n$ times. Then these
form $n$ separate channels that simultaneously measure different
(usually adjacent) parts of the spectrum.
%\subsection{Fourier and Autocorrelation Spectrometers}

\subsection{Fourier, Autocorrelation and Cross Correlation Spectrometers}
One method is to Fourier Transform (FT) the input,  $v(t)$,  to obtain $v(\nu)$  and then  square  $v(\nu)$   to obtain the 
Power Spectral Density\index{Density!spectral power}. 
From the Nyquist  theorem, it is necessary to sample  at a rate equal to twice the bandwidth. 
 These are referred to as $''$FX$''$ autocorrelators\index{autocorrelator!FX}. 
Recent developments at the Jodrell Bank Observatory have led
to the building of COBRA (Coherent Baseband Receiver for
Astronomy). COBRA can analyze  a 100 MHz bandwidth.  A similar device with a 1 GHz bandwidth has been built at the Max-Planck-Institut in 
Bonn for use in the mm/sub-mm range on APEX.

%\subsubsection{Autocorrelation and Cross Correlation Spectrometers}
%*****************************************************************************************************************put Weinreb ref here on 14 Apr. 08 TLW
 For Autocorrelators, or XF systems, the input $v(t)$ is correlated with a delayed signal $v(t -\tau)$ to obtain the autocorrelation\index{Function!autocorrelation} function $R(\tau)$ function.
$R(\tau)$ is then Fourier Transformed to obtain the spectrum. For an XF system the time delays are performed in a set of serial digital shift registers  
 with a sample delayed by a time $\tau$.  Autocorrelation can also  be carried out 
with the help of analog devices using a  series of cable delay lines. Such analog correlators have been developed at the University of 
Maryland together with NRAO for use on the Green Bank Telescope (GBT); these are used to provide very large bandwidths.

The two significant advantages of digital  spectrometers are: (1)
flexibility and (2) a noise behavior that follows $1/\sqrt{t}$
after many hours of integration. The flexibility allows one to select
many different frequency resolutions and bandwidths or even to employ a
number of different spectrometers, each with different bandwidths,
simultaneously. The second advantage follows
directly from their digital nature. Once the signal is digitized, it is
only mathematics. Tests on astronomical sources have shown that the noise follows a
${1}/{\sqrt{B t}}$ behavior for integration times  $>$100
hours; in these aspects, analog spectrometers are more limited.

%Corrections for one-bit  quantization can be expressed in a closed form; the details are  presented  in Appendix D. Corrections for 3 level and 4 level (2 bit) 
% follow  similar procedures, but 3 and 4 level corrections cannot be expressed  in a closed form.
%**************************************************************************************************************
%now fig 5.15; fig drawn in Aug 07 TLW, mod on 21 Sept 07, mod
%%4.24
% \begin{figure}[t]
%\sidecaption
   %\mpicplace{11.7cm}{7.6cm}
 % \includegraphics[width=11.7cm,height=7.6cm]{kap5f16e5.eps}
 %  \caption[An autocorrelation spectrometer]
% { \label{o-bit-ac}
% Schematic showing the essential functional blocks of an XF
% autocorrelation spectrometer}
% \end{figure}
%************************************************************************************************************** 

A serious drawback of  digital auto and cross correlation
spectrometers had been limited bandwidths. Previously 50 to 100\,MHz
had been the maximum possible bandwidth. This was determined  by the requirement to meet Nyquist sampling rate, so that the analog-to-digital (A/D) 
converters, samplers, shift registers and multipliers
 would have to run at a rate equal to 
twice the bandwidth. The speed of the electronic circuits was limited. However, advances in digital technology  in recent years have allowed the construction of 
autocorrelation spectrometers with several 1000
channels covering  bandwidths of several GHz. 
 
 Another improvement is the use of {\em recycling\index{Correlator!recycling}}
auto and cross correlators. These spectrometers have the property that
the product of bandwidth, $B$ times the number of channels, $N$, is a
constant. Basically, this type of system functions by having the digital part
running at a high clock rate, while the data are sampled at a much slower
rate. Then after the sample reaches the $N$th shift register
% (Fig.\ts \ref{o-bit-ac})
it is reinserted into the first register and another set of delays are
correlated with the current sample. This leads to a higher number of channels
and thus higher resolution. Such a system has the advantage of
high-frequency resolution, but is limited in bandwidth. This has the greatest
advantage for longer wavelength observations.
 Both of these developments have tended to make the use of digital
spectrometers more widespread. This trend is likely to continue.
 
%\subsubsection{Cross Correlation Systems}
\label{cr-co-rs}
 
Autocorrelation systems are used in single telescopes, and make use of
the symmetric nature
of the  autocorrelation function ACF. Thus, the number of delays gives the number of spectral channels. For cross-correlation, the current 
and delayed samples refer to different inputs. 
Cross-correlation systems are used in interferometers. %This is a generalization of (\ref{rt}). 
In the simplest case of a two-element interferometer, the output is {\em not}
symmetric about zero time delay, but can be expressed in terms of
amplitude and phase at each frequency, where both the phase and
intensity of the line signal are unknown. Thus, for interferometry the
zero delay of the ACF is placed in channel $N/2$ and is
 in general asymmetric. The number of delays, $N$, allows the
determination of $N/2$ spectral intensities, and $N/2$ phases. The cross-correlation hardware can employ 
either an XF or a FX correlator. The FX correlator has the advantage that the time delay 
is just a phase shift, so can be introduced more simply.

\subsubsection{Acousto-Optical Spectrometers}
 
Since the discovery of molecular line radiation in the
mm wavelength range there has been a need for spectrometers with
bandwidths of several GHz. At 100\,GHz, a velocity range of
300\,km\,s$^{-1}$ corresponds to 100\,MHz, while the narrowest line widths
observed correspond to 30\,kHz. Autocorrelation spectrometers can reach
such large bandwidths only if complex methods are used. %Thus 
%multichannel filter spectrometers are more common 
%in the mm and sub-mm ranges.  
    The AOS makes use of the diffraction of light by ultrasonic waves: these  cause
 periodic density variations in the medium through which it passes.
These density variations in turn cause variations in the bulk
constants $\epsilon$ and $n$ of the medium, so that a plane
electromagnetic wave passing through this medium will be affected. The most advanced AOS's have been designed and built at the 
First Physical Institute at Cologne University. 
%****************************************************************************************************************************************************practical-antennas
%\endd
%\chapter{Practical Aspects of Filled Aperture Antennas}
%\label{filled-aperture}
\section{Filled Aperture    Antennas}
 \label{desap}
%****************************************  
The material  presented in the following emphasizes  descriptive antenna parameters. These  allow a fairly accurate but rather simple description  of 
antenna properties that 
are needed for an accurate interpretation of  astronomical measurements. For more detail, see \cite{BAARS}.  
An important concept is {\em reciprocity}\index{Antennas!reciprocity}, in which the properties of  antennas are the same, irrespective whether these are 
 used for receiving or transmitting. Reciprocity is limited under some (very special) circumstances that are not encountered in astronomy. 

 \subsection{Angular Resolution}
From diffraction theory \cite{JENKINS}, the angular resolution of a reflector of diameter $D$ at a wavelength $\lambda$ is 
 \begin{equation}
   \label{diffraction}
   \fboxsep2mm\fbox{$ \displaystyle
     \theta  = \frac{\lambda}{D}  $}\quad .
\end{equation}
This simple result gives an approximate value for $ \theta$ but more detailed results must take into account details of the {\em illumination}. 
 
\subsection{The Power Pattern $P (\vartheta, \varphi)$}
 \label{ant-describ} 

Often, the {\em normalized\index{Power!normalized pattern} power
pattern}, not the power pattern is measured:
 \begin{equation}
   \label{norm-pp}
   \fboxsep2mm\fbox{$ \displaystyle
     P_{\rm n} (\vartheta, \varphi) = \frac{1}{P_{\rm max}} \,P (\vartheta,
                 \varphi)  $}\quad .
\end{equation}
%\longpage
The reciprocity theorem provides a method to
measure this quantity.
The
radiation source 
can be replaced by a small diameter
radio source. The flux densities of such sources are determined
by measurements using horn antennas at centimeter and millimeter wavelengths. At short wavelengths, one uses
planets, or moons of planets, whose surface temperatures are
determined from infrared data.
 
   If the power pattern is measured using artificial
transmitters, care should be taken that the
distance from a large antenna A (diameter $D$) to a small antenna B (transmitter) is so large that B produces plane waves across the aperture $D$ of antenna A, i.~e.~is 
in the far radiation field of A. This is the {\em Rayleigh} distance; it requires that the
curvature of a wavefront emitted by B is much less than a
wavelength across the geometric dimensions of A.
This curvature must be $k \ll 2D^2/\lambda$\index{Antenna!Rayleigh distance},
for an antenna of diameter $D$ and a wavelength $\lambda$.

 Consider the power pattern of the antenna used
as a transmitter. If the spectral 
power density,  $\mathcal{P}_\nu $ in [W Hz$^{-1}$] is fed into a lossless isotropic
antenna, this would transmit $P$ power units per solid
angle per Hertz. Then the total radiated power at frequency $\nu$ is $ 4 \pi \, P_\nu$
In a realistic, but still lossless antenna, a power $P (\vartheta, \varphi)$
per unit solid angle is radiated in the direction $(\vartheta, \varphi)$.
If we define the directive gain $G (\vartheta, \varphi)$ as the
%actor by which  $P (\vartheta, \varphi)$ exceeds $P$ , we have
\beq
     P (\vartheta, \varphi) = G (\vartheta, \varphi) \,P
 \eeq
or
 \begin{equation}
  \label{gain}
   \fboxsep2mm\fbox{$ \displaystyle
       G (\vartheta, \varphi) = \displaystyle \frac{4 \pi
     P (\vartheta, \varphi)}{\int\!\!\int P (\vartheta, \varphi) \,\diff
                \Omega}    $} \quad .
 \end{equation}
Thus the gain\index{Gain} or directivity\index{Directivity} is also a
normalized power pattern similar to (\ref{norm-pp}),
but with the difference that the normalizing factor is
$\int P (\vartheta, \varphi) \diff \Omega / 4 \pi$. This is the gain relative to a lossless isotropic
source. Since such an isotropic source
cannot be realized in practice, a measurable quantity is
the gain relative to some standard antenna such as a
half-wave dipole whose directivity is
known from theoretical considerations.
\subsection{The  Main Beam Solid Angle}
% %************************************************************************************************************************************************************************************ 
% %now fig 7.1 TLW, AUG 07
%%%5.5
\begin{figure}[b]
%%\sidecaption
%%  \mpicplace{2.6cm}{5.9cm}
    \includegraphics[width=2.6cm,height=5.9cm]{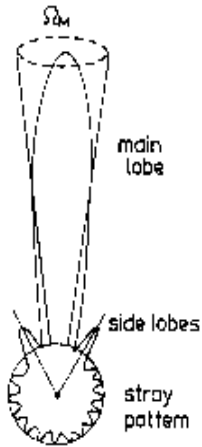}
  \caption[The polar power pattern]
 { \label{polar-pattern}
 A polar power pattern showing the main beam, and near and far
 side lobes. The weaker far side lobes have been combined to form
 the stray pattern}
\end{figure}
 %************************************************************************************************************************************************************************************ 
The {\em beam\index{Beam!solid angle} solid angle} $\Omega_{\rm A}$ of an
antenna is given by
%%*****************
 \begin{equation}
  \label{beam-sa}
  \fboxsep2mm
  \fbox{$ \displaystyle
  \Omega_{\rm A} = \parbox{8mm}
                       {$ \vspace*{-2mm}
                         {\displaystyle
                          \int\!\!\int \atop
                         {\!\!\!\!\!\scriptstyle 4 \pi}}  $  }
         P_{\rm n} (\vartheta, \varphi) \,\diff \Omega =
         \int\limits_0^{2\pi}\!\int\limits_{0}^{\pi}
        P_{\rm n} (\vartheta, \varphi) \sin \vartheta \diff \vartheta
          \diff \varphi $}
 \end{equation}
%**************
this is measured in steradians (sr). The integration is extended
over the full sphere $4\pi$, such that $\Omega_{\rm A}$ is
the solid angle of an ideal antenna having $P_{\rm n} = 1$ for
all of $\Omega_{\rm A}$ and $P_{\rm n} = 0$  everywhere else. Such
an antenna does not exist; for most antennas the (normalized) power pattern has
considerably larger values for a
certain range of both $\vartheta$ and $\varphi$ than for the remainder of
the sphere. This range is called the
main beam or main lobe of the antenna; the remainder are  the side lobes or back lobes
(Fig.\,\ref{polar-pattern}). For actual situations, the properties are well defined up to the the shortest operating wavelengths. At the shortest wavelength, there is still  a main beam, 
but much of
the power enters through sidelobes. In addition, the main beam efficiency may vary significantly  with  elevation and weather has a large effect. 
Thus, the ability to accurately calibrate the radio telescope at sub-mm 
wavelengths is challenging.  %%%%%%%%%%%%%%%%%

In analogy to (\ref{beam-sa}) we define the {\em
   main beam \index{Main beam}
solid angle} $\Omega_{\rm MB}$  by
%%********************************************************************************************************************************* 
 \begin{equation}
   \label{main-lsa}
     \fboxsep2mm\fbox{$\displaystyle
         \Omega_{\rm MB} =
                \mathop{\int\!\!\int}\limits_
               {\scriptstyle \rm main \atop
                \scriptstyle \rm  lobe}
              P_{\rm n} (\vartheta, \varphi) \,\diff \Omega
        $} \quad .
 \end{equation}
%%************************************************************************************************************************************************ 
The quality of an antenna as a direction
measuring device depends on how well the
power pattern is concentrated in the main beam. If
a large fraction of the received power
comes from the side lobes it would be rather difficult to determine the location of  the radiation source, the  so-called $''$pointing$''$. %

It is appropriate to
define a {\em main beam efficiency\index{Beam!efficiency}} or (in the slang of antenna specialists) 
{\em beam efficiency},
$\eta_{\rm B}$,  by
 \begin{equation}
    \label{mbeam-eff}
    \fboxsep2mm\fbox{$ \displaystyle
       \eta_{\rm B} = \frac{\Omega_{\rm MB}}{\Omega_{\rm A}} $} \quad .
 \end{equation}%

The main beam efficiency, 
 $\eta_{\rm B}$ is %an indication of 
the fraction of
the power is concentrated in the
main beam. The main beam efficiency can be
modified (within certain limits) for parabolic
antennas by the illumination  
of the main reflector.   
If the FWHP beamwidth is well defined, the location of an isolated source is determined to an accuracy given by  the FWHP divided by  the S/N ratio. 
Thus, it is possible to determine positions to 
small fractions of the FWHP beamwidth if noise is the only limit. 
 
   Substituting (\ref{beam-sa}) into (\ref{gain}) it is easy to see
that the maximum directive\index{Gain!directive} gain $ G_{\rm max}$ or {\em
directivity\index{Directivity}}  $\cal D $ can be expressed as
 \begin{equation}
   \label{g-max}
    \fboxsep2mm\fbox{$ \displaystyle
       {\cal D} = G_{\rm max} = \frac{4 \pi}{\Omega_{\rm A}} $} \quad .
  \end{equation}
%*****************************************************************
%*****************************************************************
 The angular extent of the main beam is usually described
by the {\em half power beam\index{Beam!width} width} (HPBW),
which is the angle between points of the main
beam where the normalized power pattern falls to $1/2$ of the maximum. 
For elliptically shaped  main beams, values for
widths in orthogonal directions are
needed. The beamwidth is related to the
geometric size of the antenna and the
wavelength used; the exact beamsize depends on details of  illumination.  
 
\subsection{ Effective Area}
Let a plane wave with the power density
$\mid\!\langle\vec{S}\rangle\!\mid$
 be intercepted by an antenna. A certain
amount of power is then extracted by the antenna from
this wave; let this amount of
power be $P_{\rm e}$. We will then call the fraction
\vspace{-6pt}
 \begin{equation}
   \label{a-e}
     A_{\rm e} = P_{\rm e} \, / \mid\!\langle\vec{S}\rangle\!\mid
 \end{equation}
the {\em effective\index{Aperture!effective} aperture} of the antenna.  $A_{\rm e}$  has the dimension of m$^2$.
Comparing this to the {\em geometric\index{Aperture!geometric} aperture}
$A_{\rm g}$ we can
define an aperture\index{Aperture!efficiency} efficiency $\eta_{\rm A}$ by
 \begin{equation}
   \label{aper-eff}
   \fboxsep2mm\fbox{$ \displaystyle
       A_{\rm e} = \eta_{\rm A} A_{\rm g} $} \quad .
 \end{equation}

Consider a receiving antenna with a normalized power
pattern $P_{\rm n} (\vartheta, \varphi)$ that is pointed at a
brightness distribution $B_\nu (\vartheta, \varphi)$ in the sky.
Then at the output terminals of the antenna, the total power per unit bandwidth,   $\cal{P}_\nu $ is  
 \begin{equation}
   \label{tp-output}
      \mathcal{P}_\nu = \fract{1}{2}\, A_{\rm e} \int\!\!\int
        B_{\nu} (\vartheta, \varphi) \,P_{\rm n} (\vartheta, \varphi)
            \,\diff \Omega \; .
 \end{equation}
By definition, we are in the Rayleigh-Jeans limit, and can therefore exchange the brightness distribution by an equivalent distribution of brightness
temperature. Using the Nyquist theorem (\ref{Nyquist}) we can introduce an
equivalent {\em antenna temperature}  $T_{\rm A}$ by
 \begin{equation}
   \label{ant-temp}
     \mathcal{P}_\nu = k \,T_{\rm A} \,.
 \end{equation}
This definition of {\em antenna\index{Temperature!antenna}
temperature}\index{Brightness temperature} relates the output of the antenna to the power
from a matched resistor. When these two power levels are equal, then the
antenna temperature is given by the temperature of the resistor.
Instead of the effective aperture $A_{\rm e}$ we can introduce
the beam solid angle $\Omega_{\rm A}$. Then (\ref{tp-output}) becomes
 \begin{equation}
  \label{a-temperature}
  \fboxsep2.5mm
  \fbox{$\displaystyle
    T_{\rm A}(\vartheta_0, \varphi_0) =  \frac{\int
     T_{\rm B}(\vartheta, \varphi) P_{\rm n}(\vartheta - \vartheta_0,
        \varphi - \varphi_0) \sin \vartheta \diff \vartheta \diff
         \varphi}{\int P_{\rm n}(\vartheta, \varphi) \diff \Omega}  $ }
 \end{equation}
which is the {\em convolution} of the brightness temperature
with the beam pattern of the telescope.
The brightness temperature\index{Temperature!brightness}
 $T_{\rm b}(\vartheta, \varphi)$
 corresponds to the thermodynamic temperature of the
radiating material only for thermal radiation in the Rayleigh-Jeans
limit from an optically thick source; in all other cases $T_{\rm B}$ is
only an convenient quantity that in general depends on the frequency. It is important to note that from (\ref{a-temperature}), 
the {\em measured} size of an extremely compact 
(i.~e.~``point'') source is the beam size. 
 
The quantity $T_{\rm A}$ in (\ref{a-temperature}) was obtained for an
antenna
with no ohmic losses, and no absorption in the
earth's atmosphere. In the mm/sub-mm range, the 
expression $T_{\rm A}$ in (\ref{a-temperature}) is actually $T_{\rm A}'$, that is, a temperature corrected for atmospheric losses.  We will use the term $T_{\rm A}'$ in 
discussions of mm/sub-mm calibration. %Sect.\ts 8.2.5.
Since $T_{\rm A}$ is the quantity measured while $T_{\rm B}$ is the one
desired, (\ref{a-temperature}) must be inverted.
(\ref{a-temperature}) is an integral equation of the first kind, which
in theory can be solved if the full range
of $T_{\rm A}(\vartheta, \varphi)$ and $P_{\rm n}(\vartheta, \varphi)$
are known. In
practice this inversion is possible only
approximately. Usually both  $T_{\rm A}(\vartheta, \varphi)$
and $P_{\rm n}(\vartheta, \varphi)$ are
known only for a limited range of $\vartheta$ and $\varphi$
 values, and the measured data are not free of errors.
Therefore usually only an
approximate deconvolution is performed. A special case is one for which
the source distribution
 $T_{\rm B} (\vartheta, \varphi)$
has a small extent compared to the telescope beam. Given a finite
signal-to-noise ratio, the best estimate for the upper limit to the
actual FWHP source size is one-half of the FWHP of the telescope beam. This point cannot be emphasized too much: we {\em cannot} assign an arbitrarily small size to a source. The best is one-half of the 
antenna FWHP!

\subsection{ Antenna Feed Horns Used Today}
 
Feed horns are needed to guide the power from the reflector (in free space conditions) into the receiver (in a waveguide); 
 details are contained in \cite{GOLDSMITH},\cite{LOVE}. The electric and magnetic field strengths
at the open face of a wave guide will
vary across the aperture. The power
pattern of this radiation depends both on the dimension of
the wave guide in units of the
wavelength, $\lambda$, and on the mode of the wave. The greater the
dimension of the wave guide in $\lambda$,
the greater is the directivity of
this power pattern. However, the
larger the cross-section of a wave guide in terms of
the wavelength, the more
difficult it becomes to restrict the wave to a single
mode. Thus
wave guides of a given size can be used only for a limited
frequency range. The aperture
required for a selected directivity\index{Directivity} is then obtained by
flaring the sides of a section of the
wave guide so that the wave guide becomes a horn.

Great advances in the design of feeds have
been made  since
1960, and most parabolic dish antennas now use hybrid\index{Horn!hybrid mode feed}
mode
feeds. % (Fig.\,\ref{hybrid-m}).
Such $''$corregated horns$''$  are also referred to as  {\em Scalar} or {\em Multi-Mode} feeds. Today such feed horns  are 
used on  all parabolic antennas. These provide much higher 
efficiencies than simple single mode horns  and are well suited for
 polarization measurements.

%%************************************************************************************************************** 
%6.6
%\begin{figure}[t]
%\sidecaption
%\mpicplace{11.7cm}{7.8cm}
%\includegraphics[width=2.6cm,height=2.6cm]{HornPdB-B2.eps}
%   \caption[hybrid-mode feed for $\lambda$ 2~mm]%
%	{\label{hybrid-m} A corrugated waveguide %hybrid mode feed for  $\lambda$ 2~mm on the Plateau de Bure interferometer. The one Euro coin in the upper left is shown to 
%give a notion of scale. (Photo courtesy of B. Lazareff (IRAM))  }
%\end{figure}
%%%************************************************************************************************************** 

\subsection{Multiple Reflector Systems}
 
If the size of a radio telescope is more than a few hundred
wavelengths,
 designs similar to those of optical telescopes are preferred. For such telescopes
Cassegrain, Gregorian and Nasmyth
systems have been used. In a
Cassegrain\index{Antenna!cassegrain } system, a convex
hyperbolic reflector is introduced into the converging beam
immediately in front of the prime focus.
This reflector transfers the converging rays to a
secondary focus which, in most
practical systems is situated close to the apex of the main
dish. A Gregorian\index{Gregory system} system
makes use of a concave reflector with an elliptical
profile. This must be
positioned behind the prime focus in the diverging beam.
In the Nasmyth\index{Antenna!nasmyth } system this secondary focus is situated in the
elevation axis of the telescope by introducing another, usually flat,
mirror. The advantage of a Nasmyth system is that the receiver front
ends remain horizontal while  when the telescope is pointed toward
different elevations. This is an advantage for receivers cooled with
liquid helium, which may become unstable when tipped. Cassegrain and Nasmyth foci are commonly used in the mm/sub-mm wavelength ranges.

%***********************************************************************************************************************July 31 TLW
In a secondary 
reflector  system,  feed illumination  
beyond the edge   receives radiation
from the sky, which has a temperature of only a few K. For low-noise systems,
this results in only a small overall system noise temperature. This is 
significantly less than for prime focus systems. This is quantified in the so-called $''$G/T value$''$\index{Antenna!G/T value}, that is, the ratio of  antenna  gain of  to 
system noise. Any telescope design must aim to minimize the excess noise at the receiver input while maximizing gain. 
For a specific antenna, this maximization  involves the design of feeds and
the choice of foci. 
%************************************************************************************************************** 
 %now fig 7.6 TLW, AUG 07
%7.7 now
\begin{figure}[t]
%\sidecaption
%\mpicplace{11.7cm}{4.6cm}
  \includegraphics[width=11.7cm,height=4.6cm]{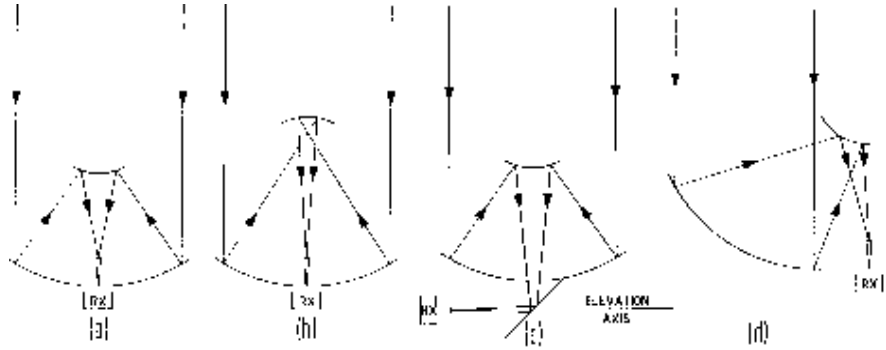}
  \caption[Geometry of Cassegrain, Gregory, Nasmyth and offset Cassegrain systems]
	{ \label{geo-cg}
   The geometry of ({\bf a}) Cassegrain, ({\bf b}) Gregorian, ({\bf c})
   Nasmyth and
({\bf d}) offset Cassegrain systems (from \cite{RW4}). 
	}
\end{figure} 
%************************************************************************************************************** 

That the secondary reflector blocks the central parts
in the main dish from reflecting the incoming radiation  causes
some interesting differences between the actual beam
pattern from that of an unobstructed telescope. Modern designs seek to minimize blockage due to the support legs and subreflector.

Realistic filled aperture antennas ($''$single dishes$''$) will have a beam pattern different from
a uniformly illuminated unblocked aperture.  First the illumination of
the reflector will not be uniform but has a taper by 10\,dB or
more at the edge of the reflector.  
The
side-lobe\index{Antenna!side lobes} level is strongly influenced by this taper: a larger taper lowers the sidelobe level. 
Second, the secondary
reflector must be supported by three or four support legs, which will
produce aperture blocking and thus  affect the shape of the beam
pattern. In particular feed leg blockage will cause deviations from circular
symmetry. For altitude-azimuth telescopes these side lobes will
change position on the sky  with hour angle. This may be a serious
defect, since these effects will be significant for maps of low
intensity regions if the main lobe is near an intense source.
The side lobe response can also  dependent on the polarization
of the incoming radiation.

   A disadvantage of on-axis systems, regardless of focus, is that they
are often more susceptible to instrumental frequency baselines, so-called
{\em baseline\index{Antenna!baseline ripple} ripples} across the receiver band
than primary focus systems. Part of
this ripple is caused by multiple reflections of noise from source or receiver 
in the antenna structure. Ripples can arise in the reciever, but these can be removed or compensated rather easily. 
Telescope baseline ripples are more difficult to eliminate: it is known that 
large amounts of blockage and  larger feed sizes lead to large baseline ripples.  
The effect is discussed in somewhat more
detail in Sect.\,\ref{lr-observ}. 
 The influence of baseline ripples on measurements  can be reduced  to a limited
extent by appropriate observing procedures.
A possible solution is the construction of
off-axis\index{Antenna!off-axis} systems such as the GBT.

\subsection{Antenna Tolerance Theory}

   It is convenient to distinguish several different kinds of
phase\index{Antenna!phase errors} errors in the current
distribution across the aperture of a two-dimensional
antenna.
%************************************************************************************************************** NASH Diagram
%6.4 now 7.7 Oct 2007
\begin{figure}[b]
%\sidecaption
%\mpicplace{5cm}{4.2cm}
  \includegraphics[width=2.6cm,height=2.6cm]{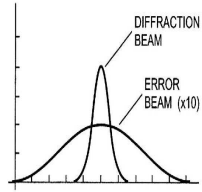}
  \caption[A sketch of main beam and error beam]
      { \label{ap-eff}
       A sketch showing the relative widths of the main beam and the error beam (shown ten times actual size). 
The width of the main beam is determined by diffraction for the entire antenna, while that  for the error beam is determined by 
scale size of the surface irregularities (from \cite{WH}).}
\end{figure}

%************************************************************************************************************inserted Sept 25, 07
 
 If the correlation distance $d$ is of the same
order of magnitude as the diameter
of the reflector, part of the phase error can be treated as
a systematic phase variation,
either a linear error resulting only in a tilt of the main
beam, or in a quadratic phase error
which could be largely eliminated by refocussing. For $d \ll D$
 the phase errors are almost
independently distributed across the aperture, while for intermediate
cases according to  a good
estimate for the expected value of the RMS\index{Antenna!surface rms} phase error is
given by: 
 \begin{equation}
   \label{rms-ph}
   \fboxsep2mm\fbox{$ \displaystyle
      \bar{\delta^2} = \left( \frac{4 \pi \epsilon}{\lambda} \right)^2
          \left[ 1 - \exp \left\{ - \frac{\Delta^2}{d^2} \right\}
           \right] $}  \quad ,
 \end{equation}
where $\Delta$ is the distance between two points in the aperture
that are to be compared and $d$ is
the correlation distance. The gain of the system now depends
both on $\bar{\delta^2}$ and on $d$. In addition,
there is a complicated dependence both on the grading of the
illumination and on the manner in which $\delta$ is
distributed across the aperture. The 
Ruze theory 
can be
expressed in the following terms: the gain of a reflector with surface   phase errors can be
approximated by an expression
 \begin{eqnarray}
   \label{gu-ruze}
    G(u) = \eta \expo^{-\bar{\delta^2}}
	 \left(\frac{\pi D}{\lambda} \right)^2
	\Lambda_1^2 \left(\frac{\pi D \,u}{\lambda} \right)+
	( 1- \expo^{-\bar{\delta^2}})
	 \left(\frac{2\pi d}{\lambda} \right)^2
      		\Lambda_1^2 \left( \frac{2 \pi d \,u }{\lambda} \right) \ts ,
\kern-100pt\nonumber\\
\phantom{XX}
 \end{eqnarray}
 
\vspace{-10pt}\noindent
where\\[1.5ex]
 \begin{tabular}{@{}lll@{}}
 \tabcolsep0.5mm
 $\eta$ & & is the aperture efficiency, \\
 $u $ &=& $\sin \vartheta$,  \\
 $\Lambda_1 (u) $ &=& $ \frac{2}{u} J_1 (u) $
            is the Lambda function,  \\
 $D$ & & the diameter of the reflector, and \\
 $d$ & & the correlation distance of the phase errors. \\[1ex]
 \end{tabular}  \\
From  are now two contributions to the beam shape of the system.
The first is that of a circular aperture with a diameter $D$,  but whose response is reduced  
due to the
random phase error $\delta$. The second
term is the so-called {\em error\index{Error beam} beam}. This can
be described as
equal to the beam of a (circular) aperture with a diameter $2 d$, its
amplitude multiplied by 
\beq
(1- e^{-{\bar\delta}^2})
\eeq 
The error beam
contribution therefore  will decrease to zero as
$\bar \delta \rightarrow 0$.
%******************************************************************************************************************************************************************************************

%***************************************************************************************************************************************************************************************** 
The gain of a filled aperture antenna
with phase irregularities $\delta$ cannot increase indefinitely with increasing frequency but
reaches a maximum at $ \lambda_{\rm m} = 4\pi \epsilon$,
and this gain is a factor of 2.7  below that of an error-free antenna of identical dimensions.
Then, if the frequency can be determined at
which the gain of a given
antenna attains its maximum value, the RMS phase error and the
surface irregularities $\epsilon$ can be
measured electrically. Experience with many radio telescopes 
shows reasonably good agreement of such values for
$\epsilon$ with direct measurements, giving empirical support for
the Ruze tolerance theory.
 
%**************************************************************************************************************************************obs-methods
%\endd
\section{Single  Dish Observational Methods}

\subsection{The Earth's Atmosphere}
 \label{observations}
For ground--based facilities, astronomical signals entering the receiver has been attenuated by the
earth's atmosphere. In addition to attenuation, the receiver noise is increased by atmospheric emission, the
signal is refracted and there are
changes in the path length. Usually these effects change slowly with
time, but there can also be rapid changes such as scintillation and anomalous
refraction. Thus propagation properties  must 
be taken into account, if the astronomical measurements are to be correctly interpreted.  In the mm/sub-mm 
ranges, tropospheric effects\index{Absorption!tropospheric} are  especially important. The
various constituents of the atmosphere absorb by different amounts.
Because the atmosphere can be considered to be in LTE, these
constituents are also radio emitters.

The total amount of precipitable water\index{Precipitable water} (usually
measured in mm) above an altitude $h_0$ is an integral along the
line-of-sight. Frequently, the amount of H$_2$O is determined by 
measurements carried out at 225\,GHz combined with models of the atmosphere. For mm/sub-mm sites, measurements of the 183 GHz 
spectral line of water vapor (see Fig.\,\ref{enlwat})  
can be used to estimate the total 
amoount of H$_2$O in the atmosphere. For sea level sites, the 22.235 GHz line of water vapor is used for this purpose.
The scale height $H_{\rm H_2 O} \approx 2\, {\rm km}$, is considerably less
than $H_{\rm air} \approx 8\, {\rm km}$ of dry air.
For this reason, sites for submillimeter radio telescopes
are usually mountain sites with elevations above $\approx 3000$\,m.

The variation of the intensity of an extraterrestrial radio source
due to propagation\index{Propagation effects} effects in the atmosphere is
given by the standard relation for radiative transfer through a uniform medium (from Eq.\,\ref{tb-isoth}).
 \begin{equation}
   \fboxsep2mm
   \fbox{$ \displaystyle
     T_{\rm B}(s) = T_{\rm B}(0) \,{\rm e}^{- \tau_\nu(s)} +
          T \, (1- \,{\rm e}^{- \tau_\nu(s)})   $} \quad .
  \end{equation}
%\begin{equation}
%\label{I(s)}
%I_\nu (s) = I_\nu (0) \expo^{-\tau_\nu (0)} +
%\int\limits_0^{\tau_\nu (0)} B_\nu (T(\tau))\expo^{-\tau} \diff \tau
%\,,
%\end{equation}
%where
%\begin{equation}
% \label{tau s}
%	\tau_\nu (s) = \int\limits_{s_0}^s \kappa_\nu (s) \diff s \,.
%\end{equation}
Here $s$ is the (geometric) path length along the line-of-sight with
$s = 0$ at the upper edge of the atmosphere and $s = s_0$ at the antenna. Both the (volume) absorption\index{Volume absorption coefficient}
coefficient $\kappa$ and the
gas temperature $T$ will vary with $s$, introducing the mass absorption
coefficient $k_\nu$ by
\begin{equation}
\label{k}
	\kappa_\nu = k_\nu \cdot \varrho  \,,
\end{equation}
where $\varrho$ is the gas density; this variation of $\kappa$ can
mainly be traced to that of $\varrho$ as long as the gas mixture remains
constant along the line-of-sight. This is a simplified relations. For a more realistic calculations, one must use a multi-layer model.

%\enlargethispage{\baselineskip}
%\vglue-6pt 
Because the variation of $\varrho$ with $s$ is so much larger that that of
$T(s)$, a useful approximation can be obtained by introducing an effective
temperature\index{Temperature!effective} for the atmosphere %so that
%\begin{equation}
% \label{TAtm}
%	T_{\rm A}(s) = T_{\rm B}(0) \expo^{-\tau(0)} + T_{\rm Atm}
%                 (1 - \expo^{-\tau(0)})  \, .
%\end{equation}
%The first term is absorption in the earth's atmosphere, while the second
%is emission.
%The physical conditions are contained in the expression for  $\tau(0)$, the total opacity along the line-of-sight, and in $T_{\rm Atm}$.
%In Fig.\,\ref{atmosphere}
%we show a model of the atmosphere used to predict attenuation from O$_2$
%and H$_2$O, and other constituents. This example gives an indication of
%the influence of the atmosphere at cm,  mm and sub-mm wavelengths. High frequency resolution  measurements
%of the atmospheric emission are possible and aid in improving models. 

%If the atmosphere is  reasonably stable so that both
%$T_{\rm Atm}$ and $\tau_0$  can be obtained by measuring the  calibration
%source repeatedly so that its radiation enters theion from the atmosphere for  different air masses $X$. The unknown atmospheric parameters then
%can be obtained from a least squares fit of $T_{\rm A}$ against $X$.
 
Refraction\index{Refraction effects!atmospheric} effects in the atmosphere
depend on the real part of the
(complex) index of refraction. Except for  the
anomalous dispersion near  water vapor lines and  oxygen lines, 
it is essentially independent of frequency. The average effect  can be calculated; fine corrections are determined from pointing corrections. 

A rapidly time variable effect is {\em anomalous refraction}
(see \cite{RW4}).  % This has been found at 1.3\,cm
%at Effelsberg and at 3\,mm and 1.3\,mm at
%Pico Veleta as well as other sites. 
If anomalous refraction
is important, the apparent positions of
radio sources appear to move away from their actual positions by up to
40$\arcsec$ for time periods of 30 seconds. This effect occurs
more
frequently in summer afternoons than during the night. Anomalous
refraction is
caused by small variations in the H$_2$O content, perhaps by single cells
of moist air. In the mm and sub-mm range, there are measurements of
rapidly time variable noise contributions, the so-called
{\em sky\index{Sky noise} noise}. This is produced by variations in
the water vapor
content in the telescope beam.  It does not depend  in an obvious
way on the transmission of the atmosphere.  This behavior is expected if the effects arise within a few km above the telescope and the cells  have limited sizes.
 %But experience has shown
%that times  of high atmospheric attenuation are often times of low
sky noise. %The fluctuations of the water content along the
%line-of-sight probably cancel reasonably well when the water content is
%high, but then the transmission is small.  As expected, sky noise increases with increasing telescope beam separation, being  larger for 
small telescopes ($D < 3$\,m) than for large telescopes
%($D > 10$\,m).
 
\vglue-10pt
\vglue-10pt

\subsection{Millimeter and Sub-mm Calibration Procedures}\label{mm-cal-proc}
 
\vglue-6pt
 
\subsubsection{General}

In radio astronomy, one usually  follows a three step practical procedure: (1)  the measurements must be corrected for 
atmospheric effects, (2)  relative  calibrations are  made using secondary standards and (3) if needed, gain versus elevation curves for the antenna must be established. 
In the mm/sub-mm ranges,  primary 
calibrators are, in many cases, planets or moons of planets; more common secondary calibrators are  non-time-variable  compact sources.

\subsubsection[Calibration of mm and sub-mm Wavelength  Heterodyne Systems]{Calibration of mm and sub-mm Wavelength  Heterodyne Systems}
 
In the mm/sub-mm wavelength range, the atmosphere has a
larger influence and can change rapidly, so one  must  make accurate
corrections to obtain well calibrated data. In the
mm range, most large telescopes are close to the limits caused by their
surface accuracy, so that the power received in the error beam may be
comparable to that received in the main beam. Thus, one must use a relevant value of 
beam efficiency.
We give an analysis of the calibration procedure which
is standard in spectral line mm astronomy following the presentations in \cite{DOWNES1989}. This calibration reference is
referred to as the {\em chopper wheel} method. The procedure consists of:
(1) the measurement of the receiver output when an ambient (room temperature)
load is placed before the feed horn, and (2) the measurement of the
receiver output, when the feed horn is directed toward cold sky at a
certain elevation. For (1), the output of the receiver while measuring
an ambient load, $T_{\rm amb}$,  is $V_{\rm amb}$:
\begin{equation}
	  \label{vamb}
	V_{\rm amb}= G\,(T_{\rm amb}+T_{\rm rx}) \, .
\end{equation}
For step (2), the load is removed; then the response to empty sky noise,
$T_{\rm sky}$ and receiver cabin (or ground), $T_{\rm gr}$, is
\begin{equation}
  \label{vsky}
	  V_{\rm sky} =
	G\,[F_{\rm eff} \,T_{\rm sky} +  (1 - F_{\rm eff}) \,T_{\rm gr} +
T_{\rm rx}] \,.
\end{equation}
$F_{\rm eff}$ is referred to as the {\em forward efficiency}. This is
basically the fraction of power in the forward beam of the feed. If we
take the difference of $V_{\rm amb}$ and $V_{\rm sky}$, we have
\begin{equation}
  \label{vcal}
 V_{\rm cal} = V_{\rm amb}-V_{\rm sky}=G \,F_{\rm eff} \,T_{\rm amb}
                \expo^{-\tau_\nu} \, ,
\end{equation}
where $\tau_\nu$ is the atmospheric absorption at the frequency of
interest. The response to the signal received from the radio
source, $T_{\rm A}$, through the earth's atmosphere, is
\beq
	\Delta V_{\rm sig}=G \,T_{\rm A}' \expo^{-\tau_\nu}
\eeq
or
\beq
	T_{\rm A}'=\frac{\Delta V_{\rm sig}}{\Delta V_{\rm cal}} \,F_{\rm eff}
             \,T_{\rm amb}
\eeq
where $T_{\rm A}'$ is the antenna
temperature\index{Temperature!antenna}\index{Temperature!brightness}
 of the source outside the
earth's atmosphere. We define
\begin{equation}
 \label{corta}
	T_{\rm A}^*=\frac{T_{\rm A}'}{F_{\rm eff}} =
	\frac{\Delta V_{\rm sig}}{\Delta V_{\rm cal}} \,T_{\rm amb} \, .
\end{equation}
The quantity $T_{\rm A}^*$ is commonly referred to as the {\em corrected
antenna temperature},\index{Temperature!antenna}
 but it is really a {\em forward beam brightness
temperature}.\index{Temperature!brightness}
This is the $T_{\rm MB}$ of a source filling a large part of
the sky, certainly more than 30\arcmin.
 
For sources (small compared to 30\arcmin), one must
still correct for the telescope beam efficiency, which is commonly
referred to as $B_{\rm eff}$.
Then
\beq
	T_{\rm MB}=\frac{F_{\rm eff}}{B_{\rm eff}} \,T_{\rm A}^*
\eeq
for the IRAM 30\,m telescope, $F_{\rm eff} \cong 0.9$ down to 1\,mm
wavelength, but $B_{\rm eff}$ varies with the wavelength.
So at $\lambda = 3$\,mm, $B_{\rm eff} = 0.65$, at 2\,mm
$B_{\rm eff} = 0.6$ and at 1.3\,mm $B_{\rm eff} = 0.45$,
for sources of diameter $ < 2\arcmin$. For
an object of size 30\arcmin, $B_{\rm eff}$ at all these wavelength is 0.65.
As usual $T_{\rm MB}$ can be considered a black body with the
temperature $T_{\rm MB}$, which just fills the beam. This analysis is
the one used at IRAM. %This uses a somewhat different notation, but
%the physics is basically the same. We give a comparison between these
%systems.
%
%
%\begin{table}
%  \small
%\noindent\begin{tabular}{@{}cc@{}}
%\kern-60.232pt{\bf Table 8.1}\\ [3.5pt]
%  \hline
%  \noalign{\vskip 3pt}
%  \kern-39.620pt Kutner \& Ulich &\kern-20pt CLASS / IRAM \\
%  \noalign{\vskip 3pt}
%  \hline
%  \noalign{\vskip 3pt}
%\kern-39.62pt  $\eta_{\rm l}  $   &\kern-20pt  $F_{\rm eff}$\\
%\kern-39.62pt  $\eta_{\rm s}  $   &\kern-20pt  $B_{\rm eff}$\\
%\kern-39.62pt  $\eta_{\rm c}  $   &\kern-20pt      --       \\
%\kern-39.62pt  $\eta_{\rm fss}$ &\kern-20pt $ B_{\rm eff} / F_{\rm
%eff}$\\
%\kern-39.62pt  $\eta_{\rm f}  $   &\kern-20pt      --       \\
%    \noalign{\vskip 3pt}
%    \hline
%    \noalign{\vskip 3pt}
%      $\eta_{\rm s} = \eta_{\rm l}   \cdot \eta_{\rm fss},\;
%       \eta_{\rm f} = \eta_{\rm fss} \cdot \eta_{\rm c} $  \\
%    \end{tabular} 
%\end{table}

%In the notation of Kutner and Ulich~(1981), $T_R^* = T_{\rm MB}$, 
In
terms of our notation 
\beq
\eta_{\rm MB}= \frac{\Omega_{\rm MB}}{\Omega_{\rm F}}=
\frac{B_{\rm eff}}{F_{\rm eff}} \, ,
\eeq
%while in that of Kutner and Ulich
%\beq
%	\eta_{\rm fss}= \frac{\Omega_{\rm D}}{\Omega_{\rm F}} \,,
%\eeq
%where $\Omega_{\rm D}$ is the solid angle of the diffraction pattern and
%$\Omega_{\rm F}$ is the forward beam solid angle.

An antenna pointing at an elevation $z$ to a position of empty sky will
deliver  an antenna temperature
\begin{equation}
 \label{TAmm}
	 T_{\rm A}(z)=T_{\rm rx}+T_{\rm atm} \,\eta_{\rm l}
  \,(1 - \expo^{-\tau_0 X(z)})  +  T_{\rm amb} (1 - \eta_{\rm l}) \,,
\end{equation}
where \\[1.5ex]
\begin{tabular}{@{}lcl@{}}
	$T_{\rm rx}$   &:& system noise temperature,\\
	$T_{\rm atm}$  &:& effective temperature of the atmosphere,\\
	$T_{\rm amb}$  &:& ambient temperature,\\
	$\eta_{\rm l}$ &:& feed efficiency (typically $\eta_{\rm l} = 0.9$),\\
\end{tabular} \\
\begin{tabular}{@{}lcl@{}}
	$\tau_0$       &:& zenith optical depth,\\
	$X(z)$         &:& air mass at zenith distance $z$. \\[2ex]
\end{tabular}\\
 
These parameters can be determined by a series of calibration
measurements. The efficiency $\eta_{\rm l}$ and the other
parameters can be determined by a least squares fit of (\ref{TAmm}),
that is a {\em skydip}\index{Skydip} giving $T_{\rm A}$ as a function of
$X(z)$. Depending on the weather conditions these measurements
have to be repeated at time intervals
from 15 minutes to hours or so, to be able to detect variations in the
atmospheric\index{Atmospheric conditions} conditions. At some observatories
a small separate instrument, a {\em taumeter} (a sky horn that measures the sky temperature at elevations 90$^{\rm o}$, 60$^{\rm o}$, 30$^{\rm o}$ and 20$^{\rm o}$)  is
available to determine the opacity $\tau$ at 10 minute intervals.

For larger mm wavelength telescopes one cannot perform tipping
measurement often. If a taumeter  is not available one must use a more
elaborate procedure. By measuring the response to a cold load, one can
determine the receiver noise, and can obtain a good estimate of the noise
from the atmosphere. Then, assuming a value of $T_{\rm atm}$ and
$\eta_{\rm l} = F_{\rm eff}$, one can then determine
$\tau = \tau_0 \,X(z)$, and can use this to correct
for atmospheric absorption.

To calibrate spectral lines, one frequently measures sources for which one has
single sideband spectra. Finally observations often have to be corrected for
yet another effect: the telescope efficiency usually depends on elevation.
Usually the telescope surface is set optimally for some intermediate zenith
distance $z \approx 40^\circ$. Both for $z \approx 0^\circ$ and $70^\circ$
the efficiency usually decreases somewhat. % by about 25\,\%.
% The apparent flux density
%corrected for the atmospheric transmission then can be represented by
%\begin{equation}
%	S_\nu = S_{\nu0} \,(A \cos (z-C)^2 + B \sin(z-C)^2) \, .
%\end{equation}
%In the wavelength  $ \lambda \sim 1300 \,\mug{\rm m}$ for the IRAM 30\,m
%telescope a set of values $A = 1.01, B = -0.2, C =52.2^{\circ}$ is obtained.%%
 
%\vglue-12pt
%\vglue-3pt

\subsubsection{Bolometer Calibrations}\label{incoherent-cals}

Since most bolometers are A.~C. coupled (i.~e.~responds to differences), the D.~C. response (i.~e.~responds to total power) to  $''$hot--cold$''$ or $''$chopper wheel$''$ 
calibration methods are not used. Instead astronomical data are calibrated in two steps: (1) measurements of atmospheric emission to 
determine the opacities at the azimuth of the target source, and (2) 
the measurement of  the response of a nearby source with a known flux density; immediately  after this, a  measurement of the target source is carried out.

\subsubsection{Compact Sources}
 
     Usually the beam of radio telescopes are well characterized by
Gaussians. As mentioned previously, Gaussians have the great
advantage that the convolution of two Gaussians is another Gaussian. For
Gaussians, the relation between the observed source size,
     $\theta_{\rm o}$, the beam size $\theta_{\rm b}$, and actual
     source size, $\theta_{\rm s}$, is given by:
     \begin{equation}
     \label{gausssize}
     \theta_{\rm o}^2 = \theta_{\rm s}^2 + \theta_{\rm b}^2 \, .
     \end{equation}
 
This is a completely general relation, and is widely
used to deconvolve source from beam sizes. Even when the
source shapes are not well represented by Gaussians these are usually
approximated by sums of Gaussians in order to have a convenient
representation of the data.  The accuracy of
     any determination of source size is limited by  (\ref{gausssize}).
     A source whose size is less than 0.5 of the beam is so
     uncertain that one can only give as an upper
     limit of $0.5 \,\theta_{\rm b}$.\index{Brightness temperature}

If the (lossless) antenna (outside the earth's atmosphere)
is pointed at a  source of
known flux density $S_\nu$ with an angular diameter that is small
compared to the telescope beam, a power $W_\nu \diff \nu$
at the receiver input terminals
\beq
   W_\nu \,\diff \nu = \fract{1}{2} A_{\rm e} \,S_\nu \diff \nu =
        k \,T_{\rm A}' \diff \nu
\eeq
is available. Here $T_{\rm A}'$ is the antenna temperature corrected for effect of the earth's atmosphere. Thus
  \begin{equation}
  \label{t-a}
  \fboxsep2mm
  \fbox{$\displaystyle
           T_{\rm A}' = \Gamma S_\nu
        $}
  \end{equation}
where $\Gamma$ is the {\em sensitivity\index{Sensitivity!telescope}} of the
telescope measured in
K Jy$^{-1}$. Introducing the aperture
efficiency $\eta_{\rm A}$ according to (\ref{aper-eff}) we find
\begin{equation}
  \label{sensit}
  \fboxsep2mm
  \fbox{$\displaystyle
         \Gamma  = \eta_{\rm A} \frac{\pi D^2}{8 k}
            $} \quad .
  \end{equation}
Thus $\Gamma$ or $\eta_{\rm A}$ can be measured with the
help of a calibrating source provided that the
diameter $D$ and the noise power scale in the receiving system
are known.
In practical work the inverse of relation (\ref{t-a})
is often used. Inserting numerical values we
find
\begin{equation}
 \label{Tcal}
	S_\nu = 3520 \,\frac{T_{\rm A}' [{\rm K}]}{\eta_{\rm A} [{\rm D/m}]^2}
\,.
\end{equation}
  The {\it brightness\index{Temperature!brightness} temperature} is defined
as the Rayleigh-Jeans
     temperature of an equivalent black body which will give the same
     power per unit area per unit frequency interval per unit solid angle
     as the celestial source. Both $T_{\rm A}'$ and T$_{\rm MB}$ are defined in the
     classical limit, and {\it not} through the Planck relation. 
     However the brightness temperature scale has been corrected for
     antenna\index{Temperature!antenna}
      efficiency.  
The conversion from source flux density
     to source brightness temperature for sources with sizes small
     compared to the telescope beam is given by (Eq.\,\ref{flux-den-tb2}):
     %\begin{equation}
     %S =\frac{ 2 \,k \,T_{\rm MB} \,\Omega }{ \lambda^2} \,.
     %\end{equation}
     %For Gaussian source and beam shapes, this relation is:
     %\begin{equation}
%\label{g-beam-RJ}
%     S = 2.65 \,\frac{T_{\rm MB} \,\theta_{\rm o}^2} {\lambda^2 } \,,
%     \end{equation}
%where the wavelength is in centimeters, and the observed source
%     size is taken to be the beam size, given in
%     arc minutes. Then the flux density is in Jy, and the brightness
%     temperature is in Kelvin.
% 
%     The expression $ T_{\rm MB}$ is still antenna dependent, in the sense
%     that the temperature is a weighted average over the telescope beam, but
%     this relation does take into account corrections for imperfections
%in the
%     antenna surface and the efficiency of feed coupling.
     For sources small compared to
     the beam, the antenna\index{Temperature!antenna}
      and main beam
     brightness temperatures\index{Temperature!brightness}
     are
     related by the main beam efficiency, $\eta_{\rm B}$:
     \begin{equation}
     \eta_{\rm B} =  \frac{T_{\rm A}'}{T_{\rm MB}} \, .
     \end{equation}
 
     The actual source brightness temperature, $T_{\rm s}$ is related to the
     main beam brightness temperature by:
     \begin{equation}
\label{g-sum}
     T_{\rm s} =  T_{\rm MB} \,\frac{(\theta_{\rm s}^2 +
                          \theta_{\rm b}^2 )} {\theta_{\rm s}^2} \,.
     \end{equation}
  Where we have made the assumption that source and beam are Gaussian
     shaped. The actual brightness temperature is a property of the
     source. To obtain this, one must determine the  actual source
     size. This is a science driver for high angular resolution (i.~e.~interferometry) 
measurements. Although the source may not be Gaussian shaped,  one normally  fits  multiple Gaussians to obtain the  effective source
     size.
 
\subsubsection{Extended Sources}
 
   For sources extended with respect to the beam, the process is vastly
     more complex, because the antenna side lobes also receive power
from
     the celestial source, and a simple relation using beam efficiency
     is not possible without detailed measurements of the
     antenna\index{Antenna!pattern} pattern. %As shown in Sect.\,8.2.5 
The error beam may be
a very significant source of calibratoin errors,  particularly  if the
measurements are carried out near the limit of telescope surface accuracy.
In principle $\eta_{\rm MB}$ could be computed by
numerical integration of $P_{\rm n}(\vartheta, \varphi)$
[cf.\,(Eq.\,\ref{beam-sa}) and (Eq.\,\ref{main-lsa})],
provided that $P_{\rm n}(\vartheta, \varphi)$ could be
measured for large range of $\vartheta$ and $\varphi$.
Unfortunately this is not possible since nearly all astronomical sources
are too weak; measurements of bright astronomical objects with known
diameters can be useful.
 
If we assume  a source has a uniform
brightness temperature over a certain
solid angle $\Omega_{\rm s}$, then the telescope measures an antenna
temperature given by (\ref{a-temperature})
which, for a constant brightness temperature across the
source, simplifies to
\beq
   T_{\rm A}' =  \, \frac{\displaystyle
        \int\limits_{\rm source} P_{\rm n}(\vartheta, \varphi) \diff \Omega}
        {\displaystyle
        \int\limits_{4\pi} P_{\rm n}(\vartheta, \varphi) \diff \Omega}
        \, \tb
\eeq
or, introducing (\ref{beam-sa}--\ref{mbeam-eff}),
 \begin{equation}
    \label{t-sml}
   T_{\rm A}' = \eta_{\rm B} \, \frac{\displaystyle
        \int\limits_{\rm source}  P_{\rm n}(\vartheta, \varphi) \diff
\Omega}
        {\displaystyle
         \parbox{8mm}
         {$ %%%\vspace*{-4mm}
              {\displaystyle
                \int \atop
                {\!\!\scriptstyle \rm main \atop
                {\!\!\scriptstyle \rm  lobe }}}  $  }
        P_{\rm n}(\vartheta, \varphi) \diff \Omega} \, \tb
= \eta_{\rm B} f_{\rm BEAM} T_{\rm B} \,,
  \end{equation}
where $f_{\rm BEAM}$ is the beam filling factor. For gaussians 

$$f_{\rm BEAM} = \frac{\theta_{\rm s}^2}{(\theta_{\rm s}^2 + \theta_{\rm b}^2 )}  $$
If the source diameter is of the same order of magnitude as
the main beam the correction
factor in (\ref{t-sml}) can be determined with high precision from
measurements of the normalized power pattern and thus (\ref{t-sml})
 gives a direct determination of $\eta_{\rm B}$, the
beam\index{Efficiency!beam} efficiency. A convenient source with constant
surface brightness in the long cm wavelength range is the moon whose
diameter of $\cong 30\arcmin$ is of the same order of
magnitude as the beams of most large
radio telescopes and whose brightness
temperature\index{Temperature!brightness}
 \begin{equation}
   \label{t-moon}
    T_{\rm b \; moon} \cong 225\,{\rm K}
 \end{equation}
is of convenient magnitude. In the mm and submillimeter range
the observed Moon temperature changes with Moon phase.
 The
planets form convenient thermal sources with known diameters that
can be used for calibration purposes \cite{ALTENHOFF1}, \cite{SANDELL}. 

\subsection{Continuum Observing Strategies}\label{cont-obs}%%
 
\vglue-3pt
\subsubsection{Point Sources}
 
\vglue-1pt\noindent

These measurements are the simplest, but in the sub-mm range the earth's atmosphere is a large
source of radiation.
    Compensation of transmission variations in the atmosphere
is possible if double \index{Double beam system} beam systems can be used. At higher frequencies, in the mm/sub-mm range, the  rapid movement  
of the telescope beam (by small movements of the sub-reflector or a mirror in the path from receiver to antenna) over small angles,  so-called $''$wobbling\index{Wobbling
scheme}$''$  is used to produce two beams on the sky from a single pixel.  This is used at all large millimeter facilities. 
In the simplest system the
individual telescope beams should be spaced by a distance of at least
3~FWHP beam widths, and the receiver should be switched between
them. The separate beams can be implemented in different ways
depending on the frequency and the technical facilities at the telescope.Observing procedures for a double beam system are
usually as follows: the source is first centered on beam one, and the
difference of the two beams is measured, optimally by wobbling the
sub reflector. Then the source is centered on beam two, and again
the difference is measured. This on--off method (better called on--on,
because the source is always in one of the beams) is often arranged in
a time symmetric fashion so that time variations of the sky noise and other
instrumental effects can be eliminated.

Multi-beam bolometer systems are now the rule.
With these, one can measure a fairly large  region simultaneously. This allows a higher mapping speed, and also provides a method to  better cancel  
sky noise due to weather. Such 
weather effects are  referred to as $''$coherent noise$''$. Some details of more recent  data methods are given in e.~g.~\cite{MOTTE}. Usually, a wobbler system is
needed for such arrays, since the bolometer outputs are usually A.~C.--coupled.

\subsubsection{Imaging of Extended Continuum Sources}
 
If extended areas are to be mapped, some kind of raster\index{Raster scan}
scan is employed: there must be reference positions at the beginning and
the end of the scan. Usually the area is measured at least twice in
orthogonal directions. After gridding, the differences of the images are
least squares minimized to produce the best result. This procedure is
called $''$basket\index{Basket weaving} weaving$''$.

Extended emission\index{Emission!extended regions} regions can also
be mapped using a double beam system, with the receiver input periodically switched between the first and second beam. In this procedure,  there is some suppression
of very extended emission. A simple summation along the scan direction has been
used to reconstruct infrared images. More sophisticated schemes  can recover most, but not all of the
information \cite{MOTTE}. Most telescopes therefore have wobbler switching in azimuth to
cancel ground radiation. By measuring a source using scans in azimuth at
different hour angles, and then combining the maps 
one can recover more information \cite{JOHNSTONE}. 
 
\subsection[Additional Requirements for Spectral Line Observations]{Additional 
Requirements\protect\newline for Spectral Line Observations}
 \label{lr-observ}
 
In addition to the requirements placed on continuum receivers, there
are three requirements specific to spectral line receiver systems.
 
%\vglue-6pt
 
\subsubsection{Radial Velocity Settings}

\vglue-2pt\noindent
If the observed frequency of a line is
compared to the known rest frequency, the
relative radial\index{Velocity!radial} velocity of the line
emitting (or absorbing) source and the
receiving system can be determined. But
this velocity contains the motion of the
source as well as that of the receiving
system. Both are measured relative to some
standard of rest. However,
usually only the motion of the source is of
interest. Thus the velocity of the receiving system must
be determined. This velocity can be separated into several
independent components: {\bf 1) Earth Rotation} with a maximum  velocity $v = 0.46 $ km\,s$^{-1}$ and {\bf 2) The Motion of the Center of the Earth}  
relative to the barycenter of the 
Solar System is  said to be reduced to the {\em heliocentric\index{Radial velocity!heliocentric}} system. Correction  algorithms are available 
for  observations of the earth relative to center of mass of the
solar system. 
The resulting radial velocities are then
as close to an inertial\index{Inertial system}
system. Results obtained by many independent
investigations 
show that the solar system
moves with a velocity given by the {\em standard
solar\index{Solar motion!standard} motion}. This is the solar
motion relative to the mode of the velocity of the stars in
the solar neighborhood. Data from which the standard
solar motion has been eliminated are said
to refer to the {\em local standard\index{Local standard of rest} of
rest}
(LSR\index{LSR}). 
 
\subsubsection{Stability of the Frequency Bandpass}
In addition to the stability of the total power of the receiver, one must also have a stable shape of the receiver bandpass. 
At millimeter and sub-mm wavelengths, it is possible that changes in the 
weather conditions between on-source and reference measurements may lead to serious baseline instabilites. If so, the time 
betwenn on-soruce and reference measurements must be shortened 
until stable conditions are reached. Such stability is easier to obtain if the bandwidth of the 
spectrometer is narrow compared to the bandwidth of those parts of the receiver in front of the spectrometer. 
 
\vglue-3pt
 
\subsubsection{Instrumental Frequency Effects}

\vglue-2pt\noindent
The result of any observing procedure should result in a spectrum  in which $T_{\rm A}(\nu) \to 0$ for $\nu$ outside the
frequency range of the line. However, quite
often this is not so because the signal response
was not completely compensated for by a reference measurement, even if receiver stability is ideal.
For larger bandwidths, there is an instrumental spectrum and  a
``baseline\index{Baseline}"
must be subtracted from the difference spectrum. Often a linear
function of frequency is sufficient,
but sometimes some curvature \index{Baseline!curvature} is
found, so that polynomials of second or
higher order must be subtracted.

Often  a sinusoidal or quasi-periodic baseline
\index{Baseline
ripple} ripple is present because  a small fraction of the signal is
reflected off obstructions in  the antenna. This reflected
signal can form a standing\index{Standing wave pattern} wave
pattern. A phase change of $2 \pi$ radians will occur
if either the distance, $d$, over
which the signals are interfering is changed by $\lambda /2$ (where
$\lambda$ is the wavelength) or if the frequency is changed by
 \begin{equation}
    \label{phchange}
      \Delta \nu = \frac{c}{2 \,d} \,.
 \end{equation}

 %***************************************************************************************************************

   There are several possible sources of
reflected \index{Reflected radiation} radiation: (1)  the receiver that injects some noise power into
the antenna, part of which
is then reflected back; or (2) strong continuum radiation from sources or the atmosphere. In both cases the partial
reflection of the radiation in the horn
aperture is the main cause of the instrumental $''$baseline
ripples$''$. 
Both changes in the position of
the telescope and small changes in the
receiving equipment can cause large changes
in the amplitude of the observed ripple. Sometimes the
amplitude of baseline ripple can be reduced considerably by
installing a cone at the apex\index{Apex cone} of the telescope that scatters
the radiation forming the standing wave pattern.

\subsubsection{Spectral Line Observing Strategies}
 \label{splotech}
 
\vglue-2pt\noindent
In radio astronomy spectral line radiation is almost
always only a small fraction of the total
power received; the signal sits on a large
pedestal of wide band noise signals
contributed by different sources: the
system noise, spillover from the antenna
and in some cases, a true background noise.
To avoid the stability problems encountered
in total power systems  the
signal of interest must be compared with another signal
that contains the same total power and
differs from the first only in
that it contains no line radiation. To
achieve this aim, mm/sub-mm  spectral line observers usually make use of  three  observing modes that
differ only in the way the
comparison signal is produced.
\enlargethispage{\baselineskip} 
 
\vglue-2pt\noindent
 
%\vspace*{3ex}
\noindent
{\bf 1) Switching Against an Absorber}
Today this method is  used only in exceptional circumstances such as  for some  studies of the 2.7~K cosmic microwave background. 
\vglue-4pt
 
\vspace*{3ex}
\noindent
{\bf 2 ) Frequency Switching}\index{Frequency switching}.
For many sources, spectral line radiation is a narrow-band
feature, that is, the emission is centered  at  $\nu_0$, present over a small frequency 
interval, $\Delta \nu$, with $\nu_0/\Delta \nu \approx 10^{-6}$. If  all other effects vary very
little over $\Delta \nu $,  changing  the frequency of a 
receiver by perhaps $10 \, \Delta \nu $  produces
a comparison signal with the  line shifted. If  other  contributions hardly differ, the final spectrum is proportional to the difference of these two measurements. 
Such $''$frequency switched$''$ measurements can
be done with almost any  rate. These   produce a particularly
good compensation for  wide-band atmospheric instabilities. Such
observations can be made for mm wave radiation even in poor
weather conditions but functions best for lines having widths
of less than a few MHz. If the spectral line is included in the
analyzing band in both the signal and the reference phases, the effective 
integration time is doubled. 

\vglue-4pt
 
\vspace*{3ex}
\noindent
{\bf 3) Position Switching and Wobbler Switching.}  \index{Position switching} The received signal
$''$on source$''$ is compared with another signal
obtained at a nearby position in the sky.
If the emission is rather  extended and the
atmospheric effects are large (for example in the case
of galactic Carbon Monoxide emission), one may need to use two reference
measurements, one at a higher, and the other at a lower elevation.
A number of conditions must be fulfilled:
(1) the receiver is stable so that any gain and
bandpass changes occur only over time scales
which are long compared to the time needed
for position change, and (2) there is
little line radiation at the comparison
region. If so, this method is efficient and
produces excellent line profiles.
A variant of this method is wobbler switching. This is very useful for compact
sources, especially in the mm and sub--mm range.
\enlargethispage{\baselineskip}  
%\vglue-4pt
 
\vspace*{3ex}
\noindent
{\bf 4) On the Fly Mapping.} This  very important observing
method is an extension of method (3)\index{On-the-fly mapping}. 
In this procedure, one
takes spectral line data at a rate of perhaps one spectrum or more per
second.
As with total power observing, usually one first takes a reference spectrum,
and then takes data along a given direction. Then one changes the position of the
telescope in the perpendicular direction, and repeats the procedure until the
entire region is sampled. Because of the short
integration times an entire image of perhaps $15\arcmin \times 15\arcmin$
taken with a $30\arcsec$ beam could be finished in roughly 20 minutes.
At each position, the S/N ratio may be low, but the procedure
can be repeated. With each data transfer, the telescope position is read out.
Even if there are absolute pointing errors, over this short time and small angle the relative positions where  spectra were taken 
are accurate. The resulting accuracy is improved  because the spectra
are oversampled and weather conditions are more uniform over the region mapped.
To produce the final image the individual spectra are
placed on a grid and then averaged.

%********************************************************************************************************************************************************interferometers
%\chapter{Interferometers and Aperture Synthesis}
 
\section{Interferometers and Aperture Synthesis  }
     \label{interferometers}
 
From  diffraction theory (\ref{diffraction}),  the angular
\index{Resolution!angular} resolution of a
radio telescope is $\theta = k\,  \lambda / D$, where $\theta$ is
the  angular resolution, $\lambda$
 is the wavelength of the radiation received, $D$ is the diameter of
the instrument and  $k$ is a factor of order unity. For a given wavelength,  to improve
this angular resolution, the diameter $D$ must be increased.The largest mm/sub-mm antenna is the Large Millimeter Telescope (LMT), with a 50 meter diameter, 
and a short wavelength 
limit of 0.8 mm, so $\theta \cong 3''$.  As shown by Michelson (\cite{JENKINS}),  
a resolving power  $\theta \approx \lambda$/D can be obtained
by coherently combining the ouput of two  reflectors of diameter $d\ll D$ separated by a distance $D$.

%***********************************************************************TLW DEC 2007
An important extension of interferometry is aperture synthesis, that is, the production of  high quality images of a source  by combining a
 number of independent  measurements in which the  antenna  spacings cover
an equivalent  aperture. In this chapter, we give an introduction to  the principles of aperture synthesis. Vastly more detailed 
accounts with lots and lots of math are to be found in   \cite{THOMPSON} or \cite{DUTREY}. 

\subsection{Two Element Interferometers}\label{teintf}

\begin{figure}[t]
%\sidecaption
%\mpicplace{7.6cm}{5.8m}
  \includegraphics[width=5.8cm,height=6.6cm]{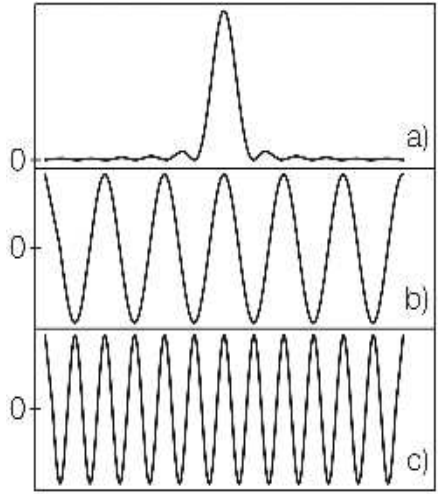}
  \caption[Comparison of double-beam and full aperture]
	{ \label{int-resolv}
	Power patterns for different antenna  configurations. The horizontal axis in this figure is angle. 
        Panel (a) shows that of a uniformly illuminated full
	aperture with a full width to half power (FWHP) of $ k \, \lambda/D$, with $k\approx 1$ .
        In  panel (b) we show the power  pattern of a two element multiplying interferometer consisting of two antennas of diameter $d$ 
spaced by a distance $D$ where $d \ll D$. 
        \break In panel (c) we show  the power pattern of the interfermeter system described in (b) but now with a spacing $2D$. 
	}
\end{figure}

The basic principles can be understood  from a consideration of 
Fig.~\ref{int-resolv}.  In panel (a) is the response of a single uniformly illuminated aperture of diameter $D$. In panels (b) and (c)  we show the response of 
a two element  interferometer consisting of two small antennas (diameter $d$) separated by a distance $D$ and $2D$, where $d \ll D$. The interferometer response 
is obtained from 
the multiplication of the outputs of the two antennas. 
The uniformly illuminated aperture has a  dominant
main beam of  width   $\theta = k \, \lambda/D$, accompanied by smaller secondary maxima, or  sidelobes. There are two differences between the case of a single 
dish response compared 
to the case of an interferometer. 
First, for an interferometer, the nomenclature is different. Instead 
of $'$main beam and sidelobes$'$ one speaks of $'$fringes$'$. There is a central fringe (or $'$white light$'$ fringe in the analogy with Young's Two Slit experiment) 
and adjacent fringes. 
Second, we will show that with the multiplication of the outputs  of two antennas, the fringes are centered about zero so that   the total
power output of each antenna is supressed. Since  some of the information (i.~e.~total power) is not used,  for a given spacing only  source structure comparable to 
(or smaller than) a fringe is recorded fully. For the  
case of an interferometer composed of two small dishes (with dish diameter $d \ll D$) there is no prominent main beam and the sidelobe
level does not markedly decrease with increasing angular offset from the axes of the antennas. 
Comparing the width of the fringes in panels (b) and (c) one finds that by doubling the separation $D$ of the  antennas, the fringe width  is halved.
For the interferometer spacing (usually referred to as {\it the baseline})  $D$, in panel (b) the resolving power of the 
filled aperture is not greatly different from the single 
dish in panel (a), 
but  the collecting area of this two element interferometer is smaller. For 
larger spacings, the interferometer angular resolution is greater. 
%7.1**************************************************************************************************
%***************************************************************************************************

If a uniform source is extended in angle by a positive and a negative fringe in Fig.~\ref{int-resolv} the response of the multiplied output is zero. 
For source structure smaller 
than a fringe, the response is not diminished.   Thus by  increasing  
$D$, finer and finer source structure is  measured.  Combining 
the outputs of  independent data sets for spacings of $D$ and $2D$
shows  that these select different structural components of the source. 
Finer source structure  can be recorded if in addition, $D\, n$ antenna  spacings are measured.
Such a series  measurements can  be made  by increasing the separation of two antennas whose outputs are coherently combined. 

A  general procedure, {\em aperture synthesis}, is now the standard method  to obtain high
quality, high angular resolution images. The first practical demonstration of aperture synthesis in radio astronomy was made by Ryle and his associates. 
Aperture synthesis allows us to reproduce the imaging properties of a large aperture by sampling the radiation field at
individual positions within the area contained within the aperture.  
In analogy with the approach used by  Michelson in the optical wavelength range, the advance in radio astronomy  was to measure   
 the {\em mutual coherence function}
and to show that these results were sufficient to produce images. Using this approach, 
 a remarkable improvement of the radio astronomical imaging  was possible.

% \section{Two-Element Interferometers}\label{teintf}

%  The coherence\index{Coherence!function} function
%$\Gamma(\vec{u}, \tau)$
%is measured  by correlating the outputs of two antenna systems. The simplest example of this process is  a
%two-element interferometer. Assume that the interferometer consists of  two antennas  $A_1$ and $A_2$
% separated by the distance $\vec{B}$
%(directed from $A_2$ to $A_1$),
%both antennas are  sensitive only to radiation of
%the same state of polarization (Fig.\,\ref{two-el-cor}).
 
Electromagnetic waves induce the voltage $U_1$
at the output of antenna $A_1$
 \begin{equation}
   \label{out-ua}
    U_1 \propto E \expo^{\,\im \omega t} \,,
 \end{equation}
while at $A_2$ we obtain
 \begin{equation}
   \label{out-ub}
    U_2 \propto E \expo^{\,\im \omega \,(t - \tau)} \,,
 \end{equation}
where $E$ is the amplitude of the incoming plane wave, $\tau$ is the geometric delay caused by the relative orientation of the
interferometer baseline $\vec{B}$ and the direction of the wave propagation.
For simplicity, in (\ref{out-ua}) and (\ref{out-ub}) we have neglected
receiver noise and instrumental phase. The outputs will be
correlated. Today, all interferometers  use direct  correlation, since the goal is to measure the correlation  accurately. 
%*********************************************************************************************************8
%7.3
\begin{figure}[b]
%\sidecaption
%\mpicplace{5.7cm}{5.8cm}
  \includegraphics[width=5.7cm,height=5.8cm]{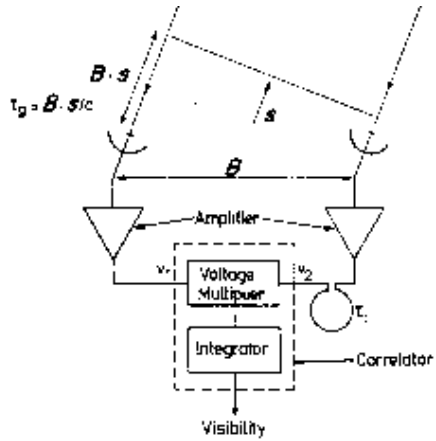}      
  \caption[Two-element correlation interferometer]
	{ \label{two-el-cor}
	A schematic diagram of a two-element correlation interferometer.
	The antenna output voltages are $V_1$ and $V_2$; the instrumental
	delay is $\tau_{\rm i}$ and the geometric delay is $\tau_{\rm g}$ (from \cite{RW4}). 
	}
\end{figure}
%**********************************************************************************************************
In a {\em correlation}\index{Interferometer!Correlation interferometer} the signals are input to a multiplying device
followed by an integrator. The output is proportional to
\beq
   R (\tau) \propto \frac{E^2}{T} \int\limits_0^T \expo^{\,\im \omega t}
                \expo^{\,-\im \omega ( t - \tau)} \,\diff t \, .
\eeq
If $T$ is a time much longer than the time of a single full
oscillation, i.e., $ T \gg 2 \pi / \omega$
then the average over time $T$ will not differ much
from the average over a single full period; that is
 \begin{eqnarray*}
  R(\tau) &\propto& \frac{\omega}{2\pi} E^2 \int\limits_0^{2\pi/\omega}
                  \expo^{\,\im \omega \tau}\,\diff t  \\
          &\propto& \frac{\omega}{2\pi} E^2 \expo^{\,\im \omega \tau}
                  \int\limits_0^{2\pi/\omega} \,\diff t   \,,
 \end{eqnarray*}
resulting in
 \begin{equation}
    \label{r-tau}
    \fboxsep2mm\fbox{$ \displaystyle
     R (\tau) \propto \fract{1}{2} E^2 \expo^{\,\im \omega \tau} $}
\quad .
 \end{equation}

The output of the correlator\,$+$\,integrator thus varies
periodically with $\tau$, the delay time.  If the  orientation of interferometer baseline
$\vec{B}$ and wave propagation direction $\vec{s}$ remain invariable,
$\tau$ remains constant, so does $R(\tau)$.
 But since $\vec{s}$ is slowly changing due to the rotation of the
earth,  $\tau$ will vary, and we will measure {\em interference fringes} as a function of time.

   In order to understand  the response of interferometers in terms of measurable quantities, we consider a
 two-element system. The basic constituents are
shown in Fig.\,\ref{two-el-cor}.
 If the radio brightness distribution is given by $I_\nu(\vec{s})$, the
power received per bandwidth $\diff \nu$ from the source element
$\diff \Omega$ is $ A(\vec{s}) I_\nu(\vec{s}) \diff \Omega \diff \nu$,
where $A(\vec{s})$ is the effective collecting area in the direction $\vec{s}$;
 we will assume the same $A(\vec{s})$ for each of the antennas.
The amplifiers are assumed to have   constant gain and phase factors (neglected for simplicity).
 
The output of the correlator for radiation from the direction $\vec{s}$
(Fig.\,\ref{two-el-cor})  is
 \begin{equation}
   \label{rot}
    r_{12} = A(\vec{s}) \,I_\nu(\vec{s}) \,\expo^{\im \omega \tau}
               \,\diff \Omega \diff \nu
 \end{equation}
where $\tau$ is the difference between the geometrical and instrumental
delays $\tau_{\rm g}$ and $\tau_{\rm i}$.
If $\vec{B}$ is the baseline vector for the two antennas
 \begin{equation}
    \label{delay}
    \tau = \tau_{\rm g} - \tau_{\rm i} = \frac{1}{c} \,\vec{B} \cdot \vec{s}
              - \tau_{\rm i}
 \end{equation}
and the total response is obtained by integrating over the source $S$
 \begin{equation}
   \label{tot-res}
   \fboxsep3mm\fbox{$ \displaystyle
     R(\vec{B}) =  \parbox{8mm}{$ \vspace*{-2mm}
                       {\displaystyle
                        \int\!\!\int \atop
                         {\!\!\!\!\!\scriptstyle \Omega }} $ }
       A(\vec{s}) I_\nu(\vec{s}) \exp \left[ \im 2 \pi \nu \left(\frac{1}{c}
            \,\vec{B} \cdot \vec{s} - \tau_{\rm i} \right) \right]
             \diff \Omega \diff \nu  $}    \quad 
 \end{equation}
This function $R(\vec{B})$, the {\em Visibility Function}  is closely related to
the mutual coherence function of the source, except for  the power pattern $A(\vec{s})$ of the
individual antennas. 
For parabolic antennas it is usually assumed that $A(\vec{s}) = 0$
outside the main beam area so that  (\ref{tot-res})
is integrated only over this region.
A one dimensional version of (\ref{tot-res}), with  a baseline $B$, $\nu=\nu_0$ and $\tau_i=0$, is
 \begin{equation}
   \label{1d-res-simple}
     R(B) =    \int   A(\theta) \, I_\nu(\theta) \exp \left[ \im 2 \pi \nu_0 \left(\frac{1}{c}  \,B \cdot \theta  \right) \right]
             \diff \theta 
 \end{equation}
Such a one dimensional relation is a very useful guide to  understand interferometer responses.  

 \subsubsection{Calibration}

Two quantities that must be calibrated for continuum measurements are  amplitude and  phase. In addition, for spectral line measurements  
 the instrument  passband  must also calibrated. 

The amplitude scale is calibrated using methods that are similar to those used for single dish measurements. This consists of using the response of each antenna to determine 
the system noise of the receiver being used. In the centimeter range, the atmosphere plays a small role while in the millimeter and sub-mm wavelength ranges, the atmospheric 
effects must be accounted for. For phase measurements, a suitable point-like source with an accurately known position is required to 
determine the instrumental phase $2\pi \nu \tau_i$ in 
(\ref{tot-res}). 
For interferometers, the calibration sources are usually unresolved or  point-like sources. Most often these are extragalactic time variable sources.  
 To calibrate the response in units of flux density or brightness temperatures,  these measurements must be referenced to a thermal calibrator. 

The calibration of the instrument passband is carried out by an integration of an intense source to determine the channel-to-channel gains and 
offsets. The amplitude, phase and passband 
calibrations are carried out before the source measurements. The passband calibration is usually carried out once per observing session. The 
amplitude and phase calibrations are made 
more often. Their frequency depends on the stability of the electronics and weather. At  millimeter wavelengths, the calibrations are usually 
made every few minutes, but may have to be made 
more often in worse weather or at shorter wavelengths. If weather demands that  frequent measurements of calibrators are required, this is 
referred to as {\em fast switching}.

%**********************************modified on Nov 25 2007
\subsection{Responses of  Interferometers}
\subsubsection{Finite Bandwidth}\label{lim-intf}
 
%In (\ref{tot-res}), we have assumed that the radiation is monochromatic. This is certainly not the case in most applications.
  Eq.\,\ref{tot-res} can be used to estimate the effect of a finite
bandwidth $\Delta \nu \,$. The geometric delay 
$\tau_{\rm g} = \frac{1}{c} \vec{B} \cdot \vec{s} \,$ is by definition
independent of frequency, but the instrumental  delay $\tau_{\rm i} \,$ may not be. Adjusting $\tau_{\rm i} \,$ the sum 
$\tau = \tau_{\rm g}-\tau_{\rm i}\,$ can be made equal to zero  for the
center of the band. Introducing the relative phase of a wave by
\beq
\varphi = \left [ \frac{c \tau}{\lambda} \right ]_{\rm fractional \; part}
\eeq
we obtain
\begin{equation} \label{def-phase}
\varphi = \frac{1}{\lambda} \vec{B} \cdot \vec{s} + \varphi_{\rm i} \; ,
\end{equation}
where $\varphi_{\rm i} \,$ is the instrumental phase corresponding to the
instrumental delay. This phase difference  varies across the  band of the interferometer 
$ \Delta \nu \,$  by
\begin{equation}
\Delta \varphi = \frac{1}{\lambda} \vec{B} \cdot \vec{s} \, \frac{\Delta \nu}{\nu} .
\end{equation}
The fringes will disappear when $\Delta \varphi \simeq 1 \,$ radian.
As can be seen the response is reduced if the frequency range, that is,
the bandwidth, is large compared to the delay caused by the
separation of the antennas. 
For large bandwidths, the loss of visibility can be minimized
by adjusting the phase  delay  the time difference  (see Fig.\,\ref{two-el-cor}) is negligible.
In effect, this is only possible if the exponential term in
(\ref{tot-res}) is kept small.
In practice, this is done by inserting a  delay between
the antennas so that $\frac{1}{c}
            \,\vec{B} \cdot \vec{s}$ equals $\tau_{\rm i}$. In
the first interferometric systems this was done by switching lengths
of cable into the system; currently this is accomplished by first
digitizing the signal after conversion to an intermediate frequency,
and then using digital shift registers. In analogy with the optical wavelength range, this
adjustment of cable length is equivalent to centering the response on  the
central, or {\em white light fringe} in 
Young's two-slit experiment.
 
The reduction of the response  caused by finite bandwidth can be
estimated by an integration of  (\ref{tot-res}) over frequency. Taking $A(\vec{s})$ and $I_\nu(\vec{s})$ to be constants, 
and integrating over a range of frequencies $\Delta \nu = \nu_1 - \nu_2$.
Then the result is an additional 
factor, $ \sin (\Delta \nu \tau)/ \Delta \nu \tau \,$ in (\ref{tot-res}).
This will reduce the interferometer response if
$\Delta \varphi \sim 1$ . For
typical bandwidths of 100\,MHz, the offset from the zero delay
must be $\ll 10^{-8}$\ts s. This adjustment of delays is
referred to as  {\em fringe\index{Fringe!stopping} stopping}.
This causes the response of
(\ref{tot-res}) to lose a component. To recover this
 input, an extra
delay of a quarter wavelength relative to the input of the
correlator is inserted, so that the sine and cosine response in
(\ref{tot-res})
can be measured. %To reduce the amount of electronics, this extra delay is switched
%in and out, but these components could both be measured simultaneously.
In digital cross-correlators, (see Sect.\,\ref{cr-co-rs}), the sine and cosine components are obtained from the positive and 
negative delays. The component with even symmetry is the cosine
component, while that with odd symmetry is the sine component. 
%*********************************************************************************TLW added Nov 25, 2007
\subsubsection{Source Size and Minimum Spacing}

Use \ref{1d-res-simple} in the following. For   an idealized  source, of shape  $I(\nu_0)=I_0$ for $\theta \, < \, \theta_0$ and  $I(\nu_0)=0$  
for $\theta \, > \, \theta_0$;  we 
take the primary beamsize of each antenna  to be much 
larger, and define the fringe width for a baseline $B$ $\theta_b$ to be $  \frac{\lambda}{B} $, 
The result is 
 \begin{equation}
   \label{1d-res-1}
 R(B) = A \, I_0 \cdot \theta_0 \, \exp \left[ \im  \pi \frac{\theta_0}{\theta_b}   \right] \, \left[ \frac{\sin{(\pi \theta_0/\theta_b)}}{{(\pi \theta_0/\theta_b)}}  \right]
 \end{equation}
The first terms are normalization and phase factors. The important  term, in the second set of brackets,  
is a $\sin{x}/x$ function. If $\theta_0  >>  \theta_b$, the interferometer 
response is reduced. This is sometimes refered to as the problem of  ``missing short spacings''. 
%make prolem with sinx/x for theta_s/theta_b=1/2

\subsubsection{Bandwidth and Beam Narrowing}\label{beam-narrowing}
In \ref{lim-intf}, we noted that on the {\em white light fringe} the compensation must reach a certain accuracy to prevent a reduction in the interferometer response. 
However for a finite  primary 
antenna beamwidth, $A$, this cannot be the case over the entire beam.  
For two different wavelengths $\lambda_l$ and  $\lambda_s$, there will be a phase difference 
\beq
\Delta \phi = 2 \pi \, d \left[ \frac{\sin{(\theta_{\rm offset})}}{\lambda_s}- \frac{\sin{(\theta_{\rm offset})}}{\lambda_l}\right]
\eeq
converting the wavelengths to frequencies, and using $\sin{\theta}=\theta$, we have
\beq
\Delta \phi = 2 \pi \, \theta_{\rm offset} \, \frac{d}{c} \, \Delta \nu
\eeq
With use of the relation $d=\frac{\lambda}{\theta_b}$, we have 
\begin{equation}
   \label{band-beam}
\Delta \phi = 2 \pi \, \frac{\theta_{\rm offset}}{\theta_b} \, \frac{ \Delta \nu}{\nu}
 \end{equation}

The effect in Eq.\,\ref{band-beam} is most important for  continuum measurements made with large bandwidths. This effect  can be reduced if the cross correlation is  carried 
out using a series of 
contiguous IF sections. For each of these IF sections, an extra delay is introduced to center the response at the value which is appropriate for 
that wavelength before correlation.

%**************************************************************************************************************************************************************************
 \subsection{Aperture Synthesis }
 \label{apsyn}
 Aperture Synthesis is a  designation for methods used to derive the
 intensity distribution $ I_\nu(\vec{s})$\, of a part of the radio sky 
  from the measured  function $R(\vec{B})$. To accomplish this we must invert the integral equation (\ref{tot-res}). This involves  
 Fourier transforms. For even simple images, a large number of computations are needed. Thus Aperture Synthesis and 
 digital computing are intimately connected. In addition, a large number of approximations to  be 
 applied. We will outline the most important steps of this development
 without, however, claiming completeness.
%*******************************************************************************************************************************************************08jan28. 

For mm/sub-mm imaging, the relevant relation is:
 \begin{equation}
   \label{i-str}
   \fbox{$ \displaystyle
     I' (x, y) = A (x, y)\, I (x, y) = \int\limits_{-\infty}^\infty
        V ( u, v, 0) \expo^{-\im 2 \pi ( u x + v y)} \diff u \diff v
        $} \quad 
 \end{equation}
where $I'(x, y)$ is
the intensity $I (x, y)$  modified by the primary beam shape $A (x, y)$.
One can easily correct $I' (x, y)$ by  dividing  by
$A(x, y)$.

Important definitions are:

\noindent
(1) {\em Dynamic Range}: The \index{Image!dynamic range}ratio of the maximum to the minimum intensity in 
an image. In images made with an interferometer array, it should be assumed tha the corrections for primary 
beam taper have been applied. If the minimum intensity is determined by the random noise in an image, the dynamic range is defined by the signal to noise ratio of the maximum feature in the image.
The dynamic range is an indication of the ability to recognize low intensity features in the presence of intense features.  
If the minimum noise is determined by artefacts, i.~e.noise in excess of the theoretical noise, the image can be improved by $'$image improvement techniques$'$.

\noindent
(2) {\em Image Fidelity}: This\index{Image!fidelity}is defined by the agreement between the measured results and the actual (so-called $''$true$''$) source structure. 
A quantitiave comparison would be 
\beq
F=|(S-R)|/R
\eeq
where $F$ is the fidelity, $R$ is the resulting image obtained from the measurement, and $S$ is the actual source structure.Of course one cannot have a priori knowledge of the correct
source structure.  In the case of simulations, $S$ is a source model, 
$R$ is the result of processing $S$ through $R$.

%**************************************************************************************************************
\begin{figure}
%\sidecaption
%\mpicplace{11.7cm}{7.9cm}
 \includegraphics[width=\linewidth]{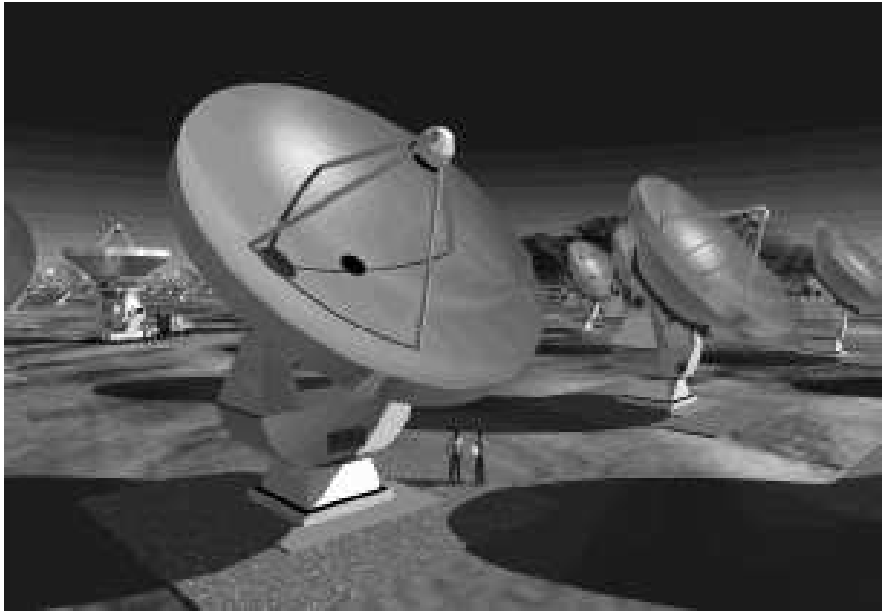}
  \caption[The ALMA Array]{ \label{ALMA}
The most ambitious construction project in radio astronomy 
is the Atacama Large Millimeter Array (ALMA). ALMA will be 
built in north Chile on a 5 km high site. It will consist 
of at least fifty-four 12-m and twelve 7-m antennas, operating in 10 bands 
between wavelength 1cm and 0.3mm. The 
ALMA project is  ambitious 
because of the superconducting receivers, the need for highly 
accurate antennas operating in the open, and the high altitude site. In addition, the 
data rates will be orders of magnitude higher than any 
existing astronomical facility. At the longest antenna spacing, 
and shortest wavelength, the angular resolution will be $\approx$5 
milli arcseconds\index{Interferometer!ALMA} (courtesy European Southern Observatory). }
\end{figure}
%************************************************************************************************************

\subsubsection{Interferometric Observations}
 \label{imagerest}
 
\vglue-5pt\noindent
Usually measurements are  carried out in 1 of 3 ways.

\begin{itemize}
\item In the first procedure,  measurements of the source of interest and a calibrator are made. This is  as in the case
of  single telescope position switching. Two significant differences with single dish measurements are that the interferometer 
measurement may have to extend over a wide range of hour angles 
to provide a better coverage of the $uv$ plane,  and that instrumental phase must be determined also. 
 One first measures a calibration
source\index{Calibration!source} or reference source, which has a known
position and size, to remove the effect of instrumental phases in the
instrument and atmosphere and to calibrate the amplitudes of the sources in
question. Sources and calibrators are usually observed alternately. 
The time variations caused by instrumental and weather effects must be slower than the time between measurements of source and calibrator.  
If, as is the case for mm/sub-mm wavelength measurements, 
weather is an important influence,  one must
 switch frequently between target source and calibration source. 
In {\em fast switching} one might spend 10 seconds on a nearby calibrator, then a few minutes on-source\index{Interferometer!fast switching}. This method will 
reduce the amount of phase fluctuations, but also the amount time available for source measurements. For more
rapid changes in the earth's atmosphere, one can correct the phase using
measurements of atmospheric water vapor, or changes in the system noise
temperature of the  individual receivers caused by atmospheric effects. The corrections for instrumental
amplitudes and phases are assumed to be constant over the times when the source
is observed. The ratio of amplitudes of source and calibrator are taken
to be the normalized source amplitudes. Since the calibrators
have known flux densities and positions, the flux densities and positions of the sources can be determined. The reference source should be as close to the 
on-source as possible, but must have a large enough intensity to guarantee a
good signal-to-noise ratio after a short time. Frequently nearby calibrators 
are time variable over months, so a more distant calibrator with a known or
fixed flux density is measured at the beginning or end of the session. This 
source is usually rather intense, so may also serve as a bandpass calibrator
for spectral line measurements. The length of time spent on  the off-source measurement is 
usually no more than few minutes. 

\item In the second  procedure, the so-called {\em
snapshots}, one makes a series of short observations (at different hour angles) of one source after another, and then repeats the measurements. 
For sensitivity reasons these are usually made
in the radio continuum or intense maser lines. As in the first observing method, one intersperses measurements of a calibration source which has a known position and size to 
remove the effect of instrumental phases in the instrument
and atmosphere and to calibrate the amplitudes of the sources in
question.
The images are affected by the shape of the synthesized
beam \index{Interferometer!synthesized beam}
of the interferometer system. If the size of the source to be  imaged is comparable
to the primary beam of the individual telescopes, the power
pattern of the primary beams will have a large effect. This 
effect can
be corrected easily.  %Still remaining are  effects  caused by 
%incomplete sampling of the  $uv$ plane.

\item In the third procedure, one aims to produce a high-resolution
image of a source where the goal is either  high dynamic range or high
sensitivity. The {\em dynamic\index{Dynamic range} range} is the
ratio of the highest to the lowest brightness level of reliable detail
in the image. This may depend on the signal-to-noise ratio for the data,
but for centimeter aperture synthesis observations, spurious features
in the image caused by the incomplete sampling \index{Interferometer!incomplete sampling} of
the $(u,v)$ plane\index{uv plane} are usually more important than the
noise. Frequently one measures the source in a number of different
interferometer configurations to better fill the $uv$ plane. These measurements are taken at different times and after calibration, 
the visibilities are entered into a common data set.

An extension of this procedure may involve the measurement of adjacent regions
of the sky. This is {\em mosaicing}. In a mosaic, the primary beams of the
telescopes should overlap, ideally this would be at the half power point. In
the simplest case, the images are formed separately and then combined to
produce an image of the larger region.

In order to eliminate the loss of source flux density due to missing short
spacings, one must  supplement the interferometer data with single dish
measurements.  The diameter of the single dish telescope should be larger than
the shortest spacing between  interferometer dishes. This single dish image
must extend to the FWHP of the smallest of the interferometer antennas. When  Fourier transformed and appropriately combined with the interferometer response, this  
data set has {\em no} missing flux density. Even in a data  set containng single dish data, there are  $``$missing spacings$''$.  
 Improvements that can be applied to images produced from  such data sets will be surveyed  next.

\end{itemize}

\subsubsection{Real Time Improvements of Visibility Functions}
 
Ideally the relation between the
measured $\widetilde{V_{ik}}$ visibility and the {\em actual}
visibility $V_{ik}$ can be considered to be linear
 \begin{equation}
   \label{Vcal}
     \widetilde{V_{ik}}(t) = g_i(t) \,g^*_k(t) \,V_{ik} + \epsilon_{ik}(t) \; .
 \end{equation}
Average values for the antenna gain factors $g_k$ and the noise term
$\epsilon_{ik}(t)$ are determined by measuring calibration sources as
frequently as possible. Actual values for $g_k$ are then computed
by linear interpolation. 
These methods make full use of the fact that the (complex) gain of the
array is obtained by multiplication
the gains of the individual
antennas. If the array consists of $n$ such antennas, $n(n-1)/2$
visibilities can be measured simultaneously, but only $(n-1)$ independent
gains $g_k$ are needed (for one antenna, one can arbitrarily set $g_k=1$ as a
reference). In an array with many antennas, the number of antenna pairs
greatly exceeds  the number of antennas. For phase, one must determine  $n$ phases. 
 
%Then 
%the individual complex gains $g_k$ can be included as free parameters
%in a CLEAN or MEM solution.   Deleted by TLW on 28 Dec. 02 
 
Often these conditions can be introduced into the solution in the form
of {\em closure \index{Error!closure} errors}. Introducing the phases
$\varphi, \theta$ and
$\psi$ by \\
\parbox{9cm}
{\begin{displaymath}
\begin{array}{rcl}
\widetilde{V_{ik}} &=& |\widetilde{V_{ik}}| \,\exp \left\{ \im
\varphi_{ik} \right\} \,, \\
G_{ik} &=& |g_i| \,|g_k| \,\exp \left\{ \im \theta_i \right\}
\exp \left\{- \im \theta_k \right\} \,, \\
V_{ik} &=& |V_{ik}| \,\exp \left\{ \im \psi_{ik} \right\} \,. \\
\end{array}
\end{displaymath} }
 \hfill
 \parbox{2cm}
  { \begin{equation}
     \label{phin}
    \end{equation}
  } \\
From (\ref{Vcal})  the visibility
phase $\psi_{ik}$ on the baseline $i k$ will be related to the
observed phase $\varphi_{ik}$ by
 \begin{equation}
   \label{obsp}
    \varphi_{ik} = \psi_{ik} + \theta_i - \theta_k + \epsilon_{ik} \,,
 \end{equation}
where $\epsilon_{ik}$ is the phase noise.
Then the {\em closure\index{Phase!closure} phase} $\Psi_{ikl}$ around a closed
triangle of baseline $i k, k l, l i$,
 \begin{equation}
   \label{cltri}
    \Psi_{ikl} = \varphi_{ik} + \varphi_{kl} + \varphi_{li} =
                 \psi_{ik} + \psi_{kl} + \psi_{li} +
                 \epsilon_{ik} + \epsilon_{kl} + \epsilon_{li} \,,
 \end{equation}
will be independent of the phase shifts $\theta$ introduced by the
individual antennas and the time variations. With this procedure, on can minimize phase errors. 
 
Closure amplitudes can also be formed.
If four or more antennas are used simultaneously, then ratios, the so-called
{\em closure amplitudes}, can be formed.  These are
independent of the antenna gain factors:
 \begin{equation}
   \label{rat-cl}
     A_{klmn} = \frac{|V_{kl}| |V_{mn}|}{|V_{km}| |V_{ln}|}  =
         \frac{|\Gamma_{kl}| |\Gamma_{mn}|}{|\Gamma_{km}| |\Gamma_{ln}|} \; .
 \end{equation}
\index{Closure amplitude}
Both phase and closure amplitudes can be used to improve the quality of the
complex visibility function.

   If each antenna introduces an unknown complex gain factor $g$ with
amplitude and phase, the total number of unknown factors in the array can be
reduced significantly by measuring closure phases and amplitudes. If four
antennas are available, 50\,\% of the phase information and 33\,\% of
the
amplitude information can thus be recovered; in a 10 antenna
configuration, these ratios are 80\,\% and 78\,\%
respectively.

\subsubsection{Multi-Antenna Array Calibrations}

For two antenna interferometers,  phase calibration can only be made pair-wise. This is referred to as $'$baseline based$'$ solutions for the calibration. 
For a multi-antenna system, there are other 
and better methods. One can use sets of three antennas to determine the best phase solutions and then 
combine these to optimize the solution for each antenna. For amplitudes,
one can combine sets of four antennas to determine the best amplitude solutions and then optimize this solution to determine the best solution.  
This process leads to $'$antenna based$'$ solutions  are used. Antenna based calibrations are used in most cases.
These are determined by applying phase and amplitude closure for subsets of antennas and then making the best fit for a given antenna. It is important 
to note that because of 
(\ref{heisenberg6}) the output of an antenna can be amplified without seriously degrading the signal-to-noise ratio. For this reason, cross-correlations 
between a large number of 
antennas is possible in the radio and  mm/sub-mm ranges. This is {\em not} the case in the optical or near-IR ranges.

\subsection{Interferometer Sensitivity}\label{intsens}
 
The random noise\index{Interferometer!noise} limit to an interferometer system is
calculated
following the method used 
for a single telescope\cite{RW4}. The RMS
fluctuations in antenna temperature are
\begin{equation}
     \Delta \Ta = \frac{M \,\Tsys}{\sqrt{t \,\Delta \nu }} \,,
\end{equation}
where $M$ is a factor of order unity used to account for extra noise
from analog to digital conversions, digital clipping etc. If we
next apply the definition of flux\index{Flux density} density, ${S_\nu}$ in
terms of antenna
temperature for a two-element system, we find:
\begin{equation}
\Delta S_{\nu} = 2\, k \,\frac{T_{\rm sys} \expo^{\tau}}{ A_{\rm e}
\sqrt{2t \,\Delta \nu }} \,,
\end{equation}
where $\tau$ is the atmospheric optical depth and $A_{\rm e}$ is the effective
collecting area of a single telescope of diameter $D$. There is  additional in this expression since a multiplying interferometer does not process all of the information
(i.~e.~the total power) that the antennas 
receive. In this case, there is an additional factor of
$\sqrt{2}$ compared to the noise in a single dish with an equivalent collecting  area since there is information not collected by a multiplying 
interferometer.   
We denote the
     system noise corrected for atmospheric absorption by
     $\Tsys' = \Tsys \exp{\tau}$, in order to simplify the following
     equations. For an array of $n$ identical telescopes, there are
     $N= n (n-1) / 2 $ simultaneous pair-wise correlations. Then
     the RMS variation in flux
     density is
     \begin{equation}
     \label{fluxrmssyn}
     \Delta S_\nu = \frac{2 \,M \,k \,\Tsys'}{ A_{\rm e} \sqrt{2 \,N
\,t \,\Delta \nu}}  \,.
     \end{equation}
     This relation can be recast in the form of brightness temperature
      fluctuations \index{Temperature!fluctuations} using the
Rayleigh-Jeans relation;
     \begin{equation}
     S = 2 \, k \,\frac{ \Tb \, \Omega_b}{\lambda^2} \,.
     \end{equation}
     Then the RMS brightness temperature, due to random noise, in
     aperture synthesis images is
     \begin{equation}
     \label{temprmssyn}
     \Delta \Tb = \frac{2 \,M \,k \,\lambda^2 \,\Tsys'}{A_{\rm e}  \Omega_{\rm b}
     \sqrt{2 \,N \,t \,\Delta \nu}} \,.
     \end{equation}
     For a Gaussian beam, $\Omega_{\rm mb}=1.133 \, \theta^2$, so  we can relate      the RMS temperature fluctuations to observed properties of a
     synthesis image. The ALMA sensitivity calculator is to be found at 

\noindent
{\em http://www.eso.org/sci/facilities/ALMA/observing/tools}

At shorter wavelengths, the
     RMS temperature fluctuations are   lower. Thus, for the same
     collecting area and system noise, if weather changes are unimportant, a millimeter image should be more
     sensitive than an image made at centimeter wavelengths. If the effective collecting area remains the same and for  a
     larger main beam solid angle, temperature fluctuations will 
     decrease. For
     this reason, smoothing an image will result in a lower RMS noise in an
     image. However, if smoothing is too extreme, this process effectively
 leads to a decrease in collecting area; then
     there will be no further improvement in sensitivity.
  
The temperature sensitivity (in Kelvins) for higher angular resolution is  worse than for a single
telescope with an equal collecting area. From the Rayleigh-Jeans relation,
the sensitivity in Jansky (Jy)  is fixed by the antenna
collecting area and the receiver noise, so only  the wavelength and the angular resolution can be varied. Thus, the increase in angular resolution is made at the
expense of temperature sensitivity. All other effects being equal, at shorter
wavelengths one gains in temperature sensitivity %(cf. The VLA versus ALMA in the previous example). 
%This is not such a great problem for the high-brightness
%radio sources, which radiate by nonthermal processes, such as synchrotron
%radiation, but would be for thermal sources, for which the maximum brightness
%temperature is  about $ 2 \times 10^4$K for regions of ionized gas surrounding
%massive stars. Our sun has a corona with a temperature of $10^6 \,{\rm K}$.
%However, even this temperature is small compared to brightness temperatures of
%up to $10^{12}$\,K, found for nonthermal sources.
 
Compared to single dishes, interferometers have the great
advantage that uncertainties such as pointing\index{Pointing} and
beam size\index{Beam!size}
depend  fundamentally on timing. Such timing
uncertainties can be made very small compared to all other uncertainties.
In contrast, the single dish measurements are critically dependent on
mechanical\index{Deformation!mechanical} deformations of the telescope.
In summary, the single dish results are easier to obtain, but
source positions and sizes on arc second scales are difficult to estimate.
The interferometer system has a much greater degree of complexity,
but allows one to measure such fine details.  The single dish system
responds to the source irrespective of the relation of source to
beam size; the correlation interferometer will not record  source
structures larger  than a few fringes.
 
 Aperture \index{Aperture synthesis} synthesis is based on
sampling the visibility function $V (u, v)$ with separate antennas to provide samples in  the $(u,v)$ plane. 
Many configurations are possible, but the goal is  the  densest possible  coverage  of the $(u,v)$ plane.
If one calculates the RMS noise in a synthesis image obtained by
simply Fourier transforming the $(u,v)$ data, one usually finds a
noise level many times higher than that given by (\ref{temprmssyn}) or
(\ref{fluxrmssyn}). There are various reasons for this. One cause is phase fluctuations due to  atmospheric or instrumental influences such as LO instabilities. Another
cause is due to incomplete 
sampling  of  the $(u,v)$ plane. This gives rise to instrumental features, such as stripe-like features in
the final images. Yet another systematic effect is the presence
of grating rings around more intense sources; these are analogous to
{\em high side lobes\index{Side lobes}} in single dish diffraction
patterns. Over the past 20 years, it has been found that these effects can
be substantially reduced  by software techniques such as CLEAN and Maximum Entropy. 

%************************************************************************************************************shifted here on Aug 12 08

\subsubsection{Post Real Time Improvements of Visibility Functions}
%\subsubsection{Gridding  $uv$ Data  }
 
Before  applying specialized techniques,  the data must be organized in a
 useful way without lowering the  signal-to-noise ratios. 
 To speed up computations for  inverting (\ref{i-str}) one uses  the
Cooley\index{Cooley-Tukey Fast Fourier Transform}-Tukey fast\index{FFT}
Fourier transform algorithm. % (FFT) are generally used.
In order to use the FFT in its simplest version, the visibility function must
be placed on a regular grid with  sizes that are powers of two of
the
sampling interval. Since the observed data seldom lie on such regular
grids, an interpolation
scheme must be used. If the measured points are randomly distributed, this
interpolation is best carried out using a convolution procedure.

 If some of the spatial
frequencies present in the intensity distribution are not present in the
$(u, v)$ plane data, then changing the amplitude or phase of the
corresponding
visibilities will not have any effect on the reconstructed intensity
distribution -- these have been eliminated. The extent of this effect is shown by the $''$dirty beam$''$\index{Interferomer!Dirty beam}.
 
Expressed mathematically, if $Z$ is an intensity distribution containing only the unmeasured
spatial frequencies, and $P_{\rm D}$ is the dirty beam, then

$
        P_{\rm D} \otimes Z = 0
$.

Hence, if $I$ is a solution of the convolution equation\,(\ref{i-xy})
then so is $I + \alpha Z$
where $\alpha$ is any number. This  shows that there is no
unique solution to the problem.
 
The solution with visibilities $V=0$ for the unsampled spatial frequencies
is usually called the {\em principal\index{Interferometer!Principal Solution} solution},
and it differs from the true intensity distribution by some unknown
{\em invisible \index{Interferometer!Invisible distribution}} or {\em ghost \index{Interferometer!Ghost
image}}
{\em distribution}. It is the aim of image reconstruction to obtain
reasonable approximations to these {\em ghosts} by using additional
knowledge or plausible extrapolations, but there is no direct way to
select the $''$best$''$ or $''$correct$''$ image from all possible images. The familiar
linear deconvolution algorithms are not adequate and nonlinear techniques
must be used.

 The result obtained from the gridded $uv$ data can be Fourier transformed to obtain an image with a resolution
corresponding to the size of the
array. However, this may still contain artifacts
caused by the details of the observing procedure, especially  the 
limited coverage of the ($uv$) plane. Therefore the
dynamic
range of such so-called {\em dirty\index{Dirty map}} maps is rather
small.
This can be improved by further data analysis, as will be described 
next. 

\enlargethispage{\baselineskip}
If the calibrated visibility function $V (u, v)$ is known for the full
$(u,v)$ plane both in amplitude and in phase, this can be used to
determine the (modified) intensity distribution $ I' (x, y)$
by performing the Fourier transformation (\ref{i-str}).
However, in a realistic situation $V (u, v)$ is only sampled at discrete
points within a radius $\cong u_{\rm max}$ along elliptical tracks,
and in some regions of the $(u,v)$ plane, $V(u, v)$ is not measured at all.

We can
weight the visibilities by a grading function, $g$. Then for a discrete
number of visibilities, we have a version of (\ref{i-str}) involving a
summation, not an integral, to obtain an image via a discrete Fourier
transform (DFT):
 \begin{equation}
   \label{i-strstr}
    I_{\rm D} (x, y) = \sum_k g (u_k, v_k) V (u_k, v_k) \expo^{-\im 2
\pi
        ( u_k x + v_k y)} \, ,
 \end{equation}
where $g (u, v)$ is a weighting function called the
grading\index{Grading} or apodisation\index{Apodisation}.
To a large extent $g(u,v)$ is arbitrary and can be used to change the
effective beam
shape and side lobe level. There are two widely used weighting
functions: uniform and natural. Uniform \index{Weighting!uniform} weighting
uses $g(u_k,v_k)=1$, while
Natural  \index{Weighting!natural} weighting uses
$g(u_k, g_k) = 1 / N_{\rm s}(k)$, where $N_{\rm s}(k)$ is
the number of data points within a symmetric region of the $(u,v)$
plane.
%\enlargethispage{\baselineskip} 
In a simple case $N_{\rm s}(k)$ would be a square centered on point $k$.
Data which are naturally weighted result in lower angular resolution but give a
better signal-to-noise ratio than uniform weighting. But these are 
only extreme cases. One can 
choose intermediate weighting schemes. These
are often referred to as {\em robust} weighting (in the nomenclature of the AIPS data reduction package).
Often the reconstructed image $I_{\rm D}$
may  not be a particularly good representation of $I'$, but
these are related. In another form, 
 (\ref{i-strstr}) is 
 \begin{equation}
   \label{i-xy}
    I_{\rm D} (x, y) = P_{\rm D} (x, y) \otimes I' (x, y) \,,
 \end {equation}
where
 \begin{equation}
   \label{pd}
    P_{\rm D}        = \sum_k g (u_k, v_k) \expo^{-\im 2 \pi
                       ( u_k x + v_k y)}
 \end{equation}
is the response to a point source. This is the {\em point spread function}
 PSF\index{Beam!point spread function} for the dirty\index{Beam!dirty} beam. Thus the {\em dirty beam}  can be understood
as a transfer function that distorts the image. (The dirty beam, $P_{\rm D}$, is produced by the Fourier
transform of a point source in the regions sampled; this is the
response of the interferometer system to a point source). That is, the
{\em dirty\index{Map!dirty} map}
$I_{\rm D}(x,y)$ contains only
those spatial frequencies $(u_k,v_k)$ where the visibility function has
been measured. The sum in (\ref{pd}) extends 
over the same positions $(u_k, v_k)$ as in (\ref{i-strstr}), and the
side lobe structure of the beam depends on the distribution of these
points.%\enlargethispage{\baselineskip} 
%\enlargethispage{\baselineskip} 

\subsection{Advanced Image Improvement Methods}\label{adsimp}

Digital computing  is clearly a crucial part of synthesis array data processing. A large part of the
advances in radio\index{Radio synthesis imaging} synthesis imaging during the
last 15--20~years relies on the progress made in the field of image
restoration. In the following we present a few schemes that are applied to improve radio images. However this is by no means an exhaustive collection. 

\subsubsection{Self-Calibration}

Amplitude and
phase errors scatter power across the image, giving the appearance of
enhanced noise. Quite often this problem can be alleviated to an impressive
extent by the method of {\em self-calibration}. This process can be applied
if there is a sufficiently intense source in the field contained
within the primary beam of the interferometer system. Basically,
self-calibration is the equivalent of focusing on the  source, analogous to
using the focus of a camera to sharpen up an object in the field of view.
One can restrict the self-calibration to an improvement of phase alone or
to both phase and amplitude. However, self-calibration is carried in the
$(u,v)$ plane. If properly used, this method leads  to a great improvement in
interferometer images of compact intense sources such as those of masering
spectral lines. If this method is used on objects with low signal-to-noise
ratios,  this method may give very wrong results by concentrating random
noise into one part of the interferometer image (see \cite{CORNWEL}).
%\enlargethispage{\baselineskip} 
%\enlargethispage{\baselineskip} 

In measurements of weak spectral lines, the self-calibration is carried out with  a
continuum source in the field. The corrections are then applied to the spectral
line data. In the case of intense lines, one of the frequency channels
containing the emission is used. If self-calibration is applied, the source
position information is usually  lost.

\subsubsection{Applying CLEAN to the Dirty Map}

{CLEAN}ing is the most commonly used technique to improve single
radio interferometer images(\cite{HOGBOM}).
The {\em dirty map} is a representation of the principal solution,
but with  shortcomings. In addition to its inherent low
dynamic range, the dirty map often contains features such as negative intensity artifacts. These cannot be 
real.
Another unsatisfactory aspect is that the principal\index{Principal solution}
solution
is quite often rather unstable, in that it can change drastically
when more visibility data are added.
Instead of a principle solution that assumes $V=0$ for all unmeasured
visibilities, values for $V$ should be adopted at these positions in the $(u, v)$
plane. These are obtained from some plausible model for the intensity
distribution. 

The CLEAN method approximates the actual but unknown intensity distribution
$ I (x, y)$ by the superposition of a finite number
of point sources with positive intensity $A_i$ placed at positions
$(x_i, y_i)$.
 It is the aim of CLEAN\index{CLEAN} to determine the $A_i( x_i, y_i)$ such that
 \begin{equation}
  \label{i-clean}
I'' (x,y) = \sum_i A_i \,P_{\rm D} (x - x_i, y - y_i) + I_\epsilon (x,
y) \,
 \end{equation}
where $I''$ is the dirty map obtained from the inversion
of the visibility function and $P_{\rm D}$ is the dirty beam (\ref{pd}).
$I_\epsilon (x, y)$ is the residual\index{Brightness distribution!residual}
brightness distribution after
decomposition. Approximation (\ref{i-clean}) is deemed
successful if $I_\epsilon$ is of the order of the noise in the measured
intensities. This decomposition cannot be done analytically, rather
an iterative technique has to be used.

The CLEAN algorithm is most commonly applied  in the image plane. This is
an iterative method which functions in the following fashion: First find
the peak intensity of the dirty image,
then subtract a fraction $\gamma$  with the shape of the dirty beam
from the image. Then repeat this $n$ times.
 This \index{Damping factor} {\em loop gain}
$0 < \gamma <  1$ helps the iteration converge, and it is usually continued
until the intensities of the remaining peaks are below some limit.
Usually the resulting point source model is convolved with a
{\em clean\index{Clean beam} beam}  of
Gaussian shape with a FWHP similar to that of the dirty beam.
Whether this algorithm produces a realistic image, and how trustworthy
the resulting point source model really is, are unanswered questions.

\subsubsection{Maximum Entropy Deconvolution Method (MEM)}
 
The Maximun Entropy Deconvolution Method (MEM)
is commonly used to produce a single optimal image from a set of
separate but contiguous images \cite{GULL}.
The problem of how to select the $``$best$''$ image from many possible images
which all agree with the measured visibilities is solved by
MEM\index{MEM}. Using MEM, \index{Maximum entropy method} those values of the
interpolated visibilities are selected, so that the resulting image
is consistent with all  previous relevant data. In addition, the MEM image has
maximum smoothness. This is obtained by
maximizing the {\em entropy} of the image. One possible definition of
entropy is given by
\begin{equation}
 \label{entropy}
   {\cal H } = - \sum_i I_i \left[ \ln \bigg(\frac{I_i}{M_i}\bigg) -1
\right] \,,
\end{equation}
where $I_i$ is the deconvolved intensity and $M_i$ is a reference image
incorporating all $``$a priori$''$ knowledge. In the simplest case $M_i$ is the
empty field $M_i = {\rm const} > 0$, or perhaps a lower angular
resolution image.
 
Additional constraints might require that all measured visibilities should
be reproduced exactly, but in the presence of noise such constraints are
often incompatible with $I_i > 0$ everywhere. Therefore the MEM image is
usually constrained to fit the data such that
\begin{equation}
 \chi ^2 = \sum \frac{|V_i - V_i'|^2}{\sigma_i^2}
\end{equation}
has the expected value, where $V_i$ is the measured visibility, $V_i'$
is a visibility corresponding to the MEM image and $\sigma_i$ is the
error of the measurement.
 %***************************************************************************************

\section{Continuum Emission from mm/sub-mm Sources}
   \label{emission-cr}
In the early days of mm/sub-mm wavelength astronomy receiver sensitivities
restricted measurements to a few continuum sources. This has improved dramatically with the use of semiconductor bolometers pioneered by F.~J.~Low. 
Subsequently spectral lines of 
molecules such as CO, HCN and CS were found. These were rather intense, and this led to a blossoming of the field. Subsequently, spectral lines of neutral 
carbon and a line of ionized carbon were detected. We  sketch continuum radiation mechanisms and then spectral line mechanisms 
that are specific to the mm/sub-mm wavelength range.

Radio sources can thus be classified into
two categories:  those which radiate by thermal\index{Radiation
mechanism!thermal}
mechanisms and the others, which radiate by
nonthermal\index{Source!nonthermal}
processes. In practice, non-thermal radiation can be explained by the synchrotron process. This is caused by relativistic electrons spiraling in a magnetic field   
(see \cite{SCHEUER} for a delightful and instructive account). 
For optically thin synchrotron radiation, most sources have a  flux density that varies as 
\begin{equation}
   \label{syncrotron}
   \fboxsep2mm\fbox{$ \displaystyle
     S_\nu = S_0  \, \left( \frac {\nu}{\nu_0} \right)^{-\alpha}
        $}
 \end{equation}
where $\alpha$ has a positive value. Synchrotron radiation often shows linear polarization, and in rare cases,  shows circular polarization. The most prominent 
example of a synchrotron source with $\alpha=0$, i.~e.~a flat spectrum is Sgr~A$^*$\index{Black Hole!Sgr~A$^*$}. This source, at 8.5 kpc from the Sun,  
is considered to be the closest  supermassive Black Hole. 
The variation  of  flux density with frequency may be an indication that the synchrotron emission is optically thick, and/or has a 
non-standard geometry (see, e.~g.~\cite{MARRONE}). Interferometry at 1.3 mm of  Sgr~A$^*$ with  baselines of up to a few 10$^3$ km is ongoing. The imaging of  
Sgr~A$^*$ with astronomical unit linear 
resolution is of high interest. Measurements of synchrotron emission give only a very limited set of source parameters. 
When combined with other data, synchrotron measurements can be used to obtain source parameters (see \cite{RW5}). 

The most famous example of black body radiation is the 2.73\,K cosmic 
microwave\index{Background!cosmic microwave} background, CMB. This source of radiation
is fit by a Planck curve to
better than 0.1\,\%.  
The
difficulty in detecting it was not due to  weak signals, but rather due to
the fact that the radiation is present in all directions, so that scanning
a telescope over the sky and taking differences will not lead to a detection. The actual discovery
was gotten from  measurements of the noise temperature from the sky, the
receiver and the ground, compared to the temperature of a helium cooled
load using Dicke switching. Studies of the CMB are conducted from satellites and balloons; these are directed at determinations of the polarization and deviations from 
the Black Body spectrum. 

As an example of thermal radiation, we consider planets. These  are black\index{Black body!radiation} bodies with  spectra
that are an almost exact representation of the Rayleigh-Jeans law
for various temperatures. 
 For \HII\ regions  such as Orion A
the spectrum is not a simple black body, but explanation is fairly straightforward. If we
consider the solution of the equation of radiation
transfer (\ref{sol-isoth}) for an isothermal object without a background
source
\beq
   I_\nu = B_\nu(T) \, (1 - \expo^{ - \tau_\nu}) \,,
\eeq
we find that $I_\nu < B_\nu$ if $\tau_\nu <  1$;
the frequency variation of $I_\nu$
depends on $\tau_\nu$ \cite{ALTENHOFF}:
 \begin{equation}
   \label{mezhen}
   \fboxsep2mm\fbox{$ \displaystyle
     \tau_\nu = 8.235 \times 10^{-2} \left(\frac{T_{\rm e}}{\rm
K}\right)^{-1.35}
                \left( \frac{\nu}{\rm GHz} \right)^{-2.1}
                \left( \frac{\rm EM}{\rm pc \, cm^{-6}} \right) \, a (\nu, T)
        $}
 \end{equation}
can be derived. The correction $a(\nu, T)$ is usually $\cong 1$. The term $EM$ is 
$$EM= \int N_e^2 \diff l$$
with units cm$^{-6}$pc. This is a complex relation and was derived only after plowing through lots and lots of math. $N_e$ cannot be directly obtained from $EM$ because of clumping.

 %\pagebreak
 %**************************************************************************************************************************************
\begin{figure}[b]%
%\sidecaptionab
%  \mpicplace{5cm}{10cm}
  \includegraphics[width=8cm,height=8cm]{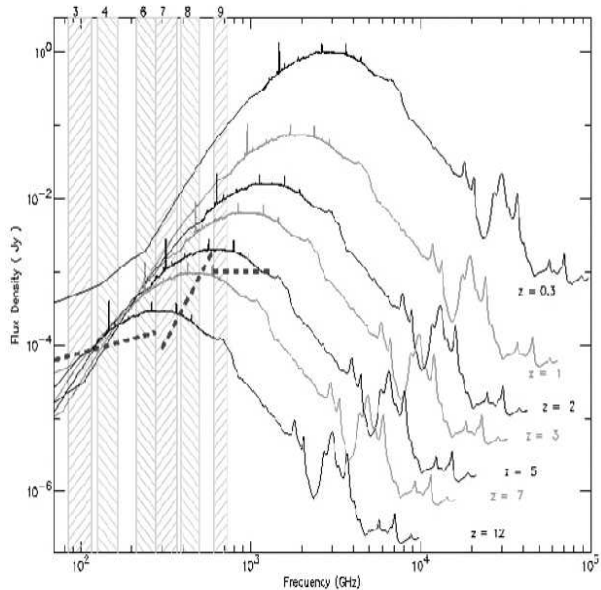}
  \caption[The spectrum of the star burst galaxy M82]
 {\label{m82-w-alma-rx}
The spectrum of the star burst galaxy M82. The flux density scale must be multiplied by $10^3$. This galaxy is 
rather close to the Sun, but has 
a star formation rate that is much higher than for the Milky Way. The large featureless maximum is 
caused by dust radiation. On top of this are various molecular and atomic lines. The different curves represent M82 for 
the labelled redshifts. Although the emission decreases inversely 
with the square of the distance, this is compensated by the shift of dust radiation to lower frequencies. These effects allow one to detect an oject such as M82 
to redshift $z=12$ with the Atacama Large Millimeter Array (ALMA). The ALMA receiver bands are indicated by numbers in the upper 
left side of this figure (P. Cox unpublished). }
\end{figure}
%********************************************************************************************************************

%**************************************************************************************************************

It appears that cold dust\index{Dust!cold}, with
temperatures 10\,K to 30\,K,  makes
up much of the mass of dust and by implication traces cold interstellar gas, given a dust-to-gas ratio. Since the average size is thought to be $\sim$0.3$\mu$m, the wavelength of
the radiation is much longer than the size of the emitter. For this reason, the efficiency of sub-mm radiation is rather low. Dust grains show linear polarization, which 
leads to the conclusion that grains are elongated and aligned by magnetic fields.  The direction but not the strength of the magnetic field can be determined from dust polarization. 
 The fractional polarization is rather small, so requires high sensitivity and care to keep
instrumental effects  small\cite{ENDRIK}. 

If we use the exact relation, we have
\begin{equation}
\label{t-dust}
T_{\rm b}(\nu)=T_0 \! \left(\frac{1}{{\rm exp}\{T_0/T_{\rm dust}\}-1}-
		\frac{1}{{\rm exp}\{T_0/2.7\}-1} \right)
\! (1-\expo^{-\tau_{\rm dust}}) \,,
\end{equation}
where $T_0 = h \nu / k$.
This is completely general; if we neglect the 2.7\,K background,
\begin{equation}
\label{mm-t-dust}
T_{\rm dust} = T_0 \left( \frac{1}{{\rm exp}\{T_0/T_{\rm dust}\}-1}  \right )
		 (1-\expo^{-\tau_{\rm dust}}) \,.
\end{equation}
If $T_{\rm dust} \gg T_0$, we can simplify this expression, but the most
important step is in making a quantitative connection between $\tau_{\rm
dust} $ and $N_{\rm H_2}$.
The relation between $\tau_{\rm dust}$ and
the gas column density must be determined empirically. Unlike the
planets, which have  measured sizes, the radiation from dust grains depends
on the surface area of the grains, which  cannot be determined directly.
If a relation between dust mass and $\tau$ can be determined,
it is simple to convert to the total mass, since dust is generally
accepted to be between 1/100 and 1/150\,th of the total mass.
All astronomical determinations are based on \cite{HILDEBRAND}.   One typical parameterized version is given by \cite{MEZGER}: 
$\lambda > 100\,{\rm\mug m}$:
\begin{equation}
\label{tau-dust}
\tau_{\rm dust} = 7 \times 10^{-21} \,\frac{Z}{Z_\odot}
		 \,b \,N_{\rm H} \,\lambda^{-2}
\end{equation}
where $\lambda$ is the wavelength in $\mug$m, $N_{\rm H}$
is in units cm$^{-2}$,
$Z$ is the metalicity as a ratio of that of the sun $Z_\odot$. The
parameter $b$
is an adjustable factor used to take into account changes in grain sizes.
Currently, it is believed that $b=1.9$ is appropriate for moderate
density gas and $b=3.4$ for dense gas (but this is not certain).
At long millimeter wavelengths, a number of observations have shown
that the optical depth of such radiation is small. Then the observed
temperature is
   \begin{equation}
     T = T_{\rm dust} \, \tau_{\rm dust} \,,
   \end{equation}
where the quantities on the right side are the dust temperature and
optical depth. Then the flux\index{Flux density} density is
   \begin{equation}
     \label{dusteq}
  S = \frac{2 \,k \,T}{\lambda^2} = 2 \,k \,T_{\rm dust}
             \lambda^{-2} \, \tau_{\rm dust} \Delta\Omega \,.
   \end{equation}
If the dust radiation is expressed in
Jy, the source in FWHP sizes, $\theta$ in arc seconds, and the
wavelength,
$\lambda$ in $\mug$m, one has for the column density of hydrogen in all
forms, $N_{\rm H}$, in the Rayleigh-Jeans approximation, the following
relation:

If the dust radiation is expressed in mJy, the source FWHP
size, $\theta$, in arc seconds, and the
wavelength, $\lambda$ in mm,
the column density of hydrogen in all forms, $N_{\rm H}$, in
the Rayleigh-Jeans approximation,
is:

\begin{equation}
\label{cold-dust-col-den}
N_{\rm H} = 1.93 \times 10^{24} \, \frac{S_\nu}{\theta^2}
	 \,\frac{\lambda^4}{Z/Z_\odot \,b \,T_{\rm dust}} \,.
\end{equation}

In the cm and mm wavelength range, the dust optical depth is
small and increases as  $\lambda^{-\beta}$ with $\beta$ values between 1 and 2; then flux density
increases as  $\lambda^{-3}$ to $\lambda^{-4}$.

Observationally, it has been determined that, in most cases, the dust
optical depth increases with $\lambda^{-2}$; then flux density
increases as $\lambda^{-4}$. Thus, dust emission will
become more important at millimeter\index{Dust!spectral index} wavelengths and
in
the infrared\index{Infrared radiation}. It might appear that Eq.\,\ref{mezhen} is more accurate than Eq.~\,\ref{cold-dust-col-den}, because of the vast amount of math. However  
for a determination of $N_e$, source clumping introduces uncertainties.

%***********************************************************************************************************************************************sp-line-fund

%\chapter{Spectral Line Fundamentals}
 %\label{line-radiation}

\section{Spectral Line Basics}
   \label{einco}
 
In local thermodynamic equilibrium (LTE\index{LTE})
the intensities of emitted and
absorbed radiation are not independent but are
related by Kirchhoff\index{Kirchhoff's law}'s law
(\ref{kirchhoff}). This  applies to both
continuous radiation and line radiation.
The Einstein\index{Einstein coefficient}
coefficients give a convenient means to
describe the interaction of radiation with
matter by the emission and absorption
of photons\cite{RYBICKI}.These are: 
%**************************************************************************** 
 \begin{figure}[b]
\sidecaption
%\mpicplace{4.8cm}{2.4cm}
   \includegraphics[width=4.8cm,height=2.4cm]{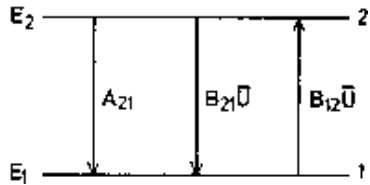}
   \caption[Transitions between the states 1 and 2]
	{ \label{trstot}
     Transitions between the states 1 and 2 and the Einstein
     probabilities}
 \end{figure}
%************************************************************************* 
 
 \begin{equation}
   \label{got}
  \fboxsep2mm\fbox{$ \displaystyle
       g_1 \,B_{12} = g_2 \,B_{21} $}
 \end{equation}

and

 \begin{equation}
  \label{atobto}
   \fboxsep2mm\fbox{$ \displaystyle
      A_{21} = \frac{8 \pi h \nu_0^3}{c^3} \,B_{21}   $} \quad 
 \end{equation}
%In deriving these relations no reference is made to
%any thermodynamic property of the cavity.
%Therefore they must  be valid for
%systems independent of the assumption
%of a TE environment.

\subsection{Radiative Transfer with Einstein Coefficients}

When the radiative transfer\index{Radiative transfer} was considered
in Sect.\,\ref{radtran}, the material properties were
expressed as the emission coefficient $\epsilon_\nu$
and the absorption coefficient $\kappa_\nu$. Both $\epsilon_\nu$
and $\kappa_\nu$ are macroscopic parameters; for
a physical theory these must to be related
to atomic properties of the matter in the
cavity. If line radiation is
considered, the Einstein\index{Einstein coefficient} coefficients
%in Sect.\,\ref{einco} 
are
very useful because these can be linked
directly to the properties of the
transition responsible for the spectral line.
For radiative\index{Radiative transfer}
transfer $\epsilon_\nu$ and $\kappa_\nu$ are needed, so we must
investigate the relation between $\kappa_\nu$
and $A_{ik}$ and $B_{ik}$. This is best done by
considering the possible change
of intensity $I_\nu$ passing through a slab of material with
thickness $\diff s$  as in Sect.\,\ref{radtran}. Now
we will use $A_{ik}$ and $B_{ik}$.
 
   According to Einstein there are three
different processes contributing to the
intensity $I_\nu$. Each system making a transition from $E_2$
to $E_1$ contributes the  energy $h \nu_0$ distributed
over the full solid angle $4 \pi$. Then the
total amount of energy emitted spontaneously\index{Emission!spontaneous} is
 \begin{equation}
   \label{espon}
     \diff E_{\rm e}(\nu) = h \nu_0 \,N_2 \,A_{21} \,
                \varphi_{\rm e}(\nu) \diff V
                \frac{\diff \Omega}{4 \pi} \diff \nu \diff t \,.
 \end{equation}
For the total energy {\em absorbed\index{Energy!absorbed}} we
similarly obtain
 \begin{equation}
   \label{teabs}
     \diff E_{\rm a}(\nu) = h \nu_0 \,N_1 \,B_{12}   \,\frac{4 \pi}{c}
\,I_\nu
                \,\varphi_{\rm a}(\nu) \diff V
                \frac{\diff \Omega}{4 \pi} \diff \nu \diff t
 \end{equation}
and for the {\em stimulated\index{Emission!stimulated}} emission
 \begin{equation}
   \label{estim}
     \diff E_{\rm s}(\nu) = h \nu_0 \,N_2 \,B_{21} \,\frac{4 \pi}{c} \,I_\nu
                \,\varphi_{\rm e}(\nu) \diff V
                \frac{\diff \Omega}{4 \pi} \diff \nu \diff t \, .
 \end{equation}
The line profiles $\varphi_{\rm a}(\nu)$ and $\varphi_{\rm e}(\nu)$
 for absorbed and emitted radiation could be different, but
in astrophysics it is usually permissible
to put $\varphi_{\rm a}(\nu) = \varphi_{\rm e}(\nu) = \varphi(\nu)$.
For the volume element we put $\diff V = \diff \sigma \diff s$,
where $\diff \sigma$ is the unit area perpendicular to the beam direction.
For a stationary situation, we find
 \begin{eqnarray*}
   & &\diff E_{\rm e}(\nu) + \diff E_{\rm s}(\nu) - \diff E_{\rm a}(\nu) =
      \diff I_\nu \diff \Omega \diff \sigma \diff \nu \diff t    \\
   & & =  \frac{h \nu_0}{4 \pi} \left[ N_2 \,A_{21} + N_2 \,B_{21} \,\frac{4 \pi}{c}
       \,I_\nu - N_1 \,B_{12} \,\frac{4 \pi}{c} \,I_\nu \right]
    \varphi(\nu) \diff \Omega \diff \sigma \diff s \diff \nu \diff t \,.
 \end{eqnarray*}
The resulting equation of transfer\index{Equation of transfer} with
Einstein coefficients is
 \begin{equation}
   \label{eqtein}
    \fboxsep2mm\fbox{$ \displaystyle
    \frac{\diff I_\nu}{\diff s} = - \frac{h \nu_0}{c} \left( N_1 \,B_{12} -
          N_2 \,B_{21} \right) I_\nu \,\varphi(\nu) + \frac{h \nu_0}{4 \pi}
          N_2 \,A_{21} \,\varphi(\nu) $} \quad .
  \end{equation}
Comparing this with (\ref{eq-of-transfer}) we obtain agreement by putting
 \begin{equation}
   \label{kneq}
   \fboxsep2mm\fbox{$ \displaystyle
      \kappa_\nu = \frac{h \nu_0}{c} \,N_1 \,B_{12} \left( 1 -
        \frac{g_1 N_2}{g_2 N_1} \right) \varphi(\nu) $}
 \end{equation}
and
 \begin{equation}
   \label{eneq}
   \fboxsep2mm\fbox{$ \displaystyle
   \epsilon_\nu = \frac{h \nu_0}{4 \pi} \,N_2 \,A_{21}
\,\varphi(\nu) $} \quad ,
 \end{equation}
 The factor in brackets in (\ref{kneq}) is
the correction for stimulated emission. In
radio astronomy, where the stimulated
emission almost completely cancels the
effect of the true absorption, this is important. How this
comes about is best seen if we investigate
what becomes of (\ref{eqtein}--\ref{eneq}) if LTE is assumed.

 From (\ref{kneq}) and (\ref{eneq}) we find  that
\beq
   \frac{\epsilon_\nu}{\kappa_\nu} = \frac{2 h \nu^3}{c^2} \left(
      \frac{g_2 N_1}{g_1 N_2} - 1 \right)^{-1} \,.
\eeq
But for LTE, according to (\ref{kirchhoff}), this
should be equal to the Planck function, resulting in
 \begin{equation}
   \label{lten}
   \fboxsep2mm\fbox{$ \displaystyle
      \frac{N_2}{N_1} = \frac{g_2}{g_1} \exp \bigg( -
        \frac{h \nu_0}{k \,T} \bigg) $} \quad .
 \end{equation}
In LTE, the energy levels  are
populated according to the same Boltzmann
distribution  for the temperature $T$
that applied to full TE. Then
 the absorption\index{Coefficient!absorption} coefficient
becomes
 \begin{equation}
   \label{knbo}
   \fboxsep2mm\fbox{$ \displaystyle
     \kappa_\nu = \frac{c^2}{8 \pi}
     \frac{1}{\nu_0^2} \frac{g_2}{g_1} \,N_1   \,A_{21}
       \left[ 1 - \exp \left( - \frac{h \nu_0}{k \,T} \right)
        \right] \varphi(\nu) $} \quad ,
 \end{equation}
where we have replaced the $B$ coefficient by the $A$ coefficient, using
\beq
  B_{12} = \frac{g_2}{g_1} \,A_{21} \,\frac{c^3}{8 \pi h \nu^3} \,.
\eeq

\subsection{Dipole Transition Probabilities}
 
The simplest sources for electromagnetic
radiation are oscillating dipoles\index{Dipole!transition probability}.
Radiating electric dipoles have already
been treated classically, but it
should also be possible to express these
results in terms of the Einstein
coefficients. There are two types of
dipoles that can be treated by quite
similar means: the electric and the magnetic dipole.
 
\vspace*{3ex}
\noindent
{\bf Electric Dipole.} Consider an oscillating electric dipole
 \begin{equation}
   \label{oed}
    d(t) = e \,x(t) = e \,x_0 \cos \omega t \,.
 \end{equation}
According to electromagnetic theory, this will
radiate. The power emitted into
a full  $4 \pi$ steradian is,
 \begin{equation}
  \label{pted}
    P(t) = \frac{2}{3}\frac{e^2 \dot{v}(t)^2}{c^3} \,.
 \end{equation}
Expressing $ x = d / e$ and $\dot{v} = \ddot{x}$, we obtain
an average power, emitted over one period of oscillation of
 \begin{equation}
  \label{aveped}
  \langle P \rangle = \frac{64 \pi^4}{3 \,c^3} \,\nu_{mn}^4
            \left(\frac{e \,x_0}{2}\right)^2 \,.
 \end{equation}
This mean emitted power\index{Emitted power} can also be
expressed in terms of the Einstein $A$ coefficient:
 \begin{equation}
   \label{aphn}
   \langle P \rangle = h \,\nu_{mn} \,A_{mn} \,.
 \end{equation}
Equating (\ref{aveped}) and (\ref{aphn}) we obtain
 \begin{equation}
   \label{amn}
   \fboxsep2mm\fbox{$ \displaystyle
    A_{mn}= \frac{64 \pi^4}{3 \,h \,c^3} \,\nu_{mn}^3 \,|\mu_{mn}|^2
      $} \quad ,
 \end{equation}
\vglue-6pt
%\pagebreak
 This was also cited in Section~\ref{hertz-larmor}.
\noindent
where
 \begin{equation}
  \label{mmn}
   \mu_{mn} = \frac{e \, x_0}{2}
 \end{equation}
is the mean electric dipole moment of the
oscillator for this transition.
 
 Expression
(\ref{amn}) is applicable only to classical
electric dipole oscillators, but is also  valid
for quantum systems.

\subsection{Simple Solutions of the Rate Equation}
 
In order to compute absorption or
emission coefficients in (\ref{kneq}) and
(\ref{eneq}), both the Einstein coefficients and
the number densities $N_i$ and $N_k$ must be known.
In the case of LTE,
the ratio of $N_2$ to $N_1$ given by the
Boltzmann function:

\begin{equation}
   \label{ggdetb}
	\frac{N_2}{N_1} =
	 \frac{g_2}{g_1} \exp \left(
      - \displaystyle \frac{ h \,\nu_0}{k \,T_{\rm K}} \right) \,,
 \end{equation}

If $C_{12}$ and $C_{21}$ are the collision
probabilities\index{Probability!collision}  per particle
(in cm$^3$\,s$^{-1}$) 
for the transitions 1 $\to$ 2 and 2 $\to$ 1,
respectively, 
$T_{\rm K}$ is  the {\em kinetic
temperature\index{Temperature!kinetic}}.
   When $T_{\rm ex}$, $T_{\rm b}$ and $T_{\rm K} \gg h \nu /
k$, and
if we use for abbreviation
\begin{equation}
	T_0 = \frac{h \nu}{k}
\end{equation}
we have
\begin{equation}
 \label{Tex}
   \fboxsep2mm\fbox{$\displaystyle
T_{\rm ex}= T_{\rm K} \frac{T_{\rm b} A_{21} + T_0 C_{21}}
{T_{\rm K} A_{21} + T_0 C_{21}} $}
\end{equation}
 If radiation dominates the rate
equation $ ( C_{21} \ll A_{21} ) $,  then  $T_{\rm ex} \to \tb$. 
If on the other hand collisions dominate $ ( C_{21} \gg A_{21} )
$, then $ T_{\rm ex} \to T_{\rm K}$.
Since $C_{ik}$ increases with
increasing $N$ collisions will dominate the distribution
in high-density situations and the
excitation temperature of the line will be
equal to the kinetic temperature. In
 low-density situations $ T_{\rm ex} \to \tb$. The density when
$A_{21} \approx C_{21} \approx N^*\langle \sigma \,v \rangle$
is called the {\em critical\index{Density!critical} density}. The
smaller $A_{21}$, the lower is $N^*$.

Radiative transfer with high optical depths can be dealt with using the Large Velocity Gradient (LVG) approximation. The LVG approximation is the simplest model for such transport.
In this,  it is assumed that
     the spherically symmetric cloud possesses large scale systematic
     motions so that the velocity is a function of distance from the
     center of the cloud, that is, $V=V_0 (r/r_0)$.
     Furthermore, the systematic
     velocity is much larger than the thermal line width.  Then the
     photons emitted by a two level system at one   position in the cloud can only interact with those that  are nearby. Then  the global problem of photon
     transport is reduced to a local problem. With LVG, one has a simple method  to estimate the the effects of photon\index{Photon trapping} trapping in the
If we neglect the 2.7\,K background and  use the relation T$_{\rm 0}= h \,\nu / k$, we have
 \begin{equation}
   \label{lvg15}
    \frac{T}{T_{\rm 0}} = \frac{T_{\rm k} / T_{\rm 0}}
        {1 + T_{\rm k}/T_{\rm 0} \, \ln \!\left[
         1 + \displaystyle\frac{A_{ji}}{3 C_{ji}
      \,\tau_{ij}} \Big( 1 - \exp{(-3 \,\tau_{ij})} \Big)  \right]} \, .
 \end{equation}
The term $(1 - \exp{(-3 \,\tau_{ij})}) / \tau_{ij}$ is caused
by `photon trapping' in the cloud. If $\tau_{ij} \gg 1$,
the case of interest, then $A_{ij}$ is replaced by $A_{ij} / \tau_{ij}$. 
%********************************************************************************************************************************************************neutral-atoms

%\endd
\section{Line Radiation from Atoms}
  \label{neutral-hydr}

Most atomic transitions give rise to spectral lines at wavelengths in the
infrared or shorter.  With the exception of radio
recombination lines,
atomic radio lines are rare.
The energy levels are described by the scheme $^{2S + 1}L_J$. In this
description, $S$ is the total spin quantum number, and
$2S + 1$ is the multiplicity of the line, that is the number of possible
spin states. $L$ is the total orbital angular momentum of the system
in question, and $J$ is the total angular momentum. For the lighter
elements, the energy levels are best described using  $LS$ coupling.
This is constructed by vectorially summing the orbital momenta to obtain the
total $\vec{L}$, then combining  the spins of the individual
electrons to obtain $\vec{S}$, and then vectorially combining  $\vec{L}$
and  $\vec{S}$ to obtain  $\vec{J}$. If the nucleus has a total spin,
 $\vec{I}$, this can be vectorially  combined with  $\vec{J}$ to form
 $\vec{F}$. For an is
olated system, all of these quantum numbers have a
constant magnitude and also a constant projection in one direction. Usually
the direction is arbitrarily chosen to be along the $z$ axis, and the
projected quantum numbers are referred to as $M_F$, $M_J$, $M_L$ and $M_S$.
 
 We give a list of
the quantum assignments together with line frequencies, Einstein $A$
coefficients and critical densities in Table\,\ref{atomicline} of millimeter and sub-millimeter atomic lines.

%**********************************************************************************************************************
The most studied mm/sub-mm  atomic lines are those of neutral carbon\index{Carbon} at 492 and
809\,GHz. These lines arise from   molecular regions that the somewhat protected from the interstellar ultraviolet radiation. In less obscured regions, ionized
carbon, ${\rm C}^+$ or C\,{\sc ii} is present. This ion has a fine structure line at
157\,$\mug$m and is expected to be a dominant cooling line in denser clouds.
Although these lines might be considered as part of infrared astronomy, the
heterodyne techniques have reached the   1.4\,THz  range (=200 $\mu$m) from the ground and will reach 150 $\mu$m with the Herschel satellite. 
The oxygen  lines must be measured from high flying 
aircraft such as SOFIA or satellites. The following Table \cite{RW4} gives a few selected atomic transitions. 
 
%%Tab.12.1
\begin{table}
\caption[Parameters of some atomic lines]{Parameters of some atomic
lines\index{Frequency!atomic lines}}%
\label{atomicline}
\setlength\tabcolsep{4.75pt}
\begin{tabular*}{11.7cm}{@{}llrrrr@{}}
\hline
  \noalign{\vskip 3pt}
       Element and  &$\displaystyle {\rm  Transition} $  &
$\displaystyle
\nu /{\rm GHz} $  &  $\displaystyle A_{ij} /{\rm s}^{-1}  $  & Critical
& Notes \\
  ionization &     &    &    & density & \\
     state      &     &    &    & $n^* $&   \\
   \noalign{\vskip 3pt}
   \hline
   \noalign{\vskip 3pt}
${\rm C} {\rm I}$&  $^3P_{1}-^3P_{0}$ & 492.16 &$7.93 \times 10^{-8}$ & $5 \times 10^{2}$ & ${\rm \scriptstyle  b}$\\
${\rm C} {\rm I}$&  $^3P_{2}-^3P_{1}$ & 809.34 &$2.65 \times 10^{-7}$ & $ 10^4$       &${\rm \scriptstyle  b}$     \\
${\rm C} {\rm II}$&  $^2P_{3/2}-^2P_{1/2}$ & 1900.54 &$ 2.4 \times 10^{-6}$ & $ 5 \times 10^3$ &  ${\rm \scriptstyle  b}$\\
${\rm O} {\rm I}$&  $^3P_{0}-^3P_{1}$  & 2060.07 &$ 1.7 \times 10^{-5}$
& $\sim 4 \times 10^5$ & ${\rm \scriptstyle  b}$\\
${\rm O} {\rm I}$&  $^3P_{1}-^3P_{2} $ & 4744.77 &$ 8.95 \times 10^{-5}$ & $\sim 3 \times 10^6$& $ {\rm \scriptstyle a, b}$ \\
${\rm O} {\rm III}$&  $^3P_{1}-^3P_{0}$  & 3392.66 &$ 2.6 \times 10^{-5}$ & $\sim 5 \times 10^2$ & ${\rm \scriptstyle a}$ \\
${\rm O} {\rm III}$&  $^3P_{2}-^3P_{1}$  & 5785.82 &$ 9.8 \times 10^{-5}$ & $\sim 4 \times 10^3 $ & ${\rm \scriptstyle a}$\\
${\rm N} {\rm II}$&  $^3P_{1}-^3P_{0}$  & 1473.2 &$ 2.1 \times 10^{-6}$ & $\sim 5 \times 10^1$ & ${\rm \scriptstyle a}$ \\
${\rm N} {\rm II}$&  $^3P_{2}-^3P_{1}$  & 2459.4 &$ 7.5 \times 10^{-6}$ & $\sim 3 \times 10^2$ & ${\rm \scriptstyle a}$\\
${\rm N} {\rm III}$&  $^2P_{3/2}-^2P_{1/2}$  & 5230.43 &$ 4.8 \times 10^{-5}$ & $\sim 3 \times 10^3 $& ${\rm \scriptstyle a, b}$ \\
   \noalign{\vskip 3pt}
   \hline
 \end{tabular*}
  \\[2ex]
  $^{\rm \scriptstyle a}$ ions or electrons as collision partners \\
$^{\rm \scriptstyle b}$ H$_2$ as a collision partner \\
\end{table}

%********************************************************************************************************************************************************HII-regions

%\chapter{Recombination Lines}

\section{Emission Nebulae, Radio Recombination Lines}
  \label{recomb-lines}
 
The physical state of the interstellar medium
varies greatly from one region to
the next because the gas temperature depends on
the local energy input. There exist large, cool
cloud complexes in which both dust grains and many
different molecular species are abundant.
Often new stars are born in these dense clouds, and since they are
sources of thermal energy the stars will heat the
gas surrounding them. If the stellar surface temperature
is sufficiently high, most of the energy will be
emitted as photons with $\lambda < 912$\,{\AA}. This
radiation has sufficient energy to ionize hydrogen. Thus
young, luminous\index{Luminous stars} stars
embedded in gas clouds will be surrounded
by emission\index{Emission region} regions in which the gas
temperature and consequently the pressure will be
much higher than in cooler clouds.
  Occasionally  an ion will recombine
with a free electron. Since the
ionization rate is rather low, the time interval
between two subsequent ionizations of the
same atom will generally be much longer than the
time  for the electron to cascade to
the ground state, and the cascading atom will emit
recombination\index{Recombination line} lines.

\subsection{Rydberg Atoms}
   \label{physlp}
 
The behavior of Rydberg atoms in the interstellar medium can show complex excitation properties; such systems give an indication of 
the excitation effects one often finds with  molecules \cite{MARKRECOMB}. When ionized hydrogen\index{Hydrogen!ionized} recombines at some level
with the principal quantum number $n > 1$, the atom will
emit recombination line emission on cascading down
to the ground state. The radius of the $n^{\rm th}$ Bohr orbit is
 \begin{equation}
   \label{bohrorb}
    a_n =  \frac{\hbar^2}{Z^2 \,m \,e^2} \,n^2 \, ,
 \end{equation}
and so for large principle quantum number $n$,
the effective radius of the atom becomes exceedingly large.
Systems in such states are generally
called Rydberg atoms. Energy\index{Energy levels} levels in these
are quite closely spaced, and since pressure effects
at large $n$ caused by atomic collision may become
important, the different lines eventually will
merge. 

 The frequency of the atomic lines of
hydrogen-like atoms are given by the Rydberg\index{Rydberg formula} formula
 \begin{eqnarray}
   \label{rydnu}
     \nu_{ki} &=& Z^2 R_M \left(\frac{1}{i^2} - \frac{1}{k^2}\right) \,,
     \quad i < k
 \end{eqnarray}
 where
 \begin{eqnarray}
   \label{rmryd}
     R_M      &=& \frac{R_\infty}{ 1 + \displaystyle\frac{m}{M}}
 \end{eqnarray}
if $m$ is the mass of the electron, $M$ that of the
nucleus and $Z$ is the effective charge of the
nucleus in units of the proton charge. For $ n > 100 $
we always have $ Z \approx 1 $ and the spectra of all atoms are
quite hydrogen-like, the only difference being a
slightly changed
 value of the Rydberg\index{Rydberg constant} constant.%%

\begin{figure}%%[b]
%\sidecaption
%\mpicplace{7.9cm}{5.7cm}
   \includegraphics[width=7.9cm,height=5.7cm]{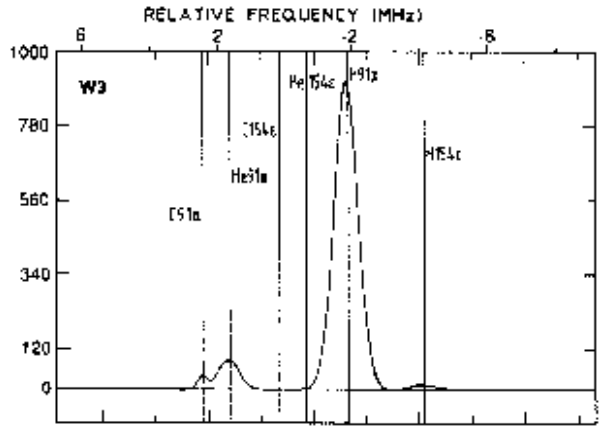}
  \caption[Recombination lines in the \HII\, region W\,3 at 8.5\,GHz]
     {\label{recorion}
       Recombination lines in the \HII\, region W\,3 at 8.5\,GHz.
       The most intense lines are
       H\,91\,$\alpha$, He\,91\,$\alpha$, C\,91\,$\alpha$, the total
       integration time, $t$, for this spectrum is 75 hours; even after this
       time, the RMS noise follows a theoretical dependence of
       $1/\sqrt{t}$ [from \cite{BALSER}}
\vglue-3pt
\end{figure}

 Lines corresponding to the
transitions $ n + 1 \to n$ are most intense and are
called $\alpha$ lines. Those for transitions $ n + 2 \to n $
are $\beta$ lines; $n + 3 \to n$ transitions are $\gamma$ lines;
etc. In the identification of a line both the element and the principle
quantum number of the lower state are given: so H\,91\,$\alpha$ is
the line corresponding to the transition $ 92 \to 91$ of H while
H\,154\,$\epsilon$ corresponds to $ 158 \to 154$ of H, Helium and Carbon (see
Fig.\ts\ref{recorion}).
%****************************************************************************************** 
%13.3
%***************************************************************************************** 

All atoms with a single electron in a highly
excited state are hydrogen-like.  The radiative
properties of these Rydberg\index{Rydberg atom} atoms differ only by
their
different nuclear masses. The Einstein
coefficients $A_{ik}$, the statistical weights $g_i$ and
the departure coefficients $b_i$ are identical for all
Rydberg atoms, if the electrons in
the inner atomic shells are not involved. The
frequencies of the recombination lines are slightly
shifted by the reduced mass of the atom. If
this frequency difference is
expressed in terms of radial velocities %(see Table\,\ref{rydcon}), 
this
difference is independent of the quantum number for a given element.
 
   The line width\index{Recombination!linewidth} of interstellar radio
recombination lines is governed by external
effects; neither the intrinsic line width nor the
fine structure of the atomic levels has observable
consequences. In normal \HII\, regions, evidence for broadening of the
lines by inelastic collisions is found for $N \ge 130$,
from the broad line wings.
For $N < 60$  the observed linewidth is fully explainable by
Doppler\index{Doppler broadening}
broadening. The purely thermal Doppler broadening  for hydrogen is:
\beq
	\Delta V_{\frac{1}{2}}=0.21 \, \sqrt{T_{\rm K}} \,.
\eeq
The electrons have a velocity
distribution that is described very closely by a
Maxwellian\index{Velocity distribution!Maxwellian} velocity distribution;
long range Coulomb forces eliminate any deviations
with a relaxation time that is
exceedingly short. This distribution is
characterized by an electron temperature $T_{\rm e}$ and,
due to the electrostatic forces, the protons should
have a similar distribution with the same
temperature. The spectral lines are observed to
have Gaussian shapes. Thermal\index{Velocity!thermal} Doppler motions
for $T_{\rm e} \cong 10^4$\,K produce a line
width of  21.4\,km\,s$^{-1}$, however a width of
$\sim 25$\,km\,s$^{-1}$ to $\sim 30$\,km\,s$^{-1}$ is
observed. Therefore it
is likely that nonthermal motions in the gas
contribute to the broadening. These motions are
usually referred to as {\em micro\index{Microturbulence}turbulence}.
If we include this effect, the half width of hydrogen is generalized
to
\begin{equation}
   \label{dopmt}
\Delta V_{\frac{1}{2}} = \sqrt{0.04576 \, T_{\rm e} + v_{\rm t}^2} \,.
\end{equation}

\subsection[LTE Line Intensities ]
    {LTE Line Intensities }

From the correspondence principle, the data
for high quantum numbers can be computed by using
classical methods, and therefore we will use for
$A_{ki}$ the expression (\ref{amn}) for the electric dipole.
For the the dipole moment in the
transition $n + 1 \to n$ we put
\beq
   \mu_{n + 1, n} = \frac{e \,a_n}{2} =
          \frac{h^2}{8 \pi^2 m \,e} \, n^2 \,,
\eeq
where $a_n = a_0 \,n^2$ is the Bohr radius of hydrogen, and
correspondingly
\beq
   \nu_{n + 1, n} = \frac{m \,e^4}{4 \pi^2 \,h^3 \,n^3} \,.
\eeq
Substituting this expression into (\ref{amn}) we obtain, for the 
limit of large $n$
 \begin{equation}
   \label{anlar}
    A_{n + 1, n} = \frac{64 \pi^6 \, m \,e^{10}}{3 \,h^6 \,c^3}
       \frac{1}{n^5} = \frac{5.36 \times 10^9}{n^5} \, {\rm s}^{-1} \,.
 \end{equation}
We adopt a Gaussian line shape, $\varphi(\nu)$.
Introducing the full line width $\Delta \nu$ at half
intensity points, we obtain for the value of $\varphi$ at
the line center
We obtain for the
optical depth in the center of a line emitted in a
region with the emission\index{Emission measure} measure for an
$\alpha$ line \\
\parbox{9.3cm}
  {\begin{displaymath}
    \fboxsep2mm
    \fbox{$
    \begin{array}{rcl}
     \displaystyle
         {\rm EM} &=& \displaystyle
             \!\int\! N_{\rm e}(s) \,N_{\rm p}(s) \diff s =
             \!\int\! \displaystyle
               \left(\frac{N_{\rm e}(s)}{\rm cm^{-3}}\right)^2 \!
               \diff \left(\frac{s}{\rm pc}\right)       \vspace*{4mm} \\
      \displaystyle
          \tau_{\rm \, L} &=& \displaystyle
         \!    1.92 \times 10^3 \! \left(\frac{T_{\rm e}}{\rm
K}\right)^{ - 5 / 2}
            \! \!   \left(\frac{\rm EM}{\rm cm^{-6} \, pc}\right)
             \!   \left(\frac{\Delta \nu}{\rm kHz}\right)^{- 1}
    \end{array}  $}  \!\quad \raisebox{-7mm}{.}
   \end{displaymath}
  }
  \hfill
%%  \raisebox{5mm}{
  \raisebox{-0.5mm}{
  \parbox{2.2cm}
    { \begin{equation}
        \label{emrec}
      \end{equation}
     \vspace*{3.5mm}
      \begin{equation}
        \label{taur}
      \end{equation}
    } } \\
Here we have assumed that $N_{\rm p}(s) \approx N_{\rm e}(s)$ which should
be reasonable due to the large abundance of H and He (= 0.1 H).
We always find that $\tau_{\rm L} \ll 1$,
and therefore that $T_{\rm L} = T_{\rm e} \, \tau_{\rm L}$, or
 \begin{equation}
  \label{tlrec}
     T_{\rm L} = 1.92 \times 10^3
       \left(\frac{T_{\rm e}}{\rm K}\right)^{ - 3 / 2}
       \left(\frac{\rm EM}{\rm cm^{-6} \, pc}\right)
       \left(\frac{\Delta \nu}{\rm kHz}\right)^{ -1} .
  \end{equation}
For $\nu >$ 1\,GHz we find that $\tau_{\rm c} < 1$ for the
continuum, so that we obtain on dividing (\ref{taur})
by (\ref{mezhen}) and using the Doppler relation
  \begin{eqnarray}
   \label{crtlc}
\kern-17.777pt   \fboxsep2mm\fbox{$ \displaystyle
       \frac{T_{\rm L}}{T_{\rm c}}
           \left(\frac{\Delta v}{\rm km\, s^{-1}}\right)
         = \frac{6.985 \times 10^3}{a(\nu, T_{\rm e})}
           \left[\frac{\nu}{\rm GHz}\right]^{1.1}
           \left[\frac{T_{\rm e}}{\rm K}\right]^{ - 1.15}
           \frac{1}{1 + N({\rm He^+}) / N({\rm H^+}) } $}  \ \ .
\kern-100pt \nonumber \\
 \end{eqnarray}
in this expresssion, $a(\nu, T_{\rm e} \cong 1$, and $T_{\rm e}$ is the LTE electron temperature, denoted as  $T_{\rm e}^*$
 The last factor is due to the fact that both $N_{\rm H^+}$ and $N_{\rm He^+}$
contribute to $N_{\rm e} = N_{\rm H^+} + N_{\rm He^+}$. 
Typical values for the \HII\, region Orion\,A are at 100 GHz are 
$\displaystyle{N({\rm H_e^+})}/{N({\rm H^+})}=0.08$,
$\displaystyle{T_{\rm L}}/{T_{\rm C}} \approx 1$ and
$\displaystyle\Delta V_{\frac{1}{2}} = 25.7 \,{\rm km\,s^{-1}}$  which give a $T_{\rm e}^*$ value of 
8200\,K.
Equation\,(\ref{crtlc}) is valid only if both the line and continuum radiation are optically thin. This  is the case for nearly all sources in the mm/sub-mm range. One 
possible exception is the  extraordinary case of MWC~349 \cite{JESUS} which shows time-variability and strong maser action in the mm/sub-mm range.

\subsection[Non LTE Line Intensities ]
      {Non LTE Line Intensities }
 
%The diameter of hydrogen\index{Hydrogen atom} atoms is strongly
%dependent on the principle quantum number [cf.\,(\ref{bohrorb})].

  The 
{\em departure\index{Departure coefficient} coefficients}, $b_n$, 
relate the true population of level, $N_n$, to the population under
LTE conditions, $N_n^\ast$, by
 \begin{equation}
   N_n = b_n \,N_n^\ast \, .
 \end{equation}
For hydrogen and helium, the $b_n$ factors are always $< 1$, since the $A$ coefficient for the lower
state is larger and the atom is smaller so collisions
are less effective. For states $i$ and $k$, with $k > i$ we have $b_n \to 1$ for LTE. For  any pair of energy levels, the upper 
level is always overpopulated relative to the lower
level. Since $h\nu \ll kT$ this overpopulation leads to a {\em negative} excitation temperature. This gives rise to recombination line masering. 
Usually the line optical depth is very small, but the background continuum could be amplified. In \hii\ regions, the dominant effect is a slightly  lower line intensity which leads to 
a slight overestimate of the electron temperature of the \hii\ region. 

%**************************************************************************************************************************************************molecular-basics

%\endd
\section{Overview of  Molecular Basics }
  \label{mollin}
 
We present the basic concepts needed to understand the radiation from  molecules that are widespread in the Interstellar Medium (ISM) \cite{TOWNES},\cite{KROTO}. 
For  linear molecules, examples  are carbon monoxide, CO, SiO, N$_2$H$^+$; for  
symmetric top molecules, these are  ammonia, NH$_3$, CH$_3$CN and CH$_3$CCH and for asymmetric top 
molecules these are water vapor, 
H$_2$O, formaldehyde, H$_2$CO and H$_2$D$^+$. Finally we present a short account  of molecules with non-zero electronic 
angular momentum in the ground state, using OH as an example, and then   
present
an account of methanol, CH$_3$OH which has hindered motion. In each section, we give  relations between  molecular energy levels, column densities and local densities. 
Although molecular line emission is complex, such measurements allow a determination of parameters in heavily obscured regions not accessible in the near-infrared or optical.

\subsection{Basic Concepts }
 
The structure and excitation of even the simplest molecules is vastly more complex than atoms.  Given the 
complicated structure, the Schr\"o\-dinger \index{Schr\"odinger}
equation of the system will be
correspondingly complex, involving positions
and moments of all constituents, both the
nuclei and the electrons. 
Because the motion of the nuclei is so
slow, the electrons make many cycles while the nuclei move  to
their new
positions. This separation of the nuclear
and electronic motion in molecular quantum
mechanics is called the Born\index{Born-Oppenheimer approximation}-Oppenheimer approximation.
 
   Transitions in a molecule can therefore
be put into three different categories according to different energies, $W$:
 \begin{itemize}
\setlength\leftskip{3pt}
\setlength\labelwidth{8pt}
   \item[a)] electronic\index{Transitions!electronic} transitions with typical
energies of a few eV --
             that is lines in the visual or UV regions of the spectrum;
   \item[b)] vibrational\index{Transitions!vibrational} transitions
caused by oscillations of the
             relative positions of nuclei with respect to their equilibrium
             positions. Typical energies are $ 0.1 - 0.01$\,eV,
             corresponding to lines in the infrared region of the spectrum;
   \item[c)] rotational\index{Transitions!rotational} transitions caused by the
rotation of the
             nuclei with typical energies of $\cong 10^{-3}$\,eV
             corresponding to lines in the cm and mm wavelength range.
  \end{itemize}
 \begin{equation}
   \label{werr}
   W^{\rm tot} = W^{\rm el} + W^{\rm vib} + W^{\rm rot} \, .
 \end{equation}
$W^{\rm vib}$ and $W^{\rm rot}$ are the vibrational and
rotational energies of the nuclei of the
molecule and $W^{\rm el}$ is the energy of the
electrons. Under this assumption, the Hamiltonian is a sum of
     $W^{\rm el} + W^{\rm vib} + W^{\rm rot}$. From quantum mechanics, the resulting wavefunction will be a product of
the electronic, vibrational and rotational wavefunctions.
 
%In  general, molecular line radiation 
%arises from  a transition between two states described by different  electronic, vibrational and rotational
%quantum numbers. If transitions involve
%different electronic states, the
%corresponding spectral lines will lie in the optical
%range. 
If we confine ourselves to the mm/sub-mm wavelength 
ranges, only transitions between different
rotational\index{Levels!rotational} levels and perhaps different
vibrational levels (e.\,g., rotational transitions of SiO or HC$_3$N from
vibrationally excited
states) will be involved. This restriction
results in a much simpler description of the molecular energy levels.
Occasionally  differences between geometrical arrangements  of the nuclei
result in a doubling of the energy levels. An example of such a case is
the inversion doubling found for the Ammonia molecule.

\subsection{Rotational Spectra of Diatomic Molecules}

Because the effective radius of even a
simple molecule is about $10^5$ times the
radius of the nucleus of an atom, the
moment of inertia $\Theta_{\rm e}$ of such a molecule
is at least $10^{10}$ times that of an atom of
the same mass. The kinetic energy of
rotation is
 \begin{equation}
   \label{kinrot}
    H_{\rm rot} = \fract{1}{2} \,\Theta_{\rm e} \, \omega^2 =
                  \vec{J}^2 / 2 \,\Theta_{\rm e} \,,
 \end{equation}
where $\vec{J}$ is the angular\index{Angular momentum} momentum.
$\vec{J}$ is a quantity that cannot be neglected compared
with the other internal energy states of
the molecule, especially if the
observations are made in the  centimeter/millimeter/sub-mm wavelength  ranges. (Note that $\vec{J}$ is {\it
not} the same as the quantum number used in atomic physics.)
 
   For a rigid  molecule consisting of two
nuclei A and B, the moment of inertia is
 \begin{eqnarray}
   \label{momin}
   \Theta_{\rm e} &=& m_{\rm A} \,r_{\rm A}^2 + m_{\rm B} \,r_{\rm B}^2 =
                m \,r_{\rm e}^2
  \end{eqnarray}
where
 \begin{eqnarray}
   \label{reab}
   \vec{r}_{\rm e} &=& \vec{r}_{\rm A} - \vec{r}_{\rm B}
  \end{eqnarray}
and
 \begin{eqnarray}
   \label{redm}
   \displaystyle
   m &=& \frac{m_{\rm A} \,m_{\rm B}}{m_{\rm A} + m_{\rm B}} \, ,
  \end{eqnarray}
 and
 \begin{eqnarray}
   \label{angmo}
   \vec{J} &=& \Theta_{\rm e} \, \vec{\omega}
  \end{eqnarray}
is the angular momentum perpendicular to
the line connecting the two nuclei. For
molecules consisting of three or more
nuclei, similar,  more complicated
expressions can be obtained. $\Theta_{\rm e}$ will depend on
the relative orientation of the nuclei and will in general be a
(three-axial) ellipsoid. In (\ref{angmo})
values of $\Theta_{\rm e}$ appropriate for the direction
of $\vec{\omega}$ will then have to be used.

 This solution of the Schr{\"o}dinger equation
then results in the {\em eigenvalues} for the rotational
energy
 \begin{equation}
   \label{rotene}
    E_{\rm rot} = W(J) =
    \frac{\hbar^2}{2 \,\Theta_{\rm e}} \,J ( J + 1 ) \,,
 \end{equation}
where $J$ is the quantum number of angular
momentum, which has integer values
\beq
   J = 0, 1, 2, \dots \,.
\eeq

%\eject

\noindent
Equation (\ref{rotene}) is correct only for a molecule that
is completely rigid\index{Rigid rotator}; for a slightly elastic
molecule, $r_{\rm e}$ will increase with the
rotational energy due to centrifugal\index{Centrifugal stretching} stretching.
(There is also the additional complication that even in the ground vibrational
     state there is still a zero point vibration; this will be
     discussed after the concept of centrifugal stretching is presented.)
For centrifugal stretching, the rotational energy is modified to first order as:
 \begin{equation}
   \label{erotel}
    E_{\rm rot} = W(J) = \frac{\hbar^2}{2 \,\Theta_{\rm e}} \,J ( J + 1 ) -
               h D \left[ J ( J + 1 ) \right]^2 \,.
 \end{equation}
      Introducing the rotational constant
 \begin{equation}
   \label{rotconst}
    B_{\rm e} = \frac{\hbar}{4 \pi \,\Theta_{\rm e}}
 \end{equation}
and the constant for centrifugal stretching
$D$, the pure rotation spectrum for electric
dipole transitions $\Delta J = +1$ (emission) or $\Delta J = - 1$
(absorption) is given by the following expression:
 \begin{equation}
   \label{nuj}
     \nu(J) = \frac{1}{h} \left[ W(J + 1) - W(J) \right] =
          2 \,B_ {\rm e} \,( J + 1 ) - 4 D \,( J + 1 )^3 \,.
 \end{equation}
%************************************************************************
%14.4--now 15.2 TLW Apr 08
%\vspace{6cm}
\begin{figure}%[b]
%\sidecaption
%\mpicplace{4.4cm}{5.0cm}
\includegraphics[width=4.4cm,height=5.0cm]{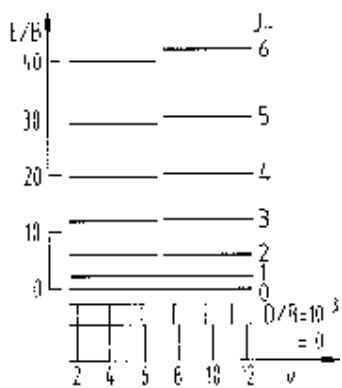}
   \caption[Rotational energy levels for a rigid rotator]
        { \label{retire}
        A schematic plot of rotational energy levels for a  rotor. The horizontal
        bars in the upper part represent  the
        rotational \index{Energy levels!rotational} energy levels for a
        rigid rotator ({\em right part}) and one deformed by centrifugal
        stretching with $D / B_{\rm e} = 10^{-3}$ ({\em left part}).
        In fact, most molecules have $D / B_{\rm e} \approx 10^{-5}$.
        The resulting line frequencies, $\nu$, are shown  in the lower
        part. The numbers next to $\nu$ refer to the $J$ values \cite{RW4}. }
\end{figure}
%**********************************************************************
 
\begin{figure}%[t]
%\sidecaption
%\mpicplace{4.3cm}{7.2cm}
\includegraphics[width=7.2cm,height=7.2cm]{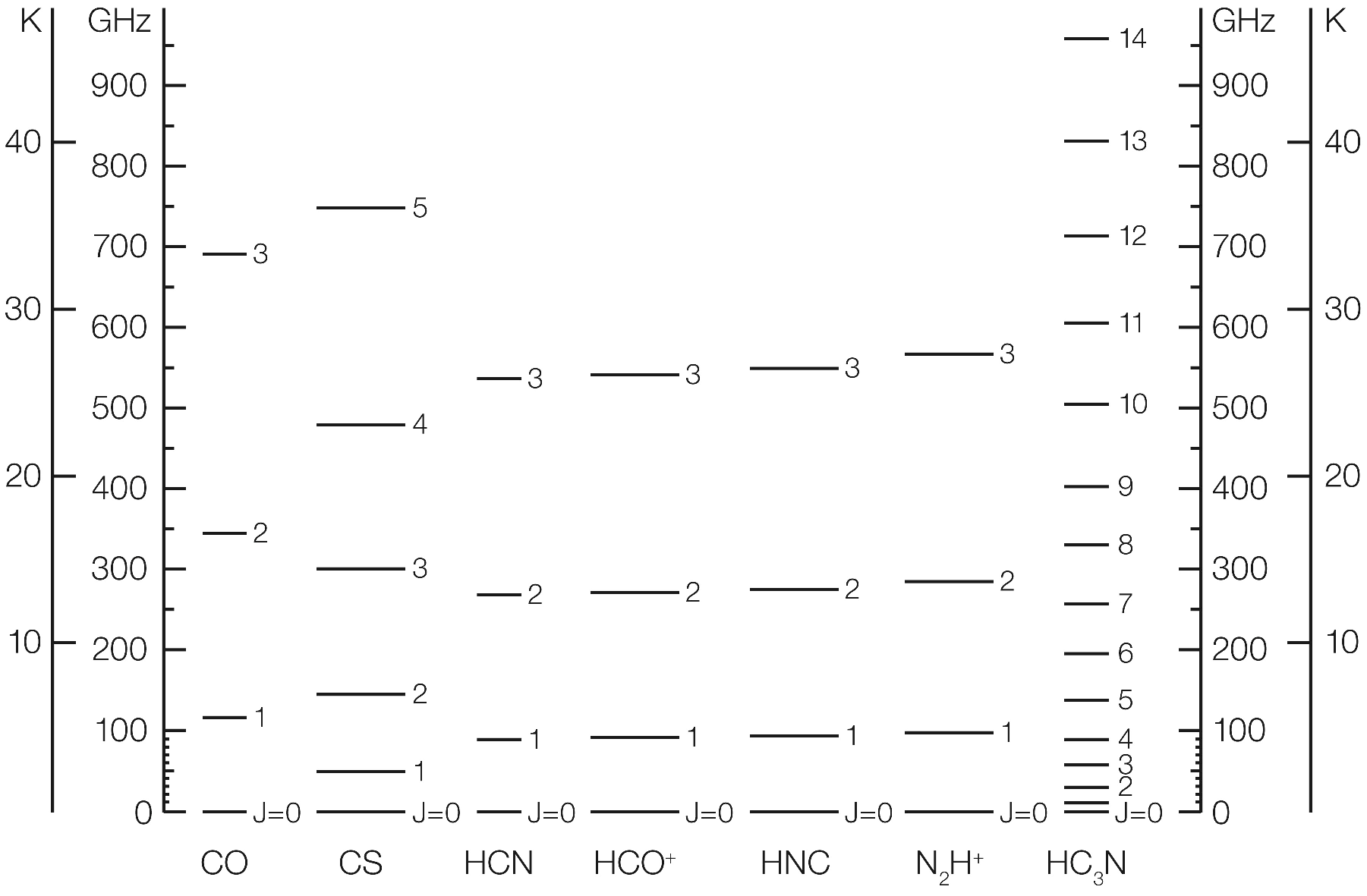}
  \caption[Rotational energy levels of ground states of some linear molecules]
        {\label{enlevels}
     Rotational energy levels\index{Molecule!linear!energy levels}
      of the vibrational ground
     states of some  linear molecules which are commonly found in the
     interstellar medium  (Taken from \cite{RW5}).}
%[adapted from Avery\,(1980)]
\end{figure}

Since $D$ is positive, the observed line frequencies will be lower than
those predicted on the basis of a perfectly rigid\index{Rigid rotator}
rotator. Typically, the
size of $D$ is about 10$^{-5}$ of the magnitude of $B_ {\rm e}$
for most molecules.
     In Fig.\,\ref{retire},
     we show a parameterized plot of the behavior of
energy above ground and line frequency of a rigid rotor with and without
the centrifugal
 distortion term. The function
plotted vertically on the left, $E_{\rm rot}/B_ {\rm e}$,  is
proportional to the energy above the molecular ground state. This
function is
given by
 \begin{equation}
   \label{erotel1}
E_{\rm rot}/B_ {\rm e}= 2\pi \hbar \,J ( J + 1 ) -
               2\pi \hbar\,D/B_ {\rm e} \left[ J ( J + 1 ) \right]^2
 \end{equation}
while those on the right are given by the first term
of (\ref{erotel1}) only. Directly below the energy
level plots is a plot of
the line frequencies for a number of transitions with quantum number
$J$. The deviation between rigid rotor and actual
frequencies becomes rapidly larger with increasing $J$, and in the sense
that the actual frequencies are always lower than the frequencies
predicted on the basis of a rigid rotor model. In Fig.\,\ref{enlevels} we
show plots of the energies above ground state for a number of diatomic
and triatomic linear molecules.
%************************************************************************************** 
%14.5
%\vspace{10cm}
%************************************************************************************** 

Allowed dipole
radiative transitions will occur between different rotational states
     only if the molecule
possesses a permanent\index{Electric dipole!permanent} electric dipole
moment; that is, the molecule must be polar. Homonuclear diatomic
molecules like H$_2$, N$_2$ or O$_2$  do not possess
 permanent electric dipole moments. 
Thus they cannot  undergo allowed transitions.
This is one reason why it was so difficult
to detect these species. 
%*******************************************************************************************************************************************************
%add tlw

For  molecules with permanent dipole moments,
a classical picture of molecular line radiation can be used to
determine  the angular distribution of the radiation. In the plane of
rotation, the dipole moment can be viewed as an antenna, oscillating as
the molecule rotates. Classically,  the acceleration of
positive and negative charges gives rise to
radiation whose frequency
     is that of the rotation frequency. For a dipole transition
     the most intense radiation occurs in the plane of
rotation of the molecule. In the
     quantum mechanical model, the angular momentum is quantized, so
that the radiation is emitted at discrete frequencies.
Dipole radiative transitions occur with a change in the angular momentum
quantum number of one
unit, that is, $\Delta J =  \pm 1$.
The parity of the initial and final states must be opposite
for dipole radiation to occur.
%*******************************************************************************************Need to insert HFS in linear molecules here
\subsection{Hyperfine Structure in Linear Molecules}
 The magnetic dipole or electric quadrupole of nuclei  interact with
electrons or other nuclei. These give rise to hyperfine\index{Hyperfine
structure} structure in molecules such as HCN, HNC and HC$_3$N.  For example, the
$^{14}$N and Deuterium nuclei have spin $I=1$ and thus a nonzero quadrupole moment. 
The
hyperfine splitting of energy levels depends on the position of the nucleus in the molecule; the effect is smaller for HNC than for HCN. In general, the effect 
is of order of a few MHz, and decreases with
increasing $J$.  For nuclei with magnetic dipole moments, such as $^{13}$C or $^{17}$O, the hyperfine splitting is smaller. 
In the case of hyperfine structure, the total quantum
number $\vec{F}=\vec{J} + \vec{I}$ is conserved. Allowed transitions
obey the selection\index{Selection rule} rule $\Delta \vec{F}= \pm 1, 0$ but 
not $\Delta \vec{F}= 0{\stackrel{\textstyle \to}{\gets}} 0$.

%*************************************************************************************************************************************************moved here on April 7, 2008
\subsection{Vibrational Transitions}
%********************************************************************************** vibrational transitions TLW April 7, 2008
\begin{figure}%[b]
%\sidecaption
%\mpicplace{4.2cm}{8.0cm}
 \includegraphics[width=4.2cm,height=8.0cm]{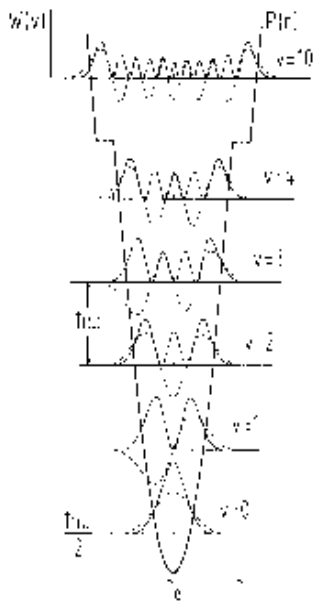}
  \caption[ Vibrational energy levels for a harmonic oscillator]
        { \label{vibrel}
       Vibrational energy levels, eigenstates (- - -) and
       probability densities (--) for a harmonic oscillators \cite{RW4}.}
\end{figure}
%*********************************************************************************
If any of the nuclei of a molecule suffers
a displacement from its equilibrium
distance $r_{\rm e}$, it will on release perform an
oscillation about $r_{\rm e}$. The Schr{\"o}dinger
equation for this is
 \begin{equation}
  \label{schrvib}
   \left( \frac{p^2}{2 m} + P(r) \right) \psi^{\rm vib}(x) =
           W^{\rm vib} \,\psi^{\rm vib}(x) \,,
 \end{equation}
where $x = r - r_{\rm e}$ and $P(r)$ is the potential
function. If  we have the simple harmonic\index{Harmonic approximation} approximation
(Fig.\,\ref{vibrel})
with the classical oscillation frequency
 \begin{equation}
   \label{closz}
 \omega = 2 \pi \nu = \sqrt{\frac{k}{m}} =
 a \, \sqrt{\frac{2 D_{\rm e}}{m}} \,,
 \end{equation}
and (\ref{schrvib}) has the eigenvalue
 \begin{eqnarray}
   \label{eigwvib}
     W^{\rm vib} &=& W(v) = \hbar \,\omega \,( v + \fract{1}{2} )
 \end{eqnarray}
 with
 \begin{eqnarray}
   \label{veig}
     v &=& 0, 1, 2, \dots \, .
 \end{eqnarray}
The solutions $\psi^{\rm vib}(x)$ can be expressed with
the help of Hermite polynomials. For the same rotational quantum numbers, lines arising from transitions in different vibrational
states, in a harmonic potential, are separated by a constant frequency interval. 
 
   For large $x$, the accuracy of the harmonic motion approximation  is no longer
sufficient  even an empirical expression, will have to
be introduced into (\ref{schrvib}). The resulting
differential equation can no longer be
solved analytically, so numerical
methods have to be used.

   A molecule consisting of only two nuclei
can vibrate only in one direction; it has
only one vibrational mode. The situation is
more complex for molecules with three or more
nuclei. In this case, a multitude of various
vibrational\index{Vibrational modes} modes may exist, each of which
will result in its own ladder of
vibrational states, some of which may be
degenerate. For a certain  molecular vibrational state, there are many
internal rotational states.
Vibrational motions along the molecular axis can be and usually are
hindered in the sense that these are subject to centrifugal forces, and
thus must overcome an additional barrier.

%added TLW
It is possible to have  transitions between rotational energy levels in a given vibrationally excited state. An example is the $J=1-0$ rotational 
transition of the SiO molecule from the $v=0, 1$ and 2 levels. The dipole moment in a vibrational state is usually the same as in the ground state. 
The dipole moment for a purely 
vibrational transition in the case of diatomic molecules is usually about 0.1
Debye.  A more complex example is the polyatomic linear molecule HC$_3$N, for
which a number of transitions have been measured in the ISM. 
%**********************************************************************************************************************************************8tlw Apr 8, 08
 
\subsection[  Line Intensities of Linear Molecules]
{ Line Intensities of Linear Molecules }
 
In this section, we will give the details needed to
 relate the
observed line intensities to column\index{Column density} densities of
the species emitting the transition. In the Born-Oppenheimer
approximation, the total energy can be written as a sum
of the electronic, vibrational and rotational energies in (\ref{werr}).
In the line spectrum of a molecule,
transitions between electronic, vibrational and rotational states are
possible. We will restrict the discussion to  rotational transitions,
and in a few cases to vibrational transitions.
 
Computations of molecular line
intensities proceed following
the principles outlined in conjunction with the Einstein coefficients. The radial part of  molecular wavefunctions is extremely complex.
For any molecular or atomic system, the spontaneous transition
probability, in s$^{-1}$,
for a  transition between an upper, $u$, and lower, $l$, level
is given
     by the Einstein\index{Einstein coefficient} $A$ coefficient $A_{ul}$.
In the CGS system of  units,  $A_{ul}$  is given by  (\ref{amn}). 
 %\begin{equation}
 %  \label{ajdipcgs}
 %   A_{ul} = \frac{64 \,\pi^4}{3 \, h \,c^3} \,\nu^3 \, |\mu_{ul}|^2 \,.
 %\end{equation}
Inserting  numerical values, we have:
\begin{equation}
\label{acoeff}
A_{ul} = 1.165 \times 10^{-11}\, \nu^3  |\mu_{ul}|^2 \,.
\end{equation}
The units of the line
     frequency $\nu$ are GHz, and the units of $\mu$ are Debyes (i.e.,
\mbox{1\,Debye} $= 10^{-18}$ e.s.u.)
Eq.\,\ref{acoeff} is a completely general relation for {\em any} transition.
The expression $|\mu|^2$ contains a term which depends on the integral over
the angular part of
the wavefunctions of the final and initial states; the radial part of the
wavefunctions is contained in the value of the dipole moment, $\mu$
(This is usually determined
from laboratory measurements).
For dipole\index{Dipole transitions} transitions  between
     two rotational levels of a linear molecule,
     $ J {\stackrel{\textstyle \to}{\gets}} J + 1$,
     there can be either absorption or emission. For the case of
     absorption,  for a dipole moment $|\mu_{ul}|^2=|\mu_J|^2$, we have:
 \begin{equation}
   \label{dipmjdown}
     |\mu_J|^2 = \mu^2 \,\frac{J + 1}{2 J + 1} \qquad {\rm for} \quad
                         J \to J + 1
\end{equation}
while for emission, the expression is given by:
 \begin{equation}
  \label{dipmj}
     |\mu_J|^2 = \mu^2 \,\frac{J + 1}{2 J + 3} \qquad {\rm for} \quad
                         J+1 \to J
 \end{equation}
where $\nu_{\rm r}$ is the spectral line frequency.
     Here, $\mu$ is the permanent electric dipole moment
of the molecule.Table \ref{lparaco} contains parameters for a few species found in the interstellar medium. 
 
%%Tab.14.2
\begin{table}%[t!]
%\vglue-10pt
  \label{lparaco}
{%\scriptsize
  \caption[Parameters of the  commonly observed short cm/mm molecular lines]
	{ Parameters of the  commonly observed short cm/mm molecular
lines\index{Frequency!molecular lines}}
\vspace*{-6pt}
\def\abs{\hspace{5.5pt}}
\def\absx{\hspace{9pt}}
 \scriptsize
  \begin{flushleft}
  \begin{tabular}{@{}l@{\abs}l@{\abs}l@{\abs}r@{\abs}r@{\absx}r@{}}
\hline
  \noalign{\vskip 3pt}
       $\displaystyle
        {\rm  Chemical}^{\rm \scriptstyle a}$ & Molecule &
       Transition  &
      $\displaystyle
       \nu/{\rm GHz} $  &
      $\displaystyle
        {\rm E_u/\rm K}^{\rm \scriptstyle b} $ &
      $\displaystyle
        A_{ij}/{{\rm s}^{-1}}^{\rm \scriptstyle c}$  \\
 formula &name &&&&\\
   \noalign{\vskip 3pt}
   \hline
   \noalign{\vskip 3pt}
${\rm H_2}{\rm O}$&ortho-water$^*$ &$J_{{K_a}{K_c}}=6_{16}-5_{23}$ &
22.235253 &640 & 1.9 $ \times 10^{-9}$ \\
${\rm N}{\rm H_3}$&para-ammonia&$(J,K)=(1,1)-(1,1)$&
23.694506 &23 & 1.7 $ \times 10^{-7}$ \\
${\rm N}{\rm H_3}$&para-ammonia&$(J,K)=(2,2)-(2,2)$&
23.722634 &64 & 2.2 $ \times 10^{-7}$ \\
${\rm N}{\rm H_3}$&ortho-ammonia&$(J,K)=(3,3)-(3,3)$&
23.870130 &122 & 2.5 $ \times 10^{-7}$ \\
${\rm Si} {\rm O}$&silicon monoxide$^*$  & $J= 1-0, v=2 $
&42.820587 &3512&3.0 $ \times 10^{-6}$\\
${\rm Si} {\rm O}$ &silicon monoxide$^*$ & $J= 1-0, v=1 $
&43.122080 &1770&3.0 $ \times 10^{-6}$\\
${\rm Si}{\rm O}$ &silicon monoxide & $J= 1-0, v=0 $
&43.423858 &2.1& 3.0 $ \times 10^{-6}$\\
${\rm C} {\rm S}$&carbon monosulfide& $J= 1-0 $
&48.990964 &2.4& 1.8 $ \times 10^{-6}$\\
${\rm D} {\rm C} {\rm O}^+$&deuterated formylium&$J= 1-0 $
&72.039331 &3.5& 2.2 $ \times 10^{-5}$\\
${\rm Si} {\rm O}$&silicon monoxide$^*$  & $J= 2-1, v=2 $
&85.640456 &3516&2.0 $ \times 10^{-5}$\\
${\rm Si} {\rm O}$&silicon monoxide$^*$  & $J= 2-1, v=1 $
&86.243442 &1774&2.0 $ \times 10^{-5}$\\
${\rm H} ^{13}{\rm C} {\rm O}^+$& formylium&$J= 1-0 $
&86.754294 &4.2& 3.9 $ \times 10^{-5}$\\
${\rm Si} {\rm O}$ &silicon monoxide & $J= 2-1, v=0 $
&86.846998 &6.2 &2.0 $ \times 10^{-5}$\\
${\rm H} {\rm C} {\rm N}$ &hydrogen cyanide& $J=1-0, F=2-1$
&88.631847 &4.3 & 2.4 $ \times 10^{-5}$\\
${\rm H} {\rm C} {\rm O}^+$& formylium&$J= 1-0 $
&89.188518 &4.3& 4.2 $ \times 10^{-5}$\\
${\rm H} {\rm N} {\rm C}$ &hydrogen isocyanide& $J=1-0, F=2-1$
&90.663574 &4.3 & 2.7 $ \times 10^{-5}$\\
$ {\rm N_2}{\rm H}^+ $ &diazenylium& $J=1-0$, $F_1=2-1$,\\
&&\quad\kern7.768pt $F=3-2$
&93.173809 &4.3 & 3.8 $ \times 10^{-5}$\\
${\rm C} {\rm S}$&carbon monosulfide& $J= 2-1 $
&97.980968 &7.1& 2.2 $ \times 10^{-5}$\\
${\rm C} ^{18}{\rm O}$  &carbon monoxide& $J=1-0$
&109.782182 & 5.3& 6.5 $ \times 10^{-8}$  \\
$^{13}{\rm C} {\rm O}$& carbon monoxide & $J= 1-0 $
& 110.201370 & 5.3 & 6.5 $ \times 10^{-8}$ \\
${\rm C} {\rm O}$&carbon monoxide  & $J=1-0 $ & 115.271203
& 5.5& 7.4 $ \times 10^{-8}$ \\
${\rm H_2} ^{13}{\rm C} {\rm O}$&ortho-formaldehyde&$J_{{K_a}{K_c}}=
2_{12}-1_{11}$ & 137.449959 & 22 & 5.3 $ \times 10^{-5}$ \\
${\rm H_2} {\rm C} {\rm O}$&ortho-formaldehyde&$J_{{K_a}{K_c}}=
2_{12}-1_{11} $ &140.839518 & 22 & 5.3 $ \times 10^{-5}$ \\
${\rm C} {\rm S}$&carbon monosulfide& $J=3-2 $
&146.969049 &14.2& 6.1 $ \times 10^{-5}$\\
${\rm C} ^{18}{\rm O}$  &carbon monoxide& $J=2-1$
&219.560319 & 15.9 &6.2 $ \times 10^{-7}$ \\
$^{13}{\rm C} {\rm O}$ &carbon monoxide & $J= 2-1 $
& 220.398714 & 15.9&6.2 $ \times 10^{-7}$ \\
${\rm C} {\rm O}$  &carbon monoxide & $J=2-1 $
& 230.538001 & 16.6&7.1 $ \times 10^{-7}$ \\
${\rm C}{\rm S}$&carbon monosulfide& $J=5-4 $
&244.935606 &33.9& 3.0 $ \times 10^{-4}$\\
${\rm H} {\rm C} {\rm N}$ &hydrogen cyanide& $J=3-2$
&265.886432 &25.5 & 8.5 $ \times 10^{-4}$\\
${\rm H} {\rm C} {\rm O}^+$& formylium&$J= 3-2 $
&267.557625 &25.7& 1.4 $ \times 10^{-3}$\\
${\rm H} {\rm N} {\rm C}$ &hydrogen isocyanide& $J=3-2$
&271.981067 &26.1 & 9.2 $ \times 10^{-4}$\\
   \noalign{\vskip 3pt}
   \hline
 \end{tabular}
  \\[2ex]
$^{\rm \scriptstyle a}$~If isotope not explicitly given, this is the
most abundant variety, i.e., $^{12}$C is C, $^{16}$O is O, $^{14}$N\\
\ \kern4.035pt{}is N,  $^{28}$Si is Si,$^{32}$S is S\\
  $^{\rm \scriptstyle b}$ Energy of upper level above ground, in Kelvin\\
$^{\rm \scriptstyle c}$ Spontaneous transition rate, i.e., the Einstein $A$
coefficient\\
  $^*$ Always found to be a maser transition\\
$^{**}$ Often found to be a maser transition\\
 \end{flushleft}
\scriptsize}
\end{table}

%*************************************************************************************************************************************************************
 
After  inserting (\ref{dipmj}) into (\ref{acoeff}) we obtain
the expression for dipole\index{Dipole emission}
emission between two levels of a linear molecule:
\begin{equation}
\label{acoeffcgs}
 \fboxsep2mm\fbox{$ \displaystyle
    A_J = 1.165 \times 10^{-11}\, \mu^2 \,\nu^3 \,\frac{J + 1}{2 J + 3}
            \qquad {\rm for} \quad
                         J+1 \to J   $}
\end{equation}
where $A$ is in units of s$^{-1}$, $\mu_J$ is in Debyes
(i.\,e.~$10^{18}$ times  e.s.u.~values),
and $\nu$ is in GHz. This expression is
valid for a dipole transition in a linear molecule, from
a level $J+1$ to $J$.

     Inserting the expression for $A$ in (\ref{knbo}),
the general relation
between line optical depth, column density in a  level $l$ and
excitation temperature, $ T_{\rm ex}$, is:
\begin{equation}
 \label{levelpop}
 \fboxsep2mm\fbox{$ \displaystyle
        N_l =  93.5 \,\frac{g_l\, \nu^3}{  g_u \,A_{ul}}
        \frac{1}{\left[1-\exp(-4.80 \times 10^{-2} \nu / T_{\rm ex})\right]}
                \int \tau \diff v $}  \quad 
\end{equation}
where the units for $\nu$ are GHz and the linewidths are in
km\,s$^{-1}$.
 $n$ is the local density in units of cm$^{-3}$, and
$N=n \,l$ is the column density, in cm$^{-2}$.

Although this expression appears simple, this is deceptive, since
there is a dependence on $T_{\rm ex}$. The excitation process may
cause  $T_{\rm ex}$\ to take on
a wide range of values. If $T_{\rm ex}/ \nu \gg 4.80 \times 10^{-2}$\,K,
the expression becomes:
\begin{equation}
\label{coldenlv}
 \fboxsep2mm\fbox{$ \displaystyle
         N_l = 1.94 \times 10^{3} \frac{g_l\, \nu^2 \,T_{\rm ex}}
                { g_u\, A_{ul}} \int \tau \diff v $}  \quad .
\end{equation}

Values for $T_{\rm ex}$\ are difficult to obtain in the general case.
Looking ahead a bit, for the $J=1 \to 0$ and $J=2 \to 1$ transitions, CO
molecules are found to be
almost always close to LTE, so it is possible to obtain estimates of
$T_{\rm K}$ from (\ref{tcotw}).
     This result could be used in (\ref{coldenlv})
if the transition is close to LTE.
Expression\,(\ref{coldenlv}) can be  simplified even further if $\tau
\ll 1$. Then, if the source fills the main beam (this is the usual
assumption) the following relation holds:
 \begin{equation}
   \label{thindepth}
    T_{\rm ex} \, \tau \cong T_{\rm MB}
 \end{equation}
where the term $T_{\rm MB}$ represents the main beam  brightness
temperature. In the general case, we will use $T_{\rm B}$, which depends
on source size.  Inserting this in (\ref{coldenlv}), we have:
\begin{equation}
\label{coldentemp}
 \fboxsep2mm\fbox{$ \displaystyle
         N_l= 1.94 \times 10^{3} \frac{g_l\, \nu^2 }
                {g_u\, A_{ul}\,} \int T_{\rm B} \diff v $}  \quad .
\end{equation}
In this relation, $T_{\rm ex}$\ appears nowhere. Thus,
for an optically\index{Optically thin line}
thin emission line, excitation plays {\it no role} in
determining the column density in the energy levels giving rise to the
transition. The units are as before; the column density, $N_l$, is an
average over the telescope beam. 
 
\subsubsection{Total Column Densities of CO Under LTE Conditions}
 
We apply the concepts developed in the last section to carbon monoxide,
a simple molecule that is
abundant in the ISM.  Microwave
radiation from this
molecule is  rather
easily detectable because CO\index{CO excitation}
 has a permanent dipole\index{Dipole moment}
moment of $\mu = 0.112$\,Debye.
%---tlw removed MKS units
CO is a diatomic molecule with a simple
ladder of rotational levels spaced such
that the lowest transitions are in the
millimeter wavelength region.    A first approximation of the abundance
of the CO molecules can be obtained by a very
standard LTE analysis of the CO line
radiation; this is also  fairly realistic since the excitation of low rotational
transitions
is usually close to LTE. Stable isotopes
exist for both C and O and
several isotopic species of CO have been
measured in the interstellar medium; among these are $^{13}$C$^{16}$O, $^{12}$C$^{18}$O, $^{12}$C$^{17}$O, $^{13}$C$^{16}$O and $^{13}$C$^{18}$O. 

For the distribution of CO, we
adopt the simplest geometry,
that is, an isothermal slab which is much larger than the telescope beam.
Then the solution (\ref{tb-isoth}) may be used.
If we recall
that a baseline is usually subtracted from
the measured line profile, and that the 2.7\,K
microwave background radiation is present
everywhere, the appropriate formula is
 \begin{equation}
  \label{tapf}
  T_{\rm B}(\nu) = T_0 \left( \frac{1}{\expo^{T_0 / T_{\rm ex}} - 1} -
               \frac{1}{\expo^{T_0 / 2.7} - 1} \right)
             ( 1 - \expo^{ - \tau_\nu} ) \, ,
 \end{equation}
where $T_{\rm 0}= h \,\nu /k$.  On
the right side of (\ref{tapf}) there are  two 
unknown quantities: the excitation \index{Excitation temperature}
temperature of the line, $T_{\rm ex}$, and the optical
depth, $\tau_\nu$. If $\tau_\nu$ is known it is possible to
solve for the column density $N_{\rm CO}$ as in the
case of the line $\lambda = 21$\,cm of \HI. But in
the case of CO we meet the difficulty that lines of
the most abundant isotope $^{12}{\rm C}\,^{16}{\rm O}$ always seem to
be optically thick. It is therefore not
possible to derive information about the CO
column density from this line without a model for the molecular clouds.
%We give a discussion in the next Chapter.
Here we give an analysis based on the measurement of weaker isotope lines
of CO. This procedure can be applied if the following assumptions are
valid.
\begin{itemize}
  \item All molecules along the line of sight possess a uniform excitation
         temperature in the $ J = 1 \to 0$ transition.
  \item  The different isotopic species have the same excitation
         temperatures. Usually the excitation temperature is taken to be
     	 the kinetic temperature of the
         gas, $ T_{\rm K}$.
  \item  The optical depth in the
 	 $^{12}{\rm C}\,^{16}{\rm O} $ $ J = 1 \to 0$ line is
         large compared to unity.
  \item  The optical depth in a rarer isotopomer
   	 transition, such as the $^{13}{\rm C}\,^{16}{\rm O}$ $ J = 1 \to 0$
         line is small compared to unity.
  \item  The \THCO\ and CO lines are emitted from the same volume.
  %\item  Clumping on a scale much smaller than the beam is not
 	%important.
\end{itemize} 
Given these assumptions, we have
$T_{\rm ex} = T_{\rm K} = T$, where $T_{\rm K}$ is the
kinetic  temperature, which is the only parameter in the
Maxwell-Boltzmann relation for the cloud in question. In the remainder
of this section and in the following section we will use the expression
$T$, since all temperatures are assumed to be equal. This is certainly
{\it not} true in general. Usually, the molecular energy
  level\index{Energy level!populations} populations are often characterized by {\it
at least}
one other temperature, $T_{\rm ex}$. 
 
In general,  the lines of  $^{12}{\rm C}\,^{16}{\rm O}$ are
    optically thick. Then, in the absence
     of background continuum sources, the excitation
temperature can be determined from the appropriate $T_{\rm B}^{12}$ of
the optically thick
$J=1-0$ line of $^{12}{\rm C}\,^{16}{\rm O}$
at 115.271\,GHz:
 \begin{equation}
   \label{tcotw}
   \fboxsep2mm\fbox{$ \displaystyle
     T  = 5.5 \Bigg/ \ln \left( 1 +
       \frac{5.5}{T_{\rm B}^{12} + 0.82} \right) $} \quad .
 \end{equation}
The optical depth of the $^{13}{\rm C}\,^{16}{\rm O}$ line at
110.201\,GHz is obtained by solving (\ref{tapf}) for
 \begin{equation}
   \label{odct}
   \fboxsep2mm\fbox{$ \displaystyle
     \tau_0^{13} = - \ln \left[ 1 -
       \frac{  T_{\rm B}^{13} }{5.3} \left\{ \left[
        \exp \left( \frac{5.3}{T} \right) - 1
           \right]^{ - 1} - 0.16 \right\}^{ - 1} \right] $} \quad .
 \end{equation}
 
Usually the total column\index{Column density} density is the quantity of
interest. To obtain this for CO, one  must sum over all
energy levels of the molecule. This can be carried out for the LTE case in
a simple way. For non--LTE conditions, the calculation is
considerably more complicated. For more complex situations, statistical equilibrium or LVG (\ref{lvg15}) models are needed. 
In this section, we concentrate on the case of CO
populations in LTE.
 
For CO, there is no statistical weight factor due to spin degeneracy.
In a level $J$, the degeneracy is $2 J + 1$. Then the
fraction of the total population  in a
particular state, $J$, is given by:
 \begin{equation}
   \label{boltzmann}
     N(J)/N({\rm total}) = \frac{(2 J + 1)}{Z} \,
	\exp \left[ - \frac{ h \,B_{\rm e}J (J+1)}{k T} \right ]
         .
 \end{equation}
$Z$ is the sum over all states, or the Partition\index{Partition function}
function.
If vibrationally excited states are not populated, $Z$ can be expressed as:
 \begin{equation}
  \label{usumj}
   Z = \sum_{ J = 0 }^\infty \,( 2 J + 1 ) \,\exp \left[ - \frac{ h \,B_{\rm e}
      \, J ( J + 1 ) }{ k \, T} \right]  \, .
 \end{equation}
The total population, $N({\rm total})$ is given
by the  measured column density for a specific level, $N(J)$,
divided by the  calculated fraction of the total population in this level:
\begin{equation}
\label{totalsum}
 N({\rm total}) = N(J) \, \frac{Z}{ (2 J + 1)} \,\exp
\left[\frac{ h \,B_{\rm e} \, J ( J + 1 ) }{ k \, T} \right] \, .
\end{equation}
This fraction is based on the assumption that all energy
levels are populated under  LTE conditions. For a
     temperature, $T$, the population will increase as $2 J + 1$,
until the energy above the ground state becomes large compared to $T$. Then
the negative exponential becomes the significant factor and the population
will quickly decrease.
If the temperature is large compared to the separation of energy levels,
the sum can be approximated by an integral,
\begin{equation}
\label{zco}
Z \approx \frac{k \,T}{h \,B_{\rm e}}
      \qquad {\rm for} \quad h \,B_{\rm e} \ll k T \; .
\end{equation}
Here $B_{\rm e}$ is the rotation constant (\ref{rotconst}),
and the molecular population is assumed to be
characterized by a single temperature, $T$, so that the Boltzmann
distribution can be applied.
Applying (\ref{totalsum}) to the $J=0$ level,
we can obtain the total column density of \THCO\ from a measurement of
the $J=1 \to 0$ line of CO and \THCO,
using the partition function of CO, from (\ref{zco}),
and (\ref{levelpop}):
 \begin{equation}
   \label{ncoth13}
   \fboxsep3mm\fbox{$ \displaystyle
       N({\rm total})_{\rm CO}^{13} =  3.0 \times 10^{14}
       \frac{ T   \displaystyle \int\limits \tau^{13}(v)  \diff v }
                    { 1 - \exp \left\{ - 5.3 / T \right\}}
                $}  \quad .
 \end{equation}
It is often the case that in dense molecular clouds \THCO\ is optically
thick. Then we  should make use of an even rarer substitution, \CEIO.
For the
$J=1 \to 0$ line of this substitution, the expression is exactly the
same as (\ref{ncoth13}). For the $J=2 \to 1$ line, we obtain a similar
expression, using (\ref{totalsum}):
 \begin{equation}
   \label{ncoth21}
  \fboxsep3mm\fbox{$ \displaystyle
        N({\rm total})_{\rm CO}^{13} = 1.5 \times 10^{14}
           \frac{ T \displaystyle \,\exp{\{ 5.3 / T \}}
                 \displaystyle \int\limits \tau^{13}(v) \diff v}
                   { 1 - \exp \left\{ - 10.6 / T \right\} }
                $}  \quad .
    \end{equation}
In both (\ref{ncoth13}) and (\ref{ncoth21}),
the beam\index{Beam average} averaged
column density of carbon monoxide is in units of \percmsq\ the line
temperatures are in Kelvin, main beam brightness temperature and the
velocities, $v$, are in \kms. If the value of $T \gg 10.6$ or $5.3$\,K, the
exponentials can be expanded to first order and then these relations become  simpler.

In the limit of optically thin lines, integrals involving
$\tau(v)$ are equal
to the integrated line intensity $\int T_{\rm MB}(v) \diff v$, as mentioned
before. However,
there will be a dependence on $T_{\rm ex}$\ in these relations
because of the Partition function.
The relation $T \,\tau(v) = T_{\rm MB}(v)$
is only approximately true. However, optical depth effects can be
eliminated to some extent by using the approximation
 \begin{equation}
   \label{ttap}
    T \int\limits_{- \infty}^\infty \tau(v) \diff v
      \cong \frac{\tau_0}{1 - \expo^{ - \tau_0}}
      \,\int\limits_{ - \infty}^\infty T_{\rm MB}(v) \diff v \, .
 \end{equation}
This formula is accurate to 15\,\% for $\tau_0 < 2$, and it always
overestimates $N$ when $\tau_0 > 1$. The formulas (\ref{tcotw}),
(\ref{odct}), (\ref{ncoth13}) and (\ref{ncoth21}) permit an evaluation of
the column density $ N_{\rm CO}^{13}$ {\it only} under the
assumption of LTE.

The most extensive and complete survey in the $J=1-0$ line of \THCO was carried out by the 
Boston University FCRAO group \cite{JACKSON}. This covers the inner part of the northern galaxy with full sampling; there are nearly $2\, \times \, 10^6$ spectra.

For other linear molecules, 
the expressions for the dipole moments and the partition functions
are similar to that for
CO and the treatment is similar. There is one very
important difference however. The  simplicity in the treatment of the CO
molecule arises because of the assumption of LTE or near-LTE condtions.  This may not be the case
for molecules such as HCN or CS  since these species have
dipole moments of order 2 to 3\,Debye. Thus populations of high $J$ levels (which
have faster spontaneous decay rates) may have populations lower than
predicted by LTE calculations. Such populations are said to be
{\it subthermal\index{Subthermal excitation temperature}}, because the excitation
temperature characterizing the
populations would be $T_{\rm ex} < T_{\rm K}$.

%********************************************************************************************************************************************

%**********************************************************************************************************************************88moved here on April 7, 2008
 
\subsection{Symmetric  Top Molecules}\index{Symmetric
molecules}
\subsubsection{ Energy Levels}
 
Symmetric and asymmetric top molecules are vastly more complex than linear molecules.   The rotation of a rigid
molecule with an arbitrary shape can be considered to be the
superposition of three free rotations about
the three principal\index{Principal axes} axes of the inertial
ellipsoid. Depending on the symmetry of the
molecule these principal axes can all be different:
in that case the molecule is an asymmetric\index{Asymmetric molecules}
top. If  two  principal axes
are  equal, the molecule is a symmetric\index{Symmetric top} top. If
all three principal axes are
equal, it is a  spherical\index{Spherical top} top. In order
to compute the angular parts of the wavefunction, the proper
Hamiltonian operator  must be solved in the
Schr{\"o}dinger equation and
the stationary state eigenvalues determined.

In general, for any rigid rotor asymmetric top molecule in a
     stable state, the total momentum $\vec{J}$
will remain constant with respect to both
its absolute value and its direction. As is known
from atomic physics, this means that both
$(\vec{J})^2$ and the projection of $\vec{J}$ into an
arbitrary but fixed direction, for example
$J_z$, remain constant. If the molecule is in
addition symmetric, the projection of $\vec{J}$ on
the axis of symmetry will be constant also.
 
   For the {\em symmetric\index{Symmetric top} top
molecule}, $\vec{J}$ is
inclined with respect to the axis of
symmetry $z$. Then the figure axis $z$ will
precess  around the direction $\vec{J}$ forming a
constant angle with it, and the molecule
will simultaneously rotate around the
$z$ axis  with the constant angular
momentum $J_z$. From the definition of a symmetric top,
$\Theta_x = \Theta_y$. Taking  $\Theta_x = \Theta_y = \Theta_\perp$ and
$\Theta_z = \Theta_\|$, we obtain a  Hamiltonian operator:
 
 \begin{eqnarray}
   \label{hamil}
   H =
	 \frac{J_x^2 + J_y^2} { 2 \,\Theta_\perp} + 
	\frac{J_z^2}{ 2 \,\Theta_\|}	
     = 
 \frac{\vec{J}^2}{ 2 \,\Theta_\perp} + J_z^2 \cdot 
       \left ( \frac{1}{2\,\Theta_\|} - \,\frac{1}{ 2\,\Theta_\perp} 
       \right )
  	\,. 
%	\kern-100pt \nonumber \\
 \end{eqnarray}

Its eigenvalues are:
 \begin{equation}
   \label{eigenv}
   W(J, K) = J (J + 1) \,\frac{\hbar^2}{2 \,\Theta_\perp} +
   K^2 \, \hbar^2 \left(\frac{1}{2 \,\Theta_\|} - \frac{1}{2 \,\Theta_\perp}
                  \right)
 \end{equation}
where $K^2$ is the eigenvalue from the operator $J_z^2 $ and
$J^2 = J_x^2 + J_y^2 + J_z^2 $ is the eigenvalue from the operator
$J_x^2 + J_y^2 + J_z^2 $.
 
The analysis of linear molecules is a subset of that for symmetric
molecules.  For linear molecules, $\Theta_\| \to 0$ so that $ 1 / ( 2
\,\Theta_\|) \to \infty$. Then finite energies in (\ref{eigenv}) are
possible only if $K = 0$.  For these cases the energies are given
by (\ref{rotene}). For symmetric top molecules each
 eigenvalue has a multiplicity
of $ 2 J + 1$.
 \begin{equation}
   \label{jkqn}
   J = 0, 1, 2, \dots   \quad  K = 0, \pm 1, \pm 2, \dots \pm J \,.
 \end{equation}
%----tlw change 2.9.94
From (\ref{eigenv}), the energy is independent of the sign of $K$, so
levels with the same $J$ and absolute value of $K$ coincide. Then
levels with $K>0$ are doubly degenerate.

It is usual to express $\displaystyle\frac{\hbar}{4\pi \,\Theta_\perp}$ as $B$,
and $\displaystyle\frac{\hbar}{4\pi \,\Theta_\|}$ as $C$.
The units of these rotational constants, $B$ and $C$ are usually
either MHz or GHz. Then (\ref{eigenv}) becomes
 \begin{equation}
   \label{eigenv1}
   W(J, K) / h = B \,J (J + 1) + K^2 \,(C - B) \,.
 \end{equation}

 %****************************************************************************************************************************************************************

%\section{Further Details of Molecular Structure}
\subsubsection{Spin Statistics} \label{spinstat}
 
In the case of molecules containing identical nuclei, the exchange of
such nuclei, for example by the rotation about an axis, has a
spectacular effect on the selection rules.
Usually there are no interactions between electron spin\index{Spin statistics}  and rotational
motion. Then
the total wavefunction is the product of the spin and rotational
wavefunctions.
Under an interchange of  fermions,
the total wave function must
be antisymmetric (these identical nuclei could be protons or have an
uneven number of nucleons). The symmetry of the spin
wavefunction of the molecule will depend on the
relative orientation of the spins. If the spin wavefunction is symmetric,
this is the {\em ortho\index{Ortho-modification}}-modification of
the molecule; if antisymmetric it is the 
{\em para\index{Para-modification}}-modification.
In thermal equilibrium in the ISM, collisions with
the exchange of identical particles will
change one modification into the other only very slowly, on time
scales of $ > 10^6$ years. This could occur much more
     quickly on grain surfaces, or with charged particles.
If the exchange is slow, the ortho and para modifications of
a particular species behave like 
different molecules; a
comparison of ortho and para populations might give an estimate
of temperatures in the distant past, perhaps at the time
of molecular\index{Molecular formation} formation.
 
For the \MOLH\ molecule, 
the symmetry of the rotational wavefunction depends on the total angular
momentum $J$ as $(-1)^J$. In the $J=0$ state
the rotational wavefunction is symmetric. However, the total
wavefunction must be antisymmetric since protons are
fermions. Thus, the $J = 0, 2, 4$, etc., rotational levels are
para-\MOLH, while the $J =1, 3, 5$, etc., are
ortho-\MOLH. Spectral lines can connect only one modification.
In the case of \MOLH, dipole rotational transitions are not allowed, but quadrupole rotational transitions ($\Delta J=\pm 2$) are. Thus, the
$28\,\mug$m line of \MOLH\ connects the $J=2$ and $J=0$ levels of
para\index{Para-\MOLH}-\MOLH. Transitions between the ground and vibrational 
states are are also possible.

Finally, as a more complex example of the relation of identical nuclei, we
consider the case of three identical nuclei. This is the case for \AMM,
CH$_3$CN and  CH$_3$C$_2$H.
Exchanging two of the nuclei is equivalent to a rotation by $120^{\rm o} $.
An exchange as was used for the case of two nuclei
would not, in general,
lead to a suitable symmetry. Instead combinations of spin states must be
used.
These lead to the result that the ortho to para ratio is
two to one if the identical nuclei are protons. That is, \AMM, CH$_3$CN
or
CH$_3$C$_2$H the ortho form has
$S(J, K)=2$, while the para form has $S(J, K)=1$.
In summary, the division of molecules with identical nuclei into ortho and
para species determines  selection rules for radiative transitions and also
rules the for collisions (see e.~g.~\cite{TOWNES}).

%*********************************************************************************************************************************col den NH3 here TLW April 7, 2008
 %changed on 25 April 08
\subsubsection{Hyperfine Structure}

For symmetric top molecules, the simplest  hyperfine\index{Hyperfine spectra} spectra
is found for the inversion doublet
transitions of  \AMM.  Since both the upper and lower levels have the
same quantum numbers $(J, K)$,  there will be 5 groups of hyperfine
components separated by a few MHz. Because of interactions between the
spins of H nuclei there will be an additional splitting, within each group, of 
order a few kHz. In Table 15.1 we give the relative intensities of the NH$_3$
satellites for the case of low optical depth and LTE. 
\begin{table}[b]
%\vglue-2.25pt
%\vglue-1pt
%\vglue-2.5pt
%\vglue2.5pt
 % \tabcolsep5mm
  \begin{petit}
  \caption[Intensities of satellite groups relative to the Main Component]
      {\label{hyperfinenh3} Intensities of satellite groups relative to the 
Main Component (see \cite{RW4}).}
  \begin{flushleft}
  \begin{tabular}{@{}c|rrrrrrr@{}} \hline
  \noalign{\vskip 3pt}
  (J,K)&(1,1)&(2,2)& (3,3) & (4,4) & (5,5) &(6,6)&(2,1)  \\
   \noalign{\vskip 3pt}
   \hline
   \noalign{\vskip 3pt}
I$_{\rm inner}$&0.295&0.0651&0.0300&0.0174&0.0117&0.0081&0.0651\\
I$_{\rm outer}$&0.238&0.0628&0.0296&0.0173&0.0114&0.0081&0.0628\\
 \noalign{\vskip 3pt}
   \hline
 \end{tabular}
 \end{flushleft}
 \end{petit}
\end{table}%%
For a molecule such as OH, one of the electrons
is unpaired. The interaction of the nuclear
magnetic moment with the magnetic moment of an unpaired electron is
described as magnetic hyperfine structure. This splits a specific line
into a number of components. In the case of the OH molecule, this
interaction gives rise to a  hyperfine splitting of the energy levels,  in
addition to the much larger $\Lambda$ doubling. Together with the
$\Lambda$ doublet splitting, this gives rise to a quartet of energy
levels in the OH ground state. Transitions between these energy
levels produces the four ground state lines of OH at 18\,cm
wavelength (see Fig.\ts \ref{loweoh}).

%*********************************************************************************************************************moved here on April 7, 2008

\begin{figure}
%\mpicplace{9.1cm}{11.9cm}
\includegraphics[width=9.1cm,height=11.9cm]{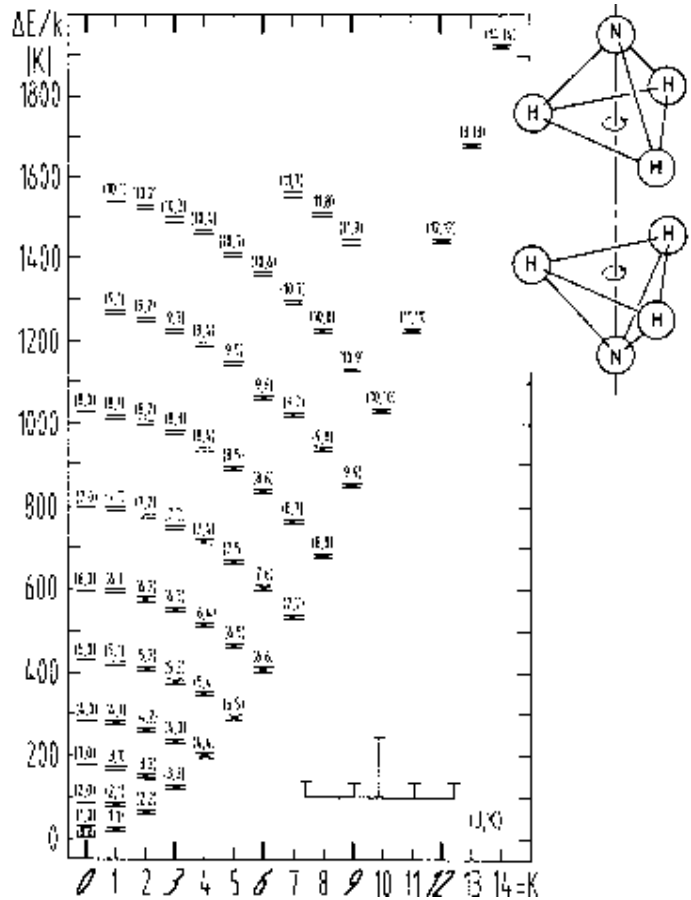}
  \caption[Energy-level diagrams of NH$_3$]
     { \label{enlnh}
        The energy-level\index{NH$_3$ energy levels}
         diagram of the vibrational ground state of
        NH$_3$, a prolate symmetric top molecule.
        Ortho-NH$_3$ has $K=0, 3, 6, 9, \cdots$, while para-NH$_3$ has all
        other $K$ values (see text).
        Rotational transitions with $\Delta J=1, \Delta K=0$, give
        rise to lines in the far IR. This molecule also has
        transitions with $\Delta J=0, \Delta K=0$  between
        inversion doublet levels. The interaction of the
        nuclear spin of $^{14}$N with the electrons causes quadrupole hyperfine
        structure. In the $\Delta J=0, \Delta K=0$  transitions, the line is 
        split into 5 groups of components. A sketch of the structure of the
        groups of hyperfine components of the $(J, K)=(1,1)$ inversion
        doublet line is indicated in the lower right;
        the separation is of order of MHz. In the upper right
        is a sketch of the molecule before and after an inversion
        transition, which gives rise to a  1.3\,cm photons \cite{RW4}. }%
\end{figure}

\AMM\ is an example of an oblate symmetric top molecule commonly found in the ISM. A
diagram of the lower energy levels of \AMM\ are shown in
Fig.\,\ref{enlnh}. A prolate\index{Prolate top} top molecule has a cigar-like 
shape. Then  $A$ replaces $C$, and $A>B$.  
  The energy-level diagrams for prolate symmetric top molecules found in the 
ISM, such as  CH$_3$CCH and CH$_3$CN, follow this rule.        
However, since these molecules are much heavier than
\AMM, the rotational transitions give rise to lines in the millimeter
wavelength range.

%**********************************************************************************************************8 
%14.6
%\vspace{10cm}
%\ignorespaces
%*********************************************************************************************************** 

Differences in the orientation
of the nuclei can be of importance. If a
reflection of all particles about the
center of mass leads to a
configuration which cannot be obtained by a
rotation of the molecule, so these reflections represent two different states.
     For \AMM, we show this situation in the upper part
of Fig.\,\ref{enlnh}. Then there are two separate,
degenerate states which exist
for each value of $(J, K)$ for $J \ge 1$. (The
$K=0$ ladder is an exception because of the Pauli principle.) These
states are  doubly degenerate as long as the potential barrier
separating the two configurations is infinitely high. However, in
molecules such as NH$_3$ the two configurations are
separated only by a small potential barrier. This gives rise to
a  measurable splitting of the degenerate energy levels, which
is referred to as  inversion\index{Inversion doubling} doubling. For \AMM,
transitions between these inversion doublet levels  are caused
by the
quantum mechanical tunneling of the nitrogen nucleus through the
plane of the
three protons. The wavefunctions of the two inversion doublet 
states have opposite parities, so that dipole transitions
are possible. Thus dipole 
transitions {\em can} occur between states with the same $(J, K)$ quantum numbers.
The splitting of
the $(J, K)$ levels for \AMM\ shown in Fig.\,\ref{enlnh} is exaggerated;
the inversion transitions give rise to spectral lines in the wavelength
range near 1\,cm. For CH$_3$CCH or CH$_3$CN, the splitting
caused by inversion doubling is very small since the barrier is much
higher than for \AMM.

     The direction of the dipole moment of symmetric
top molecules is parallel to the $K$ axis.  Spectral line radiation can
be emitted only by a changing dipole moment. Since radiation will be emitted
perpendicular to the direction of the dipole moment, there can be no
radiation along the symmetry axis.  Thus the $K$ quantum number
{\it cannot} change in dipole radiation, so allowed dipole transitions cannot
connect different $K$ ladders. The different $K$ ladders are connected
by octopole radiative transitions which require $\Delta K=\pm 3$. These are 
very slow, however, and collisions are far more likely to cause an exchange
of population between different $K$ ladders. This is used to estimate  $T_{\rm K}$ from the
ratio of populations of different $(J, K)$ states in symmetric top
molecules.

%*****************************************************************************************************

\subsubsection{ Line Intensities and Column Densities}
 
     The extension of
this analysis to symmetric\index{Symmetric top} top molecules is only slightly
more complex.
The dipole moment for an allowed transition between energy level $J+1, K$
and $J, K$ for a symmetric top such as CH$_3$CN or CH$_3$C$_2$H  is
\begin{equation} \label{symmdip}
  |\mu_{JK}|^2 = \mu^2 \,\frac{(J + 1)^2 - K^2}{(J + 1 )(2 J + 3)}
\qquad {\rm for} \quad (J + 1, K) \to (J, K)      \, .
 \end{equation}
For these transitions, $J \ge K$ always.
 
For \AMM, the most commonly observed spectral lines are the inversion
transitions at 1.3\,cm between levels $(J,K)$ and $(J,K)$. The dipole moment is 
\begin{equation}
\label{symmdipinv}
  |\mu_{JK}|^2 = \mu^2 \,\frac{K^2}{J(J + 1)} \qquad {\rm for} \quad
                         \Delta J=0, \,\Delta K=0 \, .
 \end{equation}
When these relations are
inserted in (\ref{acoeff}),
the population of a specific level can be calculated following
(\ref{levelpop}).
If we follow the analysis used for CO, we can use the LTE
assumption to
obtain the entire population
\begin{equation}
\label{totalsumsym}
 N({\rm total}) =  N(J, K) \, \frac{Z}{ (2 J + 1)\, S(J, K)}
  \,\exp \left[\frac{ W(J, K) }{ k \,T} \right] \, ,
\end{equation}
where $W$ is the energy of the level above the ground state, and the
nuclear spin statistics are accounted for through the factor $S(J,K)$
for the energy level corresponding to the transition measured.  For symmetric
top molecules, we have, using the expression for the energy of the level
in question (\ref{eigenv1}),
 \begin{equation}
   \label{gjlrsymm}
N({\rm total})=\frac{Z\, N(J, K)}{(2 J + 1)\, S(J,K)} \exp \left[
\frac{ B J ( J + 1) + (C - B)K^2}
       { k \,T} \right] \, .
 \end{equation}
For prolate tops, $A$ replaces $C$ in (\ref{gjlrsymm}), and in 
(\ref{usumjsymm}) to (\ref{usumjddc}).
If we sum
over all energy levels, we obtain $N$(total), the
partition\index{Partition function} function,
$Z$ in the following:
 \begin{eqnarray}
  \label{usumjsymm}
   Z = \sum_{ J = 0 }^\infty \sum_{ K = 0 }^{K = J } \,( 2 J + 1 )
 \,S(J, K)  \,\exp \left[ - \frac{ B J ( J + 1) + (C - B)K^2}{ k \,T}
\right]   \,.
\kern-100pt\nonumber \\
 \end{eqnarray}
 If the temperature is large compared to
the spacing between energy levels, one can replace the sums by integrals,
 so that:
 \begin{equation}
  \label{intusumj}
Z \approx \sqrt{ \frac{\pi (k \, T)^3}{ h^3 \,B^2 \, C}}   \, .
\end{equation}
If we assume that $h\nu \ll kT$, use CGS units for the physical
     constants, and GHz for
the rotational constants $A$, $B$ and $C$,  the partition function, $Z$,
becomes
\begin{equation}
\label{usumsymmetric}
Z  \approx 168.7 \, \sqrt{\frac{T^3} {B^2 \, C}} \, .
\end{equation}
Substituting into (\ref{usumjsymm}), we have:
\begin{equation}
\label{fintotalsumsym}
N({\rm total}) =  N(J, K) \, \frac{\displaystyle 168.7 \,
         \sqrt{\frac{T^3} {B^2 \, C}}}{ (2 J + 1)\, S(J,K)}
 \, \exp \left[\frac{ W(J,K) }{ k \,T} \right] \, .
\end{equation}
Here, $N(J, K)$ can be calculated from (\ref{levelpop}) or (\ref{coldenlv}),
using the appropriate expressions for the dipole moment, (\ref{symmdipinv}),
in the Einstein $A$ coefficient relation, (\ref{acoeff}) and $W$ is the
energy of the level above the ground state. In the ISM,  ammonia inversion lines up to  (J,K)=(18,18) have been detected \cite{WILNH3}.
 
We now consider a situation in which the \AMM\ population is {\it not}
thermalized. This is typically the case for dark dust clouds.
We must use some concepts presented in the next few sections for this
analysis. If $n$(\MOLH)$\sim10^4$\,\percc, and the infrared field intensity is
small, a symmetric top molecule such as \AMM\ can have a number of
excitation temperatures. The excitation
temperatures of the populations in doublet levels are
usually between 2.7\,K and $T_{\rm K}$.
The rotational temperature,
T$_{\rm rot}$, which describes populations for metastable
levels ($J=K$) in different $K$ ladders,
 is usually close to $T_{\rm K}$. This is because radiative transitions between states with a different K value are forbidden to first order.   
The excitation temperature which describes the populations with
different $J$ values within a
given $K$ ladder will be close to 2.7\,K, since radiative decay with $\Delta$K=0, $\Delta$J=1 is allowed. Then the
non--metastable energy levels, $(J>K)$, are
not populated. In this case, $Z$ is simply given by the sum over the
populations of metastable  levels:
 \begin{eqnarray}
  \label{usumjddc}
        \lefteqn{  Z(J=K)  } \nonumber \\
        &=& \sum_{ J = 0 }^\infty \,( 2 J + 1 )\,S(J,K=J)
\,\exp \left[ - \frac{ B J ( J + 1) + (C - B)J^2}{ k \,T} \right] \,.
 \end{eqnarray}
For the \AMM\ molecule in  dark dust clouds,
where $T_{\rm K} =10$\,K and $n$(\MOLH)$ =10^4$\,\percc,
we can safely restrict the sum to the three lowest metastable levels:
 \begin{equation}
	Z(J=K) \approx N(0,0) + N(1,1) + N(2,2) + N(3,3) \,.
 \end{equation}
Substituting the values for \AMM\
metastable\index{Metastable energy levels} levels:
\begin{eqnarray}
  \label{eqlowden}
        \lefteqn{Z(J=K) } \nonumber \\
        &\approx& N(1,1)\left[\frac{1}{3}\exp\left(\frac{23.1}{k \, T}
        \right) 	+ 1 + \frac{5}{3}\exp\left(- \frac{41.2}{k \, T}
        \right) +  \frac{14}{3}\exp\left(- \frac{99.4}{k \, T}
        \right) \right]   \, .
\kern-100pt\nonumber \\
 \end{eqnarray}
For \AMM\ we have given two  extreme situations: in the first case,
described by  (\ref{eqlowden}),  is a low-density cloud for which only
the few lowest metastable levels are populated. The second case is the LTE
     relation,
given in  (\ref{fintotalsumsym}). This represents a cloud in which the
populations of the molecule in question are thermalized.  More
complex are those situations for which the populations of some of the
levels are thermalized, and others not. Such a situation requires the use of a statistical equilibrium  or an LVG model.

\subsection{Asymmetric Top Molecules}\index{Asymmetric molecules}
\subsubsection{ Energy Levels}

%\vglue-4pt
%\clearpage
%********************************************************************************** 
%
%14.8
\begin{figure}%[t]
%\mpicplace{9.7cm}{13.5cm}
\includegraphics[width=9.7cm,height=13.5cm]{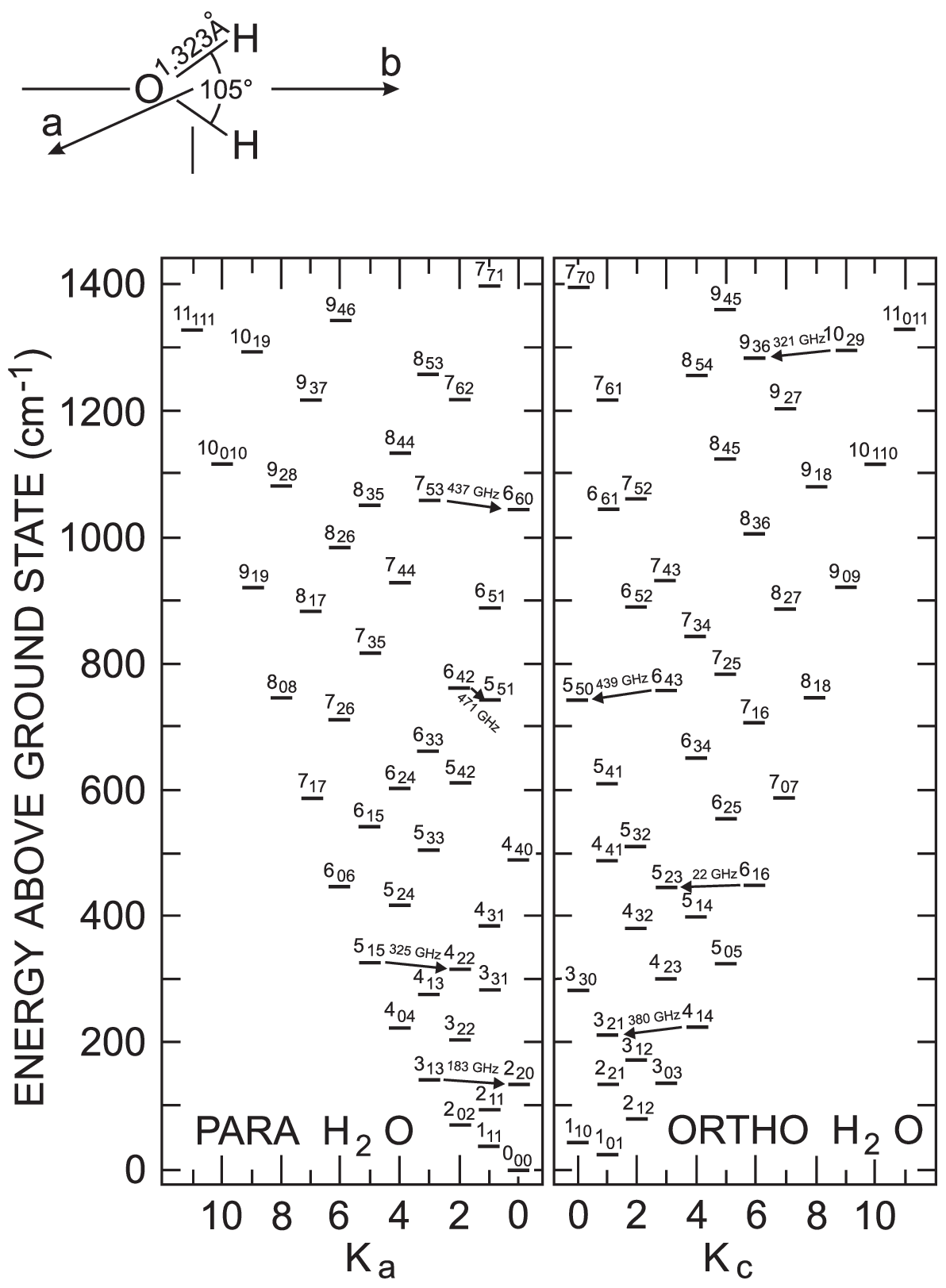}
  \caption[Energy-level diagram for H$_2$O]
       {\label{enlwat}
Energy level diagrams for ortho- and para-H$_2$O.\index{Molecules!H$_2$O}
This is an asymmetric top molecule, with the dipole moment along the $B$ axis,
that is, the axis  with an intermediate moment of inertia. Because of the two
identical nuclei, the energy level diagram is split into  ortho and
para, that show almost no interaction under interstellar conditions. The
transitions marked by arrows are well-known masers \cite{DELUC}, \cite{RW4}.}
\end{figure}%
%\clearpage
%******************************************************************************

For an {\em asymmetric\index{Asymmetric molecules} top molecule} there
are no internal molecular axes with a time-invariable
component of angular momentum.
So only the total angular\index{Angular momentum} momentum is
conserved and we have only $J$ as a good
quantum number. The moments of inertia about each axis
are different; the rotational constants are referred to as $A, B$ and $C$,
with $A>B>C$. The
prolate symmetric top ($B=C$) or oblate symmetric top ($B=A$)
molecules can be considered as the limiting cases.
But neither the
eigenstates nor the eigenvalues are easily
expressed in explicit form.
Each of the levels must be characterized by  three
quantum numbers. One choice is
$J_{{K_a}{K_c}}$, where $J$ is the total angular momentum,
$K_a$ is the component of $J$ along the $A$ axis and $K_c$ is the
component
along the $C$ axis. If the molecule is a prolate symmetric top, $J$
and
$K_a$ are good quantum numbers; if the molecule were an oblate
symmetric top, $J$ and $K_c$ would be good quantum numbers. Intermediate
     states are characterized by a superposition of the prolate and
     oblate descriptions. In Fig.\,\ref{enlfor}, we show the energy
level diagram
for the lower levels of \FORM, which  is almost a
prolate symmetric top molecule with the dipole moment along the $A$
axis. Since radiation must be emitted perpendicular to the direction of
the dipole moment,
for \FORM\ there can be no radiation emitted along the $A$ axis, so
the quantum number  $K_a$ will not change in radiative transitions.
%\clearpage
 
\subsubsection{Spin Statistics and Selection Rules}\label{select-form}
The case of a planar molecule with two equivalent nuclei, such as \FORM,
shows is a striking illustration of these effects (see Fig.\,\ref{enlfor}).
The dipole moment lies along the $A$ axis. A rotation by $180^{\rm o}$
about this axis
will change nothing in the molecule, but will exchange the two protons.
Since the protons are fermions, this exchange must lead to an antisymmetric 
wavefunction. Then the symmetry of the spin wave function and the wave function
describing the rotation about the $A$ axis must be antisymmetric. The
rotational
symmetry is $(-1)^{K_{a}}$. If the proton spins
are parallel, that is ortho-\FORM, then the wave function for  $K_{a}$ must
be anti-symmetric, or  $K_{a}$ must take on an odd value. If the proton
spins are anti-parallel, for para-\FORM,  $K_{a}$
must have an even value (Fig.\,\ref{enlfor}).
For ortho\index{Ortho-\FORM}-\FORM, the parallel spin case, there are
three possible
spin orientations. For para-\FORM, there is only one possible orientation,
so the ratio of ortho-to-para states is three. Such an effect
is taken into account in partition functions (\ref{usumj})
by spin degeneracy factors, which are denoted by the symbol
$S(J, K)$. For ortho-\FORM,
$S(J, K) = 3$, for para-\FORM, $S(J, K) = 1$.
This concept will be applied in Sect.\,\ref{toplinerad}.

Allowed transitions can occur only between energy levels of either
the ortho or the para species. For example, the 6\,cm H$_2$CO line is
emitted from the ortho-modification
only (see Fig.\,\ref{enlfor}). Another example is the interstellar
H$_2$O maser line at $\lambda = 1.35$\,cm which arises from ortho-H$_2$O
(see Fig.\,\ref{enlwat}). The \WAT\ molecule is a more complex case since
the dipole moment is along the $B$ axis. Then in a radiative transition,
both $K_{a}$ and $K_{c}$ must change between the initial and final state.

\subsubsection[ Line Intensities and Column Densities]
{ Line Intensities and Column Densitiess}
\label{toplinerad}

For asymmetric \index{Asymmetric molecules} molecules
the moments of inertia for the three axes are all different; there is no
symmetry, so three quantum numbers are needed to define an energy level.
 The relation between
energy above the ground state and quantum numbers is given in
\mbox{Appendix\,IV} of \cite{TOWNES} or in databases 
for specific molecules (see figure captions for references). The relation
for the dipole moment of a
specific transition is more complex; generalizing from
(\ref{acoeff}), we have, for a spontaneous transition from a higher
state, denoted by $u$ to a lower state, denoted by $l$:
\begin{equation}
\label{acoeffasymm}
 A = 1.165 \times 10^{-11}\, \nu^3_{\rm x} \, \mu_x^2 \,
      \frac{{\cal S}(u;l)}{2J' + 1} \,.
\end{equation}
This involves  a dipole moment in a direction $x$.
     As before, the units of $\nu$ are GHz, and the units of
$\mu$ are Debyes (i.\,e., 1\,Debye $ =10^{-18}$ times the e.s.u. value).
     The value of the
quantum number $J'$ refers to the lower state.
     The  expression for  ${\cal S}(u;l)$, the line strength
is an indication of the complexity of the physics of asymmetric
top molecules. The expression ${\cal S}(u;l)$ is the angular
part of the dipole moment  between the
initial and final state. The dipole moment can have a
     direction which is {\it not}
along a single axis. In this case there are different values of the
dipole moment along different molecular axes. In
contrast, for symmetric top or linear molecules, there is {\it a} dipole
moment for rotational transitions.  
Methods to evaluate transition probabilities  for
asymmetric molecules
 are discussed at length in \cite{TOWNES}. There is a table of ${\cal S}(u;l)$ 
in their Appendix\,V. We give  references for ${\cal S}(u;l)$ 
in our figure captions. From the expression for the
Einstein $A$ coefficient, the column density in a
given energy level can be related to the line intensity by
(\ref{levelpop}). Following the procedures used for symmetric top
molecules, we can use a relation similar to (\ref{usumjsymm}) to sum
over all levels, using the appropriate energy, $W$, of the level
$J_{K_{a}K_{c}}$ above the
ground state and the factor $S(J_{K_{a}K_{c}})$ for spin statistics:
\begin{equation}
 \label{totalsumasymm}
  N({\rm total}) =  N(J_{K_{a}K_{c}}) \, \frac{Z}{ (2 J + 1)\,
   S(J_{K_{a}K_{c}})} \,\exp \left(\frac{ W }{ k \,T} \right) \, .
\end{equation}
If the  populations are in LTE, one can follow a process
similar to that used to obtain (\ref{usumsymmetric}). Then we obtain
     the appropriate expression for the
partition \index{Partition function} function:
     \begin{equation}
     \label{usumunsymmetric}
    Z = 168.7 \, \sqrt{\frac{T^3}{A \, B \, C}}  \,.
     \end{equation}
When combined with
the Boltzmann expression for a molecule in a specific energy
level, this gives a simple expression for the fraction of the population
in a specific rotational state if LTE conditions apply:
\begin{equation}
\label{totalsumasymmfin}
    N({\rm total}) \approx  N(J_{K_{a}K_{c}}) \, \frac{ 168.7 \,
\displaystyle
     \sqrt{\frac{T^3}{A \, B \, C}}} { (2 J + 1)\, S(J_{K_{a}K_{c}})} \,\exp
       \left(\frac{ W }{ k \,T} \right) \, .
\end{equation}
In this expression, $S(J_{K_{a}K_{c}})$ accounts for spin statistics for
energy level $J_{K_{a}K_{c}}$,
and $A$, $B$ and $C$ are the molecular rotational constants in GHz. $W$,
the energy of the level above the ground state, and
$T$, the temperature, are given in Kelvin. Given the total molecular
column density and the value of $T$, the feasibility of
detecting a specific line can be obtained when the appropriate
$A$ coefficient value is inserted into (\ref{levelpop})
or  (\ref{coldenlv}).
 
As pointed out in connection with \AMM, $T$ need not be $T_{\rm K}$.
In reality, a number of different values of $T$ may be
needed to describe the populations. We will
 investigate the influence
of excitation conditions on molecular populations and
observed line intensities next.

%***************************************************************************************
%14.7
%\vspace{10cm}
\begin{figure}%[t]
%\sidecaption
%%\mpicplace{10.7cm}{8.9cm}
%\mpicplace{10.7cm}{8.5cm}
  \includegraphics[width=10.7cm,height=8.5cm]{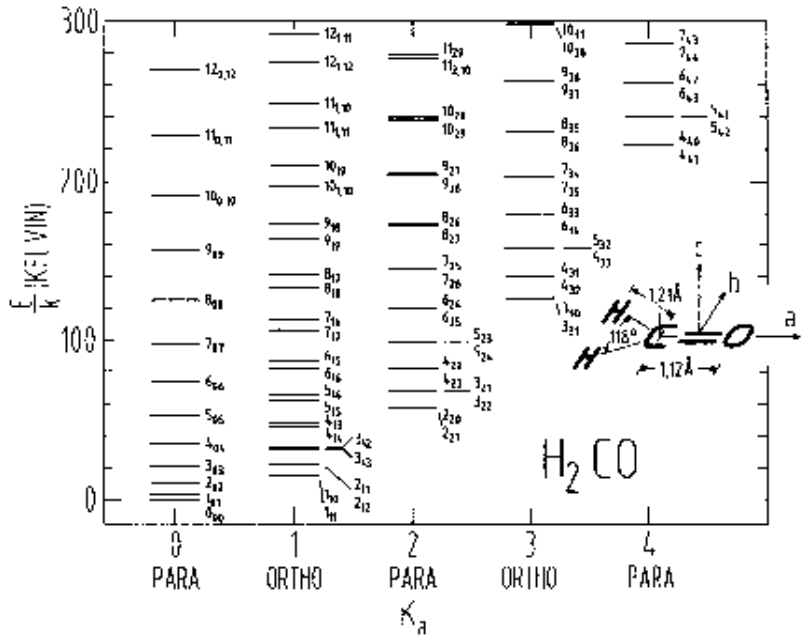}
  \caption[Energy-level diagram of formaldehyde]
   {\label{enlfor}
    Energy-level diagram of formaldehyde, H$_2$CO.\index{Molecules!H$_2$CO}
     This is a planar
asymmetric top, but the asymmetry is very small.
The energy-level structure is typical
of an almost prolate symmetric top molecule. In the lower
right is a sketch of the structure of the molecule \cite{MANGUM}}
\end{figure}
%*************************************************************************************  

Two important interstellar molecules are H$_2$CO and H$_2$O. Here we summarize
the dipole selection rules. Rotating the molecule
about the axis along the direction of the dipole moment, we effectively
exchange two identical particles. If these are fermions, under this exchange
the total wavefunction must be antisymmetric. For H$_2$CO, in
Sect.~\ref{spinstat} we reviewed the spin statistics. Since the dipole moment is
along the A axis, a dipole transition must involve a change in the quantum
numbers along the B or C axes. From Fig. \ref{enlfor}, the K$_{a}$=0 ladder is
para-H$_2$CO, so to have
a total wavefunction which is antisymmetric, one must have a space wavefunction
which is symmetric. For a dipole transition, the parities of the initial and
final states must have different parities. This is possible if the C quantum
number changes. For H$_2$O, the dipole moment is along the B axis, from
Fig.~\ref{enlwat}. In a dipole transition, the quantum number for the B
direction will not change. For ortho-H$_2$O, the spin wavefunction is
symmetric, so the symmetry of the space wavefunction must be antisymmetric. In
general, this symmetry is determined by the product of K$_a$ and K$_c$. For
ortho-H$_2$O, this must be K$_a$K$_c$=(odd)(even), i.e.~$oe$, or $eo$. For
allowed transitions, one can have $oe-eo$ or $eo-oe$. For para-H$_2$O, the rule
is $oo-ee$ or $ee-oo$. Clearly H$_2$S follows the selection rules for \WAT.
These rules will be different for SO$_2$ since the exchanged particles are
bosons. More exotic are D$_2$CO, ND$_3$ and   D$_2$O.  
%****************************************************************************
%**********************************************************tlw April 7, 2008
\begin{figure}%[t]
%\sidecaption
%%\mpicplace{10.7cm}{8.9cm}
%\mpicplace{10.7cm}{8.5cm}
  \includegraphics[width=3.8cm,height=3.8cm]{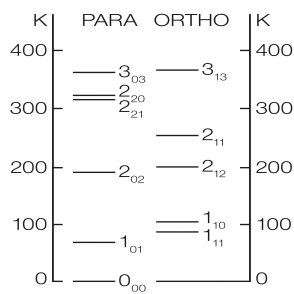}
  \caption[Energy-level diagram of H$_2$D$^+$]{\label{enlforh2dplus} The few lowest energy levels of the  H$_2$D$^+$ molecule.\index{Molecules!H$_2$D$^+$}
         This is a planar, triangular-shaped
asymmetric top \cite{GERLICH}. }
\end{figure} 
 %*****************************************************************************************************************

 The species H$_3^+$ has the shape of a planar triangle. It is a key to ion-molecule chemistry, but has no rotational transitions because of its symmetry. The deuterium 
isotopomer, H$_2$D$^+$ is  an  asymmetric top molecule with a permanent dipole moment.  The  spectral line from the 1$_{10}$-1$_{11}$ 
levels ortho species was found at 372.421 GHz. A far infrared absorption line from the  2$_{12}$-1$_{11}$ levels was also detected. 
We show an energy level diagram 
in Fig.\,\ref{enlforh2dplus}. The doubly deuterated species,  D$_2$H$^+$,has been detected in the  1$_{10}$-1$_{01}$ line at 691.660 GHz from the 
para species (see \cite{VASTEL}). 
 
\subsection{Electronic Angular Momentum}
   In many respects the description of electronic angular momentum
is similar to that of atomic fine structure
as described by Russell\index{Russell-Saunders coupling}-Saunders ($L\,S$)
%\clearpage
coupling. Each electronic state is
designated by the symbol $^{2 S + 1}\Lambda_\Omega$,
where $2 S + 1$ is the multiplicity of the state with $S$
the electron spin and $\Lambda$ is the projection
of the electronic orbital angular momentum
on the molecular axis in units of $\hbar$. The
molecular state is described as $\Sigma, \Pi, \Delta $ etc., according
to whether $\Lambda = 0, 1, 2, \dots \; .$
 
$\Sigma$ is the projection of the electron spin
angular momentum on the molecular axis in
units of $\hbar$ (not to be confused with the
symbol $\Sigma$, for $\Lambda = 0$). Finally, $\Omega$ is the
total electronic angular momentum. For the Hund\index{Hund coupling case}
coupling case A, $\Omega = | \Lambda + \Sigma | $ [see e.\,g.~\cite{RW4}].

\begin{figure}%[t]
%\sidecaption
%\mpicplace{6.8cm}{10.7cm}
 \includegraphics[width=6.8cm,height=10.7cm]{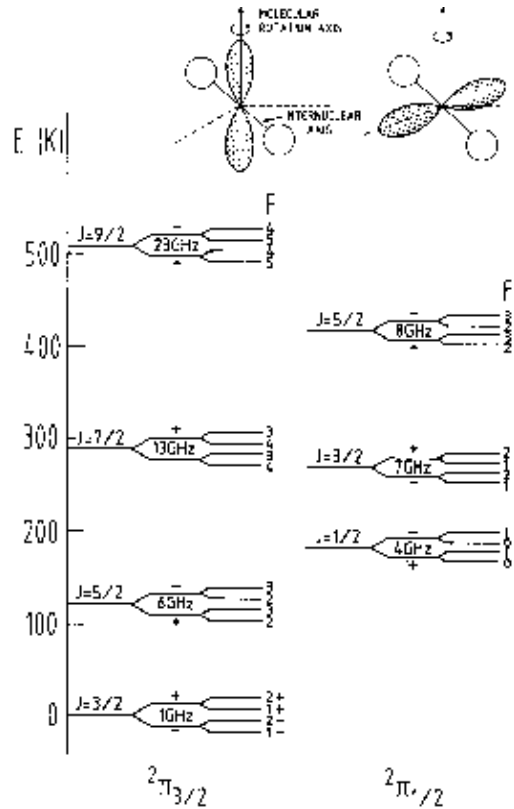}
  \caption[Energy-level diagram of OH]
        { \label{loweoh}
       The lower energy levels of OH\index{OH energy levels}
        showing $\Lambda$-doubling.
$F$ is the total angular momentum, including electron spin,
while $J$ is the rotational angular momentum due to the nuclear motion.
The quantum number $F$ includes hyperfine splitting of the
energy levels. The parities of the states are also shown under the
symbol $F$. The $\Lambda$ doubling causes a splitting of the $J$
states. In the sketch of the OH molecule, the shaded regions represent
the electron orbits in the $\Lambda$ state. The two unshaded spheres
represent the O and H nuclei. The configuration shown on the left has
the higher energy \cite{RW4}.}
     \end{figure}

   Since the frequencies emitted or
absorbed by a molecule in the optical
range are due to  electronic state changes, many of
the complications found in optical 
spectra are not encountered when considering
transitions in the cm and mm range. However the electronic state
does affect the
vibrational and rotational levels even in
the radio range. For most molecules,
the ground state has zero electronic angular momentum, that is, a
singlet sigma, $^1\Sigma$ state. For a small number of molecules such as OH, CH,
C$_2$H, or C$_3$H, this is not the case; these have ground state
electronic angular momentum.  Because of this fact, the rotational
energy levels  experience an additional energy-level splitting, which
\index{Lambda doubling}
is $\Lambda$ doubling. This is a result of the  interaction of the
rotation and the angular momentum of the electronic state. This
splitting causes the degenerate energy levels to separate.  This
splitting can be quite important for $\Pi$ states; for $\Delta$ and
higher states it is usually negligible. The OH molecule is a prominent
example for this effect. Semi-classically, the $\Lambda$ doubling of OH
can be viewed as the difference in rotational energy of the (assumed
rigid) diatomic molecule when the electronic wave function is oriented
with orbitals in a lower or higher moment of inertia state. We show a
sketch of this in the upper part of Fig.\,\ref{loweoh}.
Since the energy is directly proportional to
the total angular momentum quantum number and inversely proportional to
the moment of inertia,  the molecule shown on the left has  higher
energy than the one shown on the right.
%*************************************************************************************** 
%14.9
%\vspace{10cm}
%***************************************************************************************** 
 
There are also a few molecules for which the orbital angular
momentum is zero, but the electron spins are parallel, so that the total
spin is unity. These molecules have triplet sigma\index{Triplet ground state}
$^3\Sigma$ ground states. The most important astrophysical example is  the SO
molecule; another species with a triplet $\Sigma$ ground state is O$_2$. These 
energy levels are characterized by the quantum number, $N$, and the orbital
angular momentum quantum number $J$. The most probable transitions are those
within a ladder, with $\Delta J = \Delta N = \pm 1$, but
there can be transitions across $N$ ladders. As with the OH molecule,
some states of SO are very sensitive to magnetic fields. One could then use the Zeeman effect to determine the magnetic field strength. This may be difficult 
since a 1 $\mu$Gauss field will cause a line splitting of only about 1 Hz in
linear polarization. Even so, measurements of the polarization of the 
$J_N=1_0-0_1$ line of SO (\cite{RW4}) may allow
additional determinations of interstellar magnetic fields.

%*************************************************************************************************************************************8moved here on April 7, 2008

 \subsubsection{Molecules with Hindered Motions}
The most important hindered motion involve quantum mechanical tunneling; such
motions cannot occur in classical mechanics because of energy considerations.
 A prime example of this phenomenon is the motion of the  hydrogen atom attached to oxygen 
in CH$_3$OH, methanol. This H atom can move between 3 positions between the
three H atoms in the CH$_3$ group. Another example is motion of the CH$_3$
group in CH$_3$COOH, methyl formate\index{Molecules!CH$_3$OH}.  These are
dependent on the energy. At low energy these motions do not occur, while at
larger energies are more important.  For both methanol and methyl
formate these motions allow a large number of transitions in the millimeter and sub-millimeter range. 

\begin{figure}%[t]
%\sidecaption
%%\mpicplace{10.7cm}{8.9cm}
%\mpicplace{10.7cm}{8.5cm}
  \includegraphics[width=10.7cm,height=8.5cm]{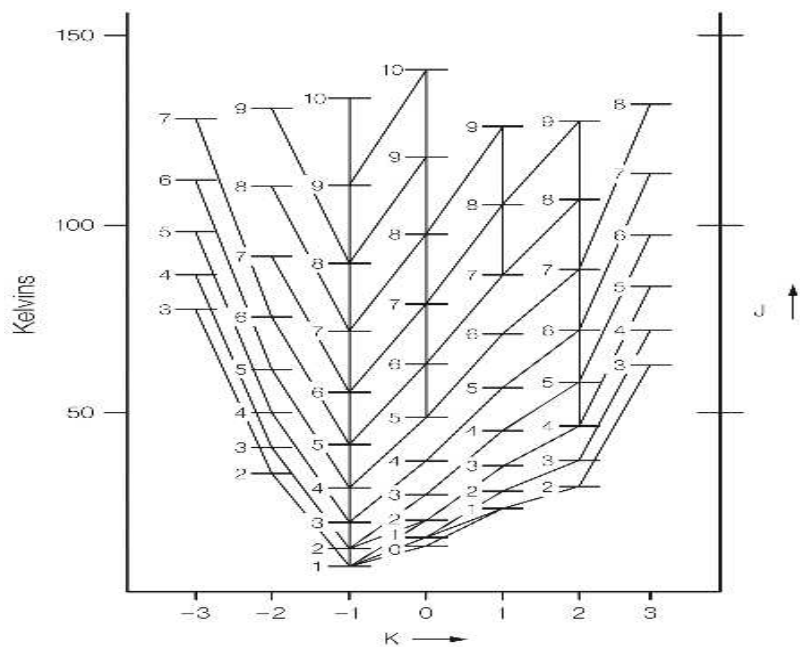}
  \caption[Energy-level diagram of E species of methanol]
   {\label{enlformeth}
    Energy-level diagram of E type methanol, CH$_3$OH.\index{CH$_3$OH energy
levels}
     This is an
asymmetric top. 
The energy-level structure is typical
of an almost prolate symmetric top molecule. The lines connecting the levels show the 
spontaneous transitions with the largest A coefficients from eac level; where the two 
largest A coefficients are within a factor of 2, both transitions are shown \cite{RW5}. }
\end{figure}

The description of energy levels of methyl formate follows the standard nomenclature. For methanol, however, this is not the case, due to historical developments.  These 
energy levels  are labelled as $J_k$, where $K$ can
take on both positive and negative values as in  Fig.\, \ref{enlformeth}. There is a similar scheme for naming energy levels of A type methanol, as $A^{\pm}_k$.  
A and E type methanol  are analogous to ortho and para species, in that these states are not normally coupled  by collisions. 

%14.7**********************************************************************************************************************************************************************
%\vspace{10cm}

Torsionally excited states of methanol have been found in the interstellar medium. These are analogous to transitions from vibrationally 
excited states to the ground state of a molecule. Another
complexity is that because of its structure, there are two dipole moments, along either the c or the a axis %(see the plot in the upper right of  Fig.\, \ref{enlformeth}.

%**************************************************************************************************

%********************************************************************************************************************************************************

%*********************************************************************************************************************************************************

 \section{Astronomical Applications}
In the following, methods to deterimine the parameters of molecular clouds are summarized

\subsection{Kinetic Temperatures}
\index{Temperature!kinetic}
\index{Kinetic temperature}
  A crucial input parameter for the LVG calculations is  $T_{\rm K}$. 
Linewidths of thermally excited species provide a definite value for  $T_{\rm K}$ if the turbulent velocity can be neglected. The relation is 
 \beq
T_{\rm k}=21.2\left({m/m}_{\rm H}\right)\, \left(\uDelta {V}_{\rm t}\right)^2
\eeq
 where $m$ is the molecule, $m_H$ is the mass of hydrogen, and $\uDelta {V}_{\rm t}$ is the FWHP thermal width. Such a relation has been applied to NH$_3$ lines 
(with success!) 
in quiescent dust clouds.

     Historically, $T_{\rm K}$ was obtained at first from the peak intensities
     of rotational transitions of CO, which has a small spontaneous decay
     rate. From the ratio of CO to $^{13}$CO line intensities,
$\tau({\rm CO}) \gg 1$.
After correcting for cloud size, from (\ref{tcotw}) $T_{\rm MB}$ and
$T_{\rm ex}$  are directly related.
In addition, the large optical depths reduce the
critical \index{Critical density}
density by a factor  $\tau$, the line optical depth, so that for the 
$J = 1 \to 0$ line, the
value of the critical density $n^* \approx 50$\,\percc.
Then, the level populations are determined by collisions, so the excitation
     temperature for the $J = 1 \to 0$ transition is $T_{\rm K}$.
Usually, it is assumed that the beam\index{Beam filling factor}
filling factor is  unity; for distant clouds or external galaxies,
the filling factor is clearly less.

An alternative to CO measurements makes use of the fact that
radiative transitions between different $K$ ladders in symmetric top
molecules are forbidden. Then the populations of the different $K$ ladders
are determined by collisions. Thus a method of
determining $T_{\rm K}$ is to use the ratio of populations in different
$K$ ladders of molecules such as \AMM\ at 1.3 cm, or \CHTHCCH\ and  \CHTHCN in the mm/sub-mm range. This is also
approximately true for different $K_a$ ladders of \FORM. Since ratios are
involved, beam filling factors play no role. Even for extended clouds,
$\TKIN$ values from CO and \AMM\ may not agree. This is because \AMM\ is more
easily dissociated so must arise from the cloud interior. Thus for a cloud
heated externally, the $\TKIN$ from CO data will
be larger than that from \AMM. 
 
For \AMM, the rotational transitions,
$(J + 1, K) \rightarrow (J, K)$,
     occur in the far infrared and have Einstein $A$ coefficients of order
     1\,s$^{-1}$; for inversion transitions,
     A values are
     $\sim10^{-7}$\,s$^{-1}$ (see, e.\,g., Table\,2).
  These non metastable states
require extremely high \MOLH\ densities or intense far infrared fields
to be populated. Thus \AMM\
     non metastable states
     ($J>K$) are not suitable for $T_{\rm K}$ determinations.
     Rather, one measures the inversion transitions in different
     metastable\index{Metastable level}
     ($J=K$) levels.  Populations
     cannot be transferred from one metastable state to another
via allowed radiative  transitions. This occurs via collisions, so the
relative populations of metastable levels
are directly related to $T_{\rm K}$.
The column densities are obtained from  the inversion transitions from
different  metastable states, and convert these to column densities using
(\ref{levelpop}) to (\ref{coldentemp}).
 If the \AMM\
     lines are optically thick, one can use the
     ratios of satellite components to main quadrupole
     hyperfine components, in most cases, to determine optical
     depths.
     A large number of  $T_{\rm K}$\ determinations have
     been made using \AMM\ in
     dark dust clouds. Usually these
     involve the ($J,K$)=(1,1) and (2,2) inversion
     transitions.

\subsection{ Linewidths, Radial Motions and Intensity Distributions}
      From the spectra themselves, the linewidths, \dvel, and radial
velocities, \vlsr, give an estimate of motions in the clouds. The
\dvel\  values are a combination
of thermal and turbulent motions. Observations show that the widths
are supersonic
in most cases.  In cold dense cores, motions barely exceed
Doppler thermal
     values. Detailed measurements
     of lines with moderate to large optical depths show that the shapes
are nearly Gaussian. However, simple models
     in which unsaturated line shapes are Gaussians would give flat
topped shapes at high optical depths. This is not found.
     More realistic  models of clouds are those in which shapes
     are determined by the relative
     motion of a large number of small condensations, or clumps, which emit
     optically thick line radiation. If the motions of
 such small clumps are balanced by gravity, one can apply the virial theorem.
Images of isolated sources can be used for comparing with models. One example
is the attempt to characterize the kinetic temperature and the \MOLH\ density
distributions from spectral line or thermal dust emission data.

\subsection{Determinations of H$_2$ Densities}
If a level is populated by collisions\index{Collision} with \MOLH,
the main collision partner, a first approximation
     to the \MOLH\ density can be had if one sets the collision rate,
      $n({\rm H}_2)\langle \sigma \,v \rangle$ equal to the $A$
 coefficient. The brackets
indicate an average over velocities of \MOLH, which are
assumed to be Boltzmann distributed.  The \MOLH\
density for a given transition which will bring $T_{\rm ex}$
midway between the radiation
temperature and $T_{\rm K}$\ is referred to as the
{\em critical \index{Critical density} density}, and is denoted by $n^*$.
That is $n^* \langle \sigma \,v \rangle = A$.
However, this can be at best only an approximate estimate.
For reliable determination of H$_2$ densities n(H$_2$) one must measure at least two spectral lines of a given species, estimates of the kinetic temperature, 
collision rates, and a radiative transport model. One can assume that  the lines are optically thin, and use a statistical equilibrium model, but the 
present approach is to apply the LVG 
model \cite{RW4}.  Clearly, the more lines measured, the more reliable the result. For linear molecules such as 
CO or CS, it is not possible to separate  kinetic temperature and density effects. For example, CO  with  $T_{\rm K}$=10K will have a J=2-1 line much 
weaker than the J=1-0 line no matter how high 
the density. However, if the plot of normalized CO column density versus energy above the ground state  
shows a {\em turn over}, that is a decrease in intensity, it is possible to find a 
unique combination of 
 $T_{\rm K}$ and n(H$_2$).

%***************************************************************8
\subsection{Cloud Masses}
\subsubsection{Virial Masses}
If we assume that only gravity is to be balanced by the motions in a cloud,
then, for a uniform density cloud of radius $R$, in terms
of the line of sight FWHP velocity,
virial\index{Virial!equilibrium} equilibrium requires:
\begin{equation}
\label{tlwvirial}
 \fboxsep2mm\fbox{$ \displaystyle
                \frac{M}{M_\odot}= 250
                \left( \frac{\Delta v_{1/2}}{{\rm km\,s}^{-1}}\right)^2
                                \left(\frac{R}{{\rm pc}}\right)
                  $} \quad .
\end{equation}
Once again very optically
thick lines should not be used in determining masses using
(\ref{tlwvirial}).

Another probe of 
such regions is thermal emission from dust (\ref{cold-dust-col-den}). Given the dust temperature, such data  allow one to derive H$_2$ abundances and cloud masses.

\subsubsection{Masses from Measurements of CO and \THCO}
CO is by far the most widespread molecule with easily measured  transitions.
The excitation of CO is close to LTE and the chemistry
is thought to be ''well understood'', so  measurements of CO and CO isotopomers (and dust emission measurments!) are the most important tool(s) for 
estimating masses of molecular clouds. 

Even if all of the concepts presented in this section are valid,
there may  be uncertainties in the calculation of the
column densities of CO. These arise from several sources which can be
grouped under the general heading {\em non-LTE effects}.
Perhaps most important is the uncertainty in the excitation
temperature. While the $^{12}$CO emission might be
thermalized even at densities $< 100$\,\percc, the less abundant isotopes
may be  sub-thermally excited, i.e., populations
characterized by $T_{\rm ex} < T_{\rm K}$.
(this can be explained using the LVG approximation (\ref{lvg15})). Alternatively, if the cloud in question has no small scale structure, 
$^{13}$CO emission will arise primarily
from the cloud interior, which may be either hotter or cooler than the surface;  the optically thick $^{12}$CO
emission may only reflect  conditions in the cloud surface.
  Another effect is that, although $T_{\rm ex}$ may describe the
population of the $ J = 0$ and $J = 1$ states well, it may not 
for $ J > 1$. That is,  the higher rotational levels might not be thermalized
because their larger Einstein $A$ coefficients lead to a faster depopulation. This lack of information about the population of the upper states leads to an
uncertainty in the partition\index{Partition function} function. Measurements
of other transitions
and use of LVG  models  allow better accuracy.
For most cloud models, LTE gives overestimates of the
true $^{13}$CO column
densities by factors from 1 to 4
depending on the properties of the model
and of the position in the cloud. Thus a
factor of $\sim$\,two uncertainties
should be expected when using  LTE models. The final step to obtain the \MOLH\ column density is to assume a ratio of CO to \MOLH. This is 
generally taken to be 10$^{-4}$. In spite of all these uncertainties, one
most often attempts to relate measurements of the CO column density
to that of H$_2$; estimates made using lines of CO (and
isotopomers) are probably the best method to obtain the H$_2$ column density
and mass of molecular clouds. 

An LVG\index{LVG} treatment of the dependence of the total column density
on the line intensity of the $J=2 \to 1$ line shows that
a simple relation is valid for  $T_{\rm K}$\ from 15\,K to 80\,K, and
$n$(\MOLH) from $\sim 10^3$ to $\sim 10^6$\,\percc. An assumption used in
obtaining this relation is that the ratio of C$^{18}$O to \MOLH\ is $1.7 \times
10^{-7}$, which corresponds to (C/\MOLH) $ =10^4$, and ($^{16}$O/$^{18}$O) $=
500$. The latter ratio is obtained
from isotopic studies for molecular clouds near the Sun.
Then we have
\begin{equation}
          \label{co21colden}
         N_{\rm H_2} = 2.65 \times 10^{21} \,
        \int T_{\rm MB}({\rm C}^{18}{\rm O}, J=2 \to 1) \diff v \,.
\end{equation}
The units of $v$ are \kms, of $T_{\rm MB}$(C$^{18}$O, $J=2 \to 1$) are
Kelvin, and
of $N_{\rm H_2}$ are \percmsq.
This result can be used to determine cloud masses, if the
distance to the cloud is known, by a summation over the
cloud, position by position, to obtain the total number of \MOLH.

  The  total cloud masses obtained from the methods above,
or similar methods, is sometimes referred to as the 'CO mass'; this
terminology can be misleading, but is frequently found in the literature.

\subsubsection{Masses from the X Factor} 
%**********************************Insert X factor
In large scale surveys of the CO $J=1 \to 0$ line in our galaxy and
external galaxies, it has been found, on the basis of a comparison of
CO with \THCO\  maps, that the CO integrated line intensities measure
mass, even though this line is optically thick. The line shapes
and intensity ratios along  different
lines of sight are remarkably similar for
both $^{12}$CO and $^{13}$CO line radiation.
This can be explained
if the total emission depends primarily on
the number of clouds. If so, $^{12}$CO
line measurements can be used to obtain
estimates of $N_{\rm CO}^{12}$.
Observationally, in the disk of our galaxy, the ratio of these two quantities varies remarkably little for
different regions of the sky.
 
This empirical approach has been followed up by a theoretical analysis.
The basic assumption is that the clouds are virial
\index{Virial!objects} objects, with self-gravity
balancing the motions. If these clouds are thought to consist of
a large number of clumps, each with the same temperature, but sub thermally
excited (i.\,e., $T_{\rm ex} < T_{\rm K}$), then from an LVG analysis
of the CO excitation, the peak intensity
of the CO line will increase with $\sqrt{n({\rm H_2})}$, and
the linewidth will also increase by the same factor, as can be seen from
    (\ref{tlwvirial}).
The exact relation between the integrated intensity of the CO $J=1 \to 0$
line and the column density of \MOLH\
must be determined empirically.
Such a relation has also been applied to other galaxies and the center of our galaxy. However, the  environment, such as the ISRF, may be very
different and this may have a large effect on the cloud properties.
For the disk of our galaxy, a frequently used conversion\index{Conversion
factor} factor is:
\begin{eqnarray}
  \label{equationco}
 N_{\rm H_2} &=& X \int T_{\rm MB} ({\rm CO}, J=1 \to 0) \diff v
\nonumber \\
 &=& 2.3 \times 10^{20} \int T_{\rm MB}({\rm CO}, J=1 \to 0) \diff v
\,.
\end{eqnarray}
where $X = 2.3 \times 10^{20}$ and $ N_{\rm H_2}$ is in units
of \percmsq. By summing the intensities over the cloud,
the mass in \solmass\ is
obtained. Strictly speaking, this relation is only valid for  whole
clouds.  The exact value of the conversion factor between CO integrated
line intensity and mass, $X$, is a matter of some dispute.

\subsection{Additional Topics}
\subsubsection{Freezing Out on Grain Surfaces}
For H$_2$ densities
$> 10^6$\,\percc, one might expect a
freezing\index{Freezing out} out of molecules onto grains
for cold, dense regions.  From a  
simplified  theory for  \MOLH\ densities $>10^6$\,\percc, the time for a molecule-grain collision  is 
3000 years, short compared to  other time scales. Then CO 
and most other molecules might be condensed out of the gas phase, so that 
spectroscopic measurements cannot be used as a probe of 
very dense regions. Empirically N$_2$H$^+$, NH$_3$ and H$_2$D$^+$ appear to remain in the gas phase even at low kinetic temperatures and high densities.

%*********************************************************************************************

\subsubsection{Self Shielding of CO}
As is well established, CO is dissociated\index{Dissociation} by line
radiation. Since the  optical depth of
CO is large, this isotopomer will be self-shielded.
If there is no fine spatial structure, selective dissociation
 will cause the extent of $^{12}$CO to be greater than that of
$^{13}$CO, which will be greater than the extent of C$^{18}$O.

% BibTeX users please use
% \bibliographystyle{}
% \bibliography{}
%
% Non-BibTeX users please follow the syntax
% the syntax of "referenc.tex" for your own citations
%%%%%%%%%%%%%%%%%%%%%%%% referenc.tex %%%%%%%%%%%%%%%%%%%%%%%%%%%%%%
% sample references
% "physics"
%
% Use this file as a template for your own input.
%
%%%%%%%%%%%%%%%%%%%%%%%% Springer-Verlag %%%%%%%%%%%%%%%%%%%%%%%%%%

%
% BibTeX users please use
% \bibliographystyle{}
% \bibliography{}

%%%%%%%%%%%%%%%%%%%%%%%%%%%%%%%%%%%%%%%%%%%%%%%%%%%%%%%%%%%%%%%%%%%%%%  }

%%%%%%%%%%%%%%%%%%%%%%%%%%%%%%%%%%%%%%%%%%%%%%%%%%%%%%%%%%%%%%%%%%%%%%

\printindex
\end{document}